\documentclass[amsmath,amssymb,aps,twocolumn,superscriptaddress,prb,10pt]{revtex4-2}

\usepackage{graphicx}
\graphicspath{{figures/}}
\usepackage{hyperref}
\usepackage{MnSymbol}
\usepackage[normalem]{ulem}
\usepackage{float}

\DeclareMathOperator{\Tr}{Tr}
\DeclareMathOperator{\tr}{tr}
\DeclareMathOperator{\Prob}{Prob}
\DeclareMathOperator{\const}{const}

\DeclareMathOperator{\diag}{diag}

\begin{document}

\title{Measurement-induced transitions for interacting fermions}

\author{Igor Poboiko}
\author{Paul P{\"o}pperl}
\author{Igor V. Gornyi}
\author{Alexander D. Mirlin}
\affiliation{\mbox{Institute for Quantum Materials and Technologies, Karlsruhe Institute of Technology, 76131 Karlsruhe, Germany}}
\affiliation{\mbox{Institut f\"ur Theorie der Kondensierten Materie, Karlsruhe Institute of Technology, 76131 Karlsruhe, Germany}}

\date{\today}

\begin{abstract}
Effect of measurements on interacting fermionic systems with particle-number conservation, whose dynamics is governed by a time-independent Hamiltonian, is studied. We develop Keldysh field-theoretical framework  that provides a unified approach to observables characterizing entanglement and charge fluctuations. Within this framework, we derive a replicated Keldysh non-linear sigma model (NLSM), which incorporates boundary conditions specifically designed to produce generating functions for charge cumulants and entanglement entropies directly in the NLSM language.  By using the renormalization-group approach for the NLSM, we determine the phase diagram and the scaling of physical observables. Crucially, the interaction-induced terms in the NLSM action reduce its symmetry, which affects the physics of the problem in a dramatic way. First, this leads to the ``information-charge separation’’: charge cumulants get decoupled from entanglement entropies. Second, the interaction stabilizes the volume-law phase for the entanglement. Third, for spatial dimensionality $d=1$, the interaction stabilizes the phase  with logarithmic growth of charge cumulants (in the thermodynamic limit). Thus, in the presence of interaction, there are measurement transitions in any $d$, at variance with free fermions, for which a $d=1$ system is always in the area-law phase. 
Analytical results are supported by numerical simulations using time-dependent variational principles for matrix product states, which, in particular, confirm the separation of information and charge as a hallmark of the delocalized phase.
 
\end{abstract}

\maketitle

\section{Introduction}
\label{sec:intro}

The problem of the influence of quantum measurements on properties of a many-body quantum system (including, in particular, entanglement and charge correlations) is attracting much attention of researchers. This area of research is part of a vibrant field of quantum technologies, with the great interest in it motivated in particular by rapid developments in quantum information processing. 
 
It has been discovered that competition between unitary dynamics (that tends to increase entanglement) and non-unitary stochastic evolution due to quantum measurements of local observables (that tends to reduce entanglement) may lead to measurement-induced entanglement phase transitions. Initially, the work on these transitions was carried out in the area of quantum circuits \cite{Li2018a, Skinner2019a, Chan2019a, Szyniszewski2019a, Li2019a,  Bao2020a, Choi2020a, 
Gullans2020a, 
Gullans2020b, 
Jian2020a, Zabalo2020a, 
Iaconis2020a, 
Turkeshi2020a, Szyniszewski2020, Zhang2020c, Nahum2021a, Ippoliti2021a,    Ippoliti2021b, Lavasani2021a, 
Lavasani2021b, 
Sang2021a, Li2021, Fisher2022, 
Block2022a, 
Sharma2022, Barratt2022, Agrawal2022,
Jian2023, Kelly2023, Agrawal2023, Zabalo2023, Pixley2024, Halpern2023,Yajima2024}. It was however quickly understood that these phenomena are much more ubiquitous, with the work over the past few years including fermionic systems \cite{Cao2019a,Alberton2021a,Chen2020a,Tang2021a,Coppola2022,Ladewig2022,Carollo2022,Buchhold2022,Yang2022,Szyniszewski2022,Buchhold2021a,Buchhold2021a,VanRegemortel2021a,Mueller2022,Youenn2023,Loio2023,Turkeshi2022b, Lumia2023,Russomanno2023,Poboiko2023a, Poboiko2023b,chahine2023entanglement,Zilberberg2023,Jin2024,Turkeshi2024,Piccitto2024,starchl2024generalized,Soares2024,FavaNahum2024,Dalmonte2024,Schiro2024,Szyniszewski2024}, Majorana models \cite{Kells2023,Fava2023,Swann2023, Merritt2023,Klocke2023,Klocke2024}, spin systems \cite{Lang2020a,
Rossini2020a,
Biella2021a,Turkeshi2021,Tirrito2022,Piccitto2022,
Yang2023,
Weinstein2023,
Murciano2023,Sierant2022a,Turkeshi2022a,Xing2023,Cecile2024,Li2024,Fazio2024,Leung2024}, bosonic models \cite{Tang2020a,Goto2020a,Fuji2020a,Fuji2021,Minoguchi2022,Doggen2022a,Doggen2023,doggen2023ancilla,Yokomizo2024}, disordered systems with Anderson or many-body localization \cite{Szyniszewski2022,Lunt2020a,Yamamoto2023}, and models of Sachdev-Ye-Kitaev type~\cite{Jian2021a,Altland2022}, as well as zero-dimensional models~\cite{snizhko_2020,Paul2024,Gerbino2024,Santini2024}. 
While most of the studies were computational, analytical progress has been achieved for some models. Recent works on trapped-ion systems  \cite{Noel2022a,Agrawal2023} and superconducting quantum processors \cite{Koh2022,Hoke2023} reported experimental realizations of setups for studying measurement-induced phase transitions.
 
Much of the work on measurement-induced quantum-information physics was dealing with one-dimensional (1D) random quantum circuits
\cite{Li2018a, Skinner2019a,Szyniszewski2019a,Chan2019a,Li2019a,Choi2020a,Bao2020a, Gullans2020a, Zhang2020c,Turkeshi2020a,
Iaconis2020a,
Zabalo2020a,Jian2020a,Ippoliti2021a, Ippoliti2021b, Lavasani2021a,Sang2021a,Nahum2021a}, see recent reviews \cite{Potter2022,Fisher2022}. In most of these works, a transition between the area-law and volume-law phases was found numerically. An analogous result was also obtained analytically in a limiting case of infinite Hilbert-space dimensionality of individual elements forming a circuit, by a mapping onto known statistical mechanics models \cite{Skinner2019a,Bao2020a,Jian2020a,
Iaconis2020a,
Lavasani2021a,Fisher2022}. 
Evidences of the volume-law to area-law entanglement transition were also found numerically for interacting 1D many-body Hamiltonian models \cite{Tang2020a,Goto2020a,Fuji2020a,Doggen2022a,Doggen2023,doggen2023ancilla}.
In particular, Refs.~\cite{Doggen2022a,Doggen2023,doggen2023ancilla} used matrix-product states (MPS) to study larger systems that are not accessible via exact diagonalization.

Another class of systems that recently attracted much attention in the context of measurement-induced physics is non-interacting fermionic systems (and related Ising models), with local measurements preserving the Gaussian character of the system (i.e., the Slater-determinant form of the wave function for a pure system)
\cite{Cao2019a,Alberton2021a,Chen2020a,Tang2021a,Coppola2022,Ladewig2022,Carollo2022,Buchhold2022,Yang2022,Szyniszewski2022,Buchhold2021a,Buchhold2021a,VanRegemortel2021a,Youenn2023,Loio2023,Turkeshi2022b,Kells2023, Lumia2023, Poboiko2023a,Poboiko2023b}. While initial results were rather controversial, major progress was achieved recently. Specifically, in Ref.~\cite{Poboiko2023a}, 
the problem in $d$ spatial dimensions was mapped onto a non-linear sigma model (NLSM) field theory in $d+1$ dimensions.
On the semiclassical level, this field theory (which is a close relative of theories of Anderson localization) yields a logarithmic scaling for the entanglement entropy. 
 
However, the one-loop renormalization group (RG) analysis shows that this logarithmic growth is affected by ``weak-localization corrections'' and saturates even for very rare measurements. 
Thus, the system is in the area-law phase even for rare measurements but the crossover from the intermediate logarithmic behavior to the asymptotic area-law behavior takes place at exponentially large system sizes. This is analogous to a crossover from weak localization  to strong Anderson localization in weakly disordered two-dimensional systems. 
In $d > 1$ dimensions, this analysis  \cite{Poboiko2023b} predicts a transition (bearing similarity to Anderson transition in $d+1$ dimensions) between a phase with the area$\times$log ($\ell^{d-1} \ln \ell$) growth of the entanglement entropy with subsystem size $\ell$ and the area-law phase ($\ell^{d-1}$). 
These analytical predictions were confirmed by numerical studies in $d=1$ and $d=2$ dimensions 
\cite{Poboiko2023a,Poboiko2023b}. Closely related results were obtained in Refs.~\cite{chahine2023entanglement,starchl2024generalized,FavaNahum2024}.

In Refs.~\cite{Jian2023,Fava2023}, models of monitored Majorana fermions were studied by mapping onto 
an NLSM of a different symmetry class compared to models of complex fermions characterized by a conserved charge (particle number) discussed above. It was found that the Majorana models are characterized by ``weak antilocalization'' behavior, which should be contrasted to the ``weak-localization'' behavior for models with conserved charge. For $d=1$ Majorana models, this results in a phase transition between an area-law phase and a phase with $\ln^2 \ell$ scaling of the entanglement entropy \cite{Fava2023}. This, in particular, emphasizes the crucial role of symmetries in the problem of monitored systems.

As mentioned above, works on random quantum circuits suggest stabilization of the volume-law phase for entanglement and a logarithmic scaling of the entanglement entropy at the transition to the area-law phase. This is in contrast to the analytically predicted behavior for free fermions, where no volume-law phase exists.
Generic random quantum circuits can be viewed as intrinsically interacting models.
It is thus important to explore how interactions affect properties of the measurement-induced phases and phase transitions for a conventional model of fermions with a time-independent Hamiltonian.
The NLSM approach to Anderson localization transitions has been generalized also to problems involving electron-electron interactions
\cite{Finkelstein1990,Finkelstein2023scale-dependent,BelitzKirkpatrick1994,KamenevLevchenko}.
The analogy between the measurement problem for free fermions and the theory of Anderson localization naturally calls for generalizing the NLSM framework to the case of interacting fermions.

From the symmetry perspective, the model of free fermions is a special case, which is characterized by an extra symmetry (operative in the replica space \cite{Poboiko2023a, Poboiko2023b, starchl2024generalized, FavaNahum2024}) compared to the interacting case (and the case of generic random quantum circuits).
In particular, the stability of the area-law phase for free fermions in one dimension in the thermodynamic limit is guaranteed by the presence of Goldstone modes that give rise to ``localization'' of entanglement \cite{Poboiko2023a}. Interactions added to this model are expected to break this specific free-fermion symmetry and open the gap in some soft modes, in a certain similarity to dephasing in mesoscopic systems. As a result, emergence of the volume-law phase for the entanglement entropy at not-too-frequent measurements can be anticipated, in analogy with generic quantum circuits. For strong monitoring, the area-law phase is still present, corresponding to the quantum Zeno effect. Thus, a transition between area-law and volume-law entanglement phases could be expected for interacting fermions.

An important question that has attracted considerable attention concerns the relation between charge fluctuations and information (entanglement). 
For random quantum circuits, it was argued that there are two distinct phase transitions that were termed ``charge sharpening transition'' and ``purification transition'', respectively \cite{Agrawal2022,Barratt2022,Agrawal2023,Pixley2024} (see also \cite{Halpern2023,Oshima2023}). Contrary to this, for free fermion models with measurements preserving the Gaussian character of a state, the charge and information are in direct relation (the second cumulant of particle number is proportional to the entanglement entropy), and thus exhibit the same behavior. It is, therefore, desirable to explore the measurement-induced phases and transitions between them in an interacting Hamiltonian model, for which a tractable analytical theory can be developed.

\begin{figure}[t]
\includegraphics[width=\columnwidth]{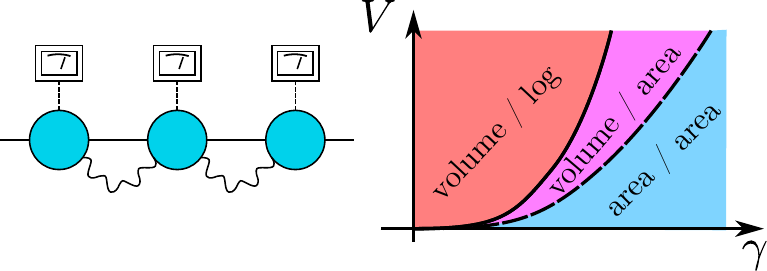}
    \caption{{\it Left panel:} Schematic representation of a $d=1$ model, which includes nearest-neighbor hoppings (straight lines) and interactions (wavy lines), as well as a particle-number measurement apparatus attached to each site. {\it Right panel:} Phase diagram in the parameter plane of measurement rate $\gamma$ and interaction strength $V$, as obtained for $d=1$. From left to right: a phase with volume-law scaling for the entanglement entropy and logarithmic scaling for charge fluctuations (red),  an intermediate phase with volume-law scaling for entanglement entropy and area-law scaling for charge fluctuations (magenta), and  an area-law phase for both quantities (blue). Detailed description is provided in Sec. \ref{sec:RG}.
    \vspace{-5mm}}
    \label{fig:scheme}
\end{figure}

The goal of this paper is to understand the evolution of entanglement and charge fluctuations when the interaction is ``switched on'' in a free-fermion model with particle-number conservation. 
Let us emphasize that charge fluctuations play a special role in models under consideration since the charge is the only conserved quantity. Indeed, both energy and momentum conservation laws are violated by quantum measurements.
On the analytical side, we derive an NLSM field theory for interacting fermions and analyze it by RG means.  This allows us to determine the phase diagram in the plane spanned by measurement rate and interaction. This phase diagram involves the transitions between an area-law phase and a phase with volume-law behavior for the entanglement entropy and between the area-law phase and the area$\times$log ($\ell^{d-1}\ln \ell$) phase for the particle-number cumulant. In particular, for the $d=1$ model, we demonstrate that interactions at not-too-frequent measurements destabilize the single area-law phase obtained previously for free fermions, giving rise to measurement-induced transitions in one dimension (Fig.~\ref{fig:scheme}).

These analytical results are corroborated by a numerical study. Exact diagonalization turns out to be insufficient for our goals, in view of system-size limitations. We thus use the time-dependent variational principle (TDVP) for MPS that allows us to study considerably larger systems. The MPS-TDVP approach reveals a clear information-charge separation as a hallmark of the phase with ``delocalized'' information and charge fluctuations. The numerically observed behavior of the entropy and the charge cumulant is consistent with analytical predictions. We also carry out time-dependent Hartree-Fock (TDHF) simulations as a complementary tool to estimate the position of the transition between the area-law phase and the logarithmic behavior for charge fluctuations. 

The organization of the paper is as follows.
In Sec.~\ref{sec:S2}, we discuss the basics of generalized measurements for fermionic systems, develop the replicated Keldysh approach for monitored fermions, analyze the symmetries of the model, and introduce the tools for calculating the entanglement entropies and statistics of charge fluctuations within the field-theoretical approach. Section~\ref{Sec3:non-int} outlines the derivation of the effective field theory---NLSM---for monitored non-interacting fermions of different symmetry classes. We further introduce the boundary conditions for the NLSM fields, which are utilized for the calculation of the relevant observables, and demonstrate the relation between entanglement and charge fluctuations for Gaussian states. The non-interacting NLSM is analyzed first in terms of a semiclassical approximation and then employing the RG approach. We include interactions between fermions into the NLSM framework in Sec.~\ref{Sec:Interaction}. The action of the interacting NLSM is presented in Sec.~\ref{sec:4A}, where we put particular focus on the symmetry breaking on the NLSM manifold caused by the interaction-induced ``effective potential''.
A semiclassical analysis of the model is used to describe the effect of interactions on charge fluctuations (Sec.~\ref{sec:Gaussian_approx}) and entanglement (Sec.~\ref{sec:EEInt}). 
In Sec.~\ref{sec:4D}, we discuss the observed phenomenon of charge-information separation. Section \ref{sec:RG} is devoted to the RG analysis of the action of the interacting NLSM, which is then used to describe the size-dependence of the observables and construct the phase diagrams
(shown in Sec.~\ref{sec:phase-dia}) for monitored interacting fermions in dimensions $d=1$ and $d>1$.
The results of our numerical simulations are presented in Sec.~\ref{sec:numerics}.
Finally, in Sec.~\ref{sec:Summary}, we summarize our main findings, compare the obtained results with those available in the literature, and outline possible directions for future studies that are opened by our work. Technical details are relegated to Appendixes \ref{sec:appendix:NLSM}--\ref{sec:appendix:numerical-details}.

\section{Microscopic description of monitored fermions}
\label{sec:S2}

We start by introducing general formalism for the microscopic description of the fermionic models subjected to monitoring. We develop this formalism in terms of quantum trajectories, whose statistics will determine various observables that are not described by the average density matrices (Lindblad formalism), in particular, those describing entanglement properties of the system. Importantly, the consideration of individual quantum trajectories is extremely advantageous for the symmetry analysis of the problem.
In what follows, this formalism will be employed for both non-interacting and interacting fermions. 

\subsection{Generalized measurements}

Consider the evolution of a generic interacting fermionic lattice system with a conserved particle number, which is subjected to quantum measurements. This evolution consists of two ingredients. The first one is the continuous unitary evolution governed by the Hamiltonian $\hat{H}$ that consists of finite-range tight-binding hopping and two-particle interaction $\hat{H}=\hat{H}_{0}+\hat{H}_{\text{int}}$, with
\begin{align}
\hat{H}_{0}&=-\frac{1}{2}\sum_{\boldsymbol{x}\boldsymbol{x}^{\prime}}\left[J(\boldsymbol{x},\boldsymbol{x}^{\prime})\hat{\psi}^{\dagger}(\boldsymbol{x})\hat{\psi}(\boldsymbol{x}^{\prime})+\text{H.c.}\right],
\label{eq:H0}
\\
\hat{H}_{\text{int}}&=\sum_{\boldsymbol{x}\boldsymbol{x}^{\prime}}V(\boldsymbol{x},\boldsymbol{x}^{\prime})\hat{\psi}^{\dagger}(\boldsymbol{x})\hat{\psi}^{\dagger}(\boldsymbol{x}^{\prime})\hat{\psi}(\boldsymbol{x}^{\prime})\hat{\psi}(\boldsymbol{x}).
\label{eq:Hint}
\end{align}
We will focus on systems with translation invariance, where $J$ and $V$ are functions of a coordinate difference $(\boldsymbol{x}-\boldsymbol{x}^\prime)$ only.

The second ingredient of the evolution is introduced by generalized measurements.  To define them, we consider a set of observables enumerated by index $i$; each measurement of an observable $i$ can produce one of the outcomes enumerated by index $\alpha$. We describe these generalized measurements by Kraus operators $\hat{\mathbb{K}}_{i, \alpha}$ that form a complete set for each observable: $\sum_\alpha \hat{\mathbb{K}}_{i, \alpha}^\dagger \hat{\mathbb{K}}_{i,\alpha} = \hat{\mathbb{I}}$. 
The probability of outcome $\alpha$ for $i$-th observable is given by Born's rule:
\begin{equation}
\label{eq:BornRule}
P_{i,\alpha}(t)\equiv\left\langle\Psi(t-0)\right|\hat{\mathbb{K}}_{i,\alpha}^{\dagger}\hat{\mathbb{K}}_{i,\alpha}\left|\Psi(t-0)\right\rangle,
\end{equation}
where $\left.|\Psi(t-0)\right\rangle$ is the pure many-body state at the time right before the measurement.
After such a measurement, the wavefunction undergoes an instantaneous quantum jump originating from the von Neumann wavefunction collapse:
\begin{equation}
\label{eq:QuantumJump}
\left|\Psi(t+0)\right>=\frac{\hat{\mathbb{K}}_{i,\alpha}}{\sqrt{P_{i,\alpha}(t)}}\left|\Psi(t-0)\right>
\end{equation}
The measurements are assumed to happen at random times: for each observable $i$, we fix its average measurement rate $\gamma_i$, such that measurement times $\{t_m\}$ are sampled from the Poissonian distribution.

At a time $t_0$, the system is prepared at some initial pure state $\left|\Psi_{0}\right>$, and then evolves according to the described protocol up to time $t_f$. For a fixed \emph{quantum trajectory} ${\cal T} \equiv \{i_m, t_m, \alpha_m\}$---i.e., for a given set of measurement times $t_m$, measured observables $i_m$, and measurement outcomes $\alpha_m$, with $m= 1,2,\ldots$---
the system at time $t_f$ is described  by a pure-state wavefunction $\left|\Psi_{\cal T}\right>$. 
Each quantum trajectory is characterized by a probability originating from randomness of measurements and their outcomes. One thus gets a statistical ensemble of final pure states $\left|\Psi_{\cal T}\right>$. 
We will be interested in various statistical properties of this ensemble, as will be described below.

Formally, we will be interested in describing the ensemble in the steady-state regime defined by $T = t_f - t_0 \to \infty$. It is worth noting that the limits $\gamma \to 0$ and $T \to \infty$ do not commute: the behavior of system at a weak measurement rate (small but finite $\gamma$) is drastically different from the behavior at $\gamma=0$ 
 because, for any non-zero $\gamma$, the system evolves at a sufficiently long time $T$  to a measurement-induced steady state.

\subsection{Observables}
\label{sec:Observables}

Within the context of measurement-induced dynamics and, in particular, measurement-induced phase transitions (MIPTs), it is crucial to separate quantum fluctuations inherent to a quantum state described by wavefunction $\left|\Psi_{\cal T}\right>$ from statistical fluctuations due to different realization of quantum trajectories. For a given quantum trajectory ${\cal T}$, we will denote a quantum average of an arbitrary observable ${\cal O}$ over the wavefunction $\left|\Psi_{\cal T}\right>$ (or, equivalently, over the pure-state density matrix $\hat{\rho}=\left|\Psi_{\cal T}\right\rangle\left\langle\Psi_{\cal T}\right|$) by angular brackets:
\begin{equation}
\left\langle {\cal O}\right\rangle \equiv\left\langle\Psi_{\cal T}\right|\hat{{\cal O}}\left|\Psi_{\cal T}\right\rangle=\Tr(\hat{\rho}\, \hat{{\cal O}}).
\label{quantum-average}
\end{equation}
To explicitly emphasize the dependence on a quantum trajectory, we will also use the notation ${\cal O}({\cal T})$ for the quantum average
$\left\langle {\cal O}\right\rangle$.  
Note that, in order to perform the quantum averaging \eqref{quantum-average} {\it in experiment}, one needs to be able to reproduce the same state $\left|\Psi_{\cal T}\right>$, i.e., the same quantum trajectory ${\cal T}$, multiple times. The trajectories, however, are random, and even for a fixed set of measurement times (which can be controlled in the experiment), there are exponentially many possible outcomes $\{\alpha_m\}$. Thus, such reproduction requires repetition of the protocol exponentially many times. This issue is infamously known as a \textit{postselection problem} (see, e.g., Ref.~\cite{Garratt2024} for a recent discussion of these issues).

Having determined the quantum average ${\cal O}({\cal T})$ for an individual quantum trajectory ${\cal T}$, we then perform averaging over trajectories, which by definition includes averaging over measurement times $t_m$ for chosen observables $i_m$, and over outcomes $\alpha_m$. We will denote this averaging over quantum trajectories by an overline. Splitting explicitly the average over the measurement outcomes with the Born-rule weight, we write
\begin{equation}
\label{eq:TrajectoryAverage}
\overline{{\cal O}({\cal T})}\equiv\sum_{{\cal T}}\text{Prob}(\{\alpha_{m}\}|\{i_{m},t_{m}\}){\cal O}({\cal T}).
\end{equation}
Here we introduced the notation 
$\sum_{{\cal T}}$ that includes summation over the measurement outcomes $\alpha_m$, as well as averaging over measurement times $t_m$ of the observables $i_m$,
\begin{equation}
\label{eq:TrajectorySummation}
    \sum_{{\cal T}}{\cal F}({\cal T})\equiv\overline{\sum_{\{\alpha_{m}\}}{\cal F}({\cal T})}^{(i_{m},t_{m})},
\end{equation}
see Eq.~\eqref{eq:sum-t-notation} in Appendix \ref{sec:appendix:Summation} for a more explicit form of Eq.~\eqref{eq:TrajectorySummation}.

In the present paper, we will mainly focus on two quantities, both requiring the partition of the whole system into a subsystem $A$ and the rest of the system $\bar{A}$. The first quantity is the $N$-th R{\'e}nyi entropy ${\cal S}_{A}^{(N)}$, which we define through the $N$-th purity $\mathcal{P}_{A}^{(N)}$  of the reduced density matrix of subsystem $A$, \begin{equation}
\hat{\rho}_{A}=\Tr_{\bar{A}}\left|\Psi_{\cal T}\right\rangle\left\langle\Psi_{\cal T}\right|,
\end{equation} as follows:
\begin{align}
\label{eq:Purity}
\mathcal{P}_{A}^{(N)}&\equiv\overline{\Tr\left(\hat{\rho}_{A}^{N}\right)}, \\
\label{eq:RenyiEntanglementEntropy}
{\cal S}_{A}^{(N)}&\equiv-\frac{1}{N-1}\ln\mathcal{P}_{A}^{(N)}.
\end{align}
In the limit $N \to 1$, it reduces to the standard entanglement entropy:
\begin{equation}
\label{eq:EntanglementEntropy}
{\cal S}_{A} \equiv -\overline{\Tr(\hat{\rho}_{A}\ln\hat{\rho}_{A})} = \lim_{N\to1}{\cal S}_{A}^{(N)}.
\end{equation}

The second object of our interest is the $N$-th particle-number generating function $Z_A^{(N)}(\lambda)$, associated with the cumulant-generating function $\chi_{A}^{(N)}(\lambda)$, which we define via the following relations:
\begin{align}
\label{eq:GeneratingFunction}
Z_{A}^{(N)}(\lambda)&\equiv\overline{\left\langle e^{i\lambda\hat{N}_{A}}\right\rangle ^{N-1}\left\langle e^{-i(N-1)\lambda\hat{N}_{A}}\right\rangle },\\
\label{eq:CumulantGeneratingFunction}
\chi_{A}^{(N)}(\lambda)&\equiv-\frac{i}{N-1}\ln Z_{A}^{(N)}.
\end{align}
Similarly to the R{\'e}nyi entropy, in the limit $N \to 1$, $Z_A^{(N)}$ reduces to the standard  generating function for full counting statistics (FCS):
\begin{equation}
    \label{eq:FCS}
    \chi_{A}(\lambda)\equiv(-i)\, \overline{\ln\left\langle e^{i\lambda\left[\hat{N}_{A}-\left\langle \hat{N}_{A}\right\rangle \right]}\right\rangle }=\lim_{N\to1}\chi_{A}^{(N)}(\lambda).
\end{equation}
Both  $S_A^{(N)}$ and $\chi_{A}^{(N)}(\lambda)$ are nonlinear functionals of the reduced density matrix, which makes averaging these objects over quantum trajectories nontrivial.
It is worth emphasizing that the ensemble-averaged density matrix has always a simple, infinite-temperature form and does not contain information about the observables of interest.

\subsection{Replica trick}

For an integer $N$, both the $N$-th purity \eqref{eq:Purity} and the $N$-th particle-number generating function \eqref{eq:GeneratingFunction} can be expressed via replicated density matrix:
\begin{equation}
\hat{\rho}_{N}\equiv\overline{\otimes_{r=1}^{N}\hat{\rho}_{r}}.
\end{equation}
One of the main analytical complications is that the non-unitary part of the evolution involves nonlinearity because of the presence of the denominator in Eq.~\eqref{eq:QuantumJump}. This complication can be overcome by introducing a non-normalized wavefunction $\left|\widetilde{\Psi}(t)\right>$ and the corresponding non-normalized density matrix 
\begin{equation}
\hat{D}({\cal T})=\left|\widetilde{\Psi}_{\cal T}(t)\right\rangle\left\langle\widetilde{\Psi}_{\cal T}(t)\right|, 
\end{equation}
whose evolution does not contain such a denominator and is thus linear. The normalization then can be restored at the end of evolution:
\begin{equation}
\left|\Psi_{\cal T}\right>\equiv\frac{\left|\widetilde{\Psi}_{\cal T}\right>}{\sqrt{\left\langle \widetilde{\Psi}_{\cal T}|\widetilde{\Psi}_{\cal T}\right\rangle }},\quad \hat{\rho}_{\cal T}\equiv\frac{\hat{D}({\cal T})}{\Tr\hat{D}({\cal T})}.
\end{equation}
Conveniently, the normalization factor can be related to the Born-rule probability that enters Eq.~\eqref{eq:TrajectoryAverage}:
\begin{equation}
\label{eq:BornRuleD}
\Prob\left(\{\alpha_{m}\}|\{i_{m},t_{m}\}\right)=\left\langle \widetilde{\Psi}_{\cal T}|\widetilde{\Psi}_{\cal T}\right\rangle =\Tr\hat{D}({\cal T}),
\end{equation}
allowing us to write
\begin{equation}
\label{eq:NReplicaDensityMatrix}
\hat{\rho}_{N}=\sum_{{\cal T}}\Tr^{1-N}\hat{D}({\cal T})\,\otimes_{r=1}^{N}\hat{D}_{r}({\cal T}),
\end{equation}
where $\hat{D}_{r}({\cal T})$ is the $r$th copy of a replicated matrix $\hat{D}({\cal T})$.

In order to calculate $\hat{\rho}_{N}$ as given by Eq.~\eqref{eq:NReplicaDensityMatrix},
we utilize the replica trick. Specifically, we define the density matrix built on $R \geq N$ copies of the matrix $\hat{D}$:
\begin{equation}
\label{eq:ReplicatedDensityMatrix}
\hat{\rho}_{R,N}=\Tr_{r=N+1,\dots,R}\sum_{{\cal T}}\otimes_{r=1}^{R}\hat{D}_{r}({\cal T})
\end{equation}
where the partial trace over $(R-N)$ replicas yields the prefactor $\Tr^{R-N}\hat{D}({\cal T})$. Analytic continuation from $R \geq N$ to $R \to 1$ then reduces the $R$-replicated density matrix to 
Eq.~\eqref{eq:NReplicaDensityMatrix}:
\begin{equation}
\hat{\rho}_{N}\equiv\lim_{R\to1}\hat{\rho}_{R,N}.
\end{equation}

The replica limit $R \to 1$ originates from the Born-rule probability \eqref{eq:BornRuleD} and should be contrasted with the more conventional replica limit $R \to 0$ for systems with quenched disorder. It should also be emphasized that the $R \to 1$ limit is independent of $N$. In particular, it is not related to the $N \to 1$ limit which has to be taken to calculate the entanglement entropy from R{\'e}nyi entropies, as in Eq.~\eqref{eq:EntanglementEntropy}, or to calculate the FCS, as in Eq.~\eqref{eq:FCS}. For this latter limit, an additional analytical continuation from integer $N \ge 2$ to $N \to 1$ is required.

\subsection{Keldysh path integral}

The non-normalized density matrix $\hat{D}({\cal T})$ can be conveniently represented using the notation of the time ordering along the standard Keldysh time contour $${\cal C}=(t_{f},t_{0})_{-}\cup(t_{0},t_{f})_{+}$$ consisting of the backward-moving branch (denoted by ``$-$'') and the forward-moving branch (``$+$''):
\begin{multline}
\label{eq:DMatrixKeldyshContour}
\hat{D}({\cal T})={\cal T}_{{\cal C}}\Bigg\{\hat{\rho}_{0}\exp\left(-i\int_{{\cal C}}dt\,\hat{H}(t)\right)\\\times\prod_{m}\hat{\mathbb{K}}_{i_{m},\alpha_{m}}^{(+)}(t_{m})\hat{\mathbb{K}}_{i_{m},\alpha_{m}}^{(-)\dagger}(t_{m})\Bigg\}.
\end{multline}
Here, ${\cal T}_{\cal C}$ denotes time ordering along the contour, the $(+)$ and $(-)$ superscripts mark the branches of the Keldysh contour where the operator is placed, and $\hat{\rho}_{0}$ is the initial density matrix.

We proceed with the standard Keldysh coherent-state path integral derivation. Since we deal with fermions, the corresponding fields will be Grassmann variables.
It is convenient to present matrix elements of the Kraus operators in the coherent-state basis in the exponential form
\begin{equation}
\label{eq:KraussOperatorCoherent}
\left<\psi\right|\hat{\mathbb{K}}_{i,\alpha}\left|\chi\right>=\exp\left[-{\cal M}_{i,\alpha}(\psi^{\ast},\chi)\right],
\end{equation}
which allows us to write
\begin{equation}
\label{eq:DMatrixPathIntegral}
\hat{D}({\cal T})=\int{\cal D}\psi^{\ast}{\cal D}\psi\,\exp\left[i\int_{{\cal C}}dt\,L(t,\psi^{\ast}(t),\psi(t), {\cal T})\right].
\end{equation}
The Lagrangian $L = L_0 + L_M$ consists of the part describing unitary evolution,
\begin{equation}
L_{0}(\psi^{\ast},\psi)=\psi^{\ast}i\partial_{t}\psi-H(\psi^{\ast},\psi),
\end{equation}
and the measurement-induced part,
\begin{eqnarray}
L_{M}(t,\psi^{\ast},\psi, {\cal T}) &=& \sum_{m}\delta(t-t_{m}) \nonumber \\ 
&\times& \begin{cases}
\phantom{-}i{\cal M}_{i_{m},\alpha_{m}}(\psi^{\ast},\psi), & t\in{\cal C}_{+} \,,\\
-i{\cal M}_{i_{m},\alpha_{m}}^{\dagger}(\psi^{\ast},\psi), & t\in{\cal C}_{-} \,.
\end{cases}
\end{eqnarray}
The information about the initial state can then be encoded in the boundary conditions at $t = t_0$.

\subsection{Boundary conditions}

The replica representation of the average density matrix, Eq.~\eqref{eq:ReplicatedDensityMatrix}, together with the path integral representation for individual non-normalized density matrices $\hat{D}_r$, implies that the fermionic action should also be replicated, so that the whole theory is defined on $R$ copies of the Keldysh contour with $4R$ fermionic fields $\{\psi_{r,\pm}^\ast, \psi_{r,\pm}\}_{r=1}^{R}$:
\begin{equation}
\label{eq:DMatrixReplicatedAction}
\otimes_{r=1}^{R}\hat{D}({\cal T})=\int{\cal D}\psi^{\ast}{\cal D}\psi\exp\left[i\sum_{r=1}^{R}S\left(\psi_{r}^{\ast},\psi_{r}, {\cal T}\right)\right],
\end{equation}
where the action for each individual replica is defined in Eq.~\eqref{eq:DMatrixPathIntegral}.

In the time domain, the Keldysh contours run from time $t_0 \to -\infty$ up to time $t_f$, so that the fermionic field theory has to be supplied with the boundary conditions at $t_f$. 
Conveniently, both quantities of interest---the purity \eqref{eq:Purity} and the FCS generating functional \eqref{eq:GeneratingFunction}---can be encoded in the boundary conditions.

Let us start with FCS and define a replicated density generating functional as follows:
\begin{multline}
Z[\hat{\lambda},{\cal T}]\equiv\prod_{r=1}^{R}\Tr\left\{\hat{D}({\cal T})\exp\left[i\sum_{\boldsymbol{x}}\lambda_{r}(\boldsymbol{x})\hat{\psi}^{\dagger}(\boldsymbol{x})\hat{\psi}(\boldsymbol{x})\right]\right\}\\=\Big\langle \exp\left(i\bar{\psi}_{-}\hat{\lambda}\psi_{+}\right)\Big\rangle.
\end{multline}
Here angular brackets denote the path-integral average with the action \eqref{eq:DMatrixReplicatedAction}, the fields are taken at time $t = t_f$, and we have introduced a source matrix 
\begin{equation}
\hat{\lambda}(\boldsymbol{x}) = \diag\left[\{\lambda_r(\boldsymbol{x})\}_{r=1}^{R}\right]
\end{equation}
that is diagonal both in replica space and in coordinate space. Such source can be gauged out of the integrand to the boundary conditions in the path integral \eqref{eq:DMatrixReplicatedAction}:
\begin{eqnarray}
\label{eq:TFCS}
  &&  \psi_{-}(\boldsymbol{x},t=t_f)=\hat{\mathbb{T}}_{\text{FCS}}(\boldsymbol{x})\psi_{+}(\boldsymbol{x},t=t_f), 
  \nonumber \\ 
  && \hat{\mathbb{T}}_{\text{FCS}}(\boldsymbol{x})\equiv \exp[i\hat{\lambda}(\boldsymbol{x})].
\end{eqnarray}
Furthermore, in accordance with Eq.~\eqref{eq:ReplicatedDensityMatrix}, the FCS \eqref{eq:GeneratingFunction} is then given by:
\begin{equation}
Z_{A}^{(N)}(\lambda)=\lim_{R\to1} Z_{\text{FCS}}[\hat{\lambda}],
\end{equation}
with the specific choice of sources:
\begin{align}
\label{eq:DensitySource:FCS}
\begin{split}\lambda_{1,\dots,N-1}(\boldsymbol{x}) & =\lambda\,{\cal I}_{A}(\boldsymbol{x}), \\
\lambda_{N}(\boldsymbol{x}) & =-\lambda(N-1){\cal I}_{A}(\boldsymbol{x}),\\
\lambda_{N+1,\dots,R}(\boldsymbol{x}) & =0,
\end{split}
\end{align}
where  ${\cal I}_{A}(\boldsymbol{x})$ is the indicator function defined as
\begin{equation}
{\cal I}_{A}(\boldsymbol{x})=\begin{cases}
1, & \boldsymbol{x}\in A \,,\\
0, & \boldsymbol{x}\notin A \,.
\end{cases}
\label{eq:indicator-I-A}
\end{equation}

Turning to the calculation of purity, we define
\begin{equation}
\label{eq:gamma-A-N-T}
\mathcal{P}_{A}^{(N)}({\cal T})\equiv\Tr\left[\left({\rm Tr}_{\bar{A}}\hat{D}(\mathcal{T})\right)^{N}\right]\Tr\left[\hat{D}^{R-N}(\mathcal{T})\right]
\end{equation}
for each quantum trajectory $\mathcal{T}$.
Outside of region $A$, the density matrices in each replica are traced out independently, which corresponds to the conventional boundary conditions \begin{equation}
\psi_{-}^{(r)}(\boldsymbol{x},t=t_{f})=\psi_{+}^{(r)}(\boldsymbol{x},t=t_{f}).
\end{equation}
The same applies to replicas $r = N+1, \dots, R$ for arbitrary $\boldsymbol{x}$. On the other hand, within the region $A$, the first $N$ (out of total $R$) matrices $\hat{D}$ are multiplied, which mixes adjacent replicas. As a consequence, $\mathcal{P}_{A}^{(N)}({\cal T})$  can be calculated as a path integral of the form \eqref{eq:DMatrixReplicatedAction} with twisted boundary conditions:
\begin{align}
\psi_{-}(\boldsymbol{x},t=t_{f})&=\hat{\mathbb{T}}_{\text{ent}}(\boldsymbol{x})\psi_{+}(\boldsymbol{x},t=t_{f}), \label{eq:TEnt}\\ \hat{\mathbb{T}}_{\text{ent}}(\boldsymbol{x}\notin A)&=\hat{\mathbb{I}},\\
\hat{\mathbb{T}}_{\text{ent}}(\boldsymbol{x}\in A)&=\hat{\mathbb{T}}_{N}\oplus\hat{\mathbb{I}}_{R-N},
\end{align}
where $\hat{\mathbb{T}}_{N}$ is the $N \times N$ cyclic permutation matrix,
\begin{equation}
\label{eq:TN}
\hat{\mathbb{T}}_{N}=\begin{pmatrix}0 & -1 & 0 & \dots & 0\\
0 & 0 & -1 & \dots & 0\\
0 & 0 & 0 & \ddots & \vdots\\
\vdots & \vdots & \ddots & 0 & -1\\
1 & 0 & \dots & 0 & 0
\end{pmatrix}.
\end{equation}
Finally, the averaged purity \eqref{eq:Purity} is obtained from quantum-trajectory purities $\mathcal{P}_{A}^{(N)}({\cal T})$,
Eq.~\eqref{eq:gamma-A-N-T}, as
\begin{equation}
\mathcal{P}_{A}^{(N)}=\lim_{R\to1}\sum_{{\cal T}}\mathcal{P}_{A}^{(N)}({\cal T}).
\end{equation}

\begin{figure}
    \centering
    \includegraphics[width=\columnwidth]{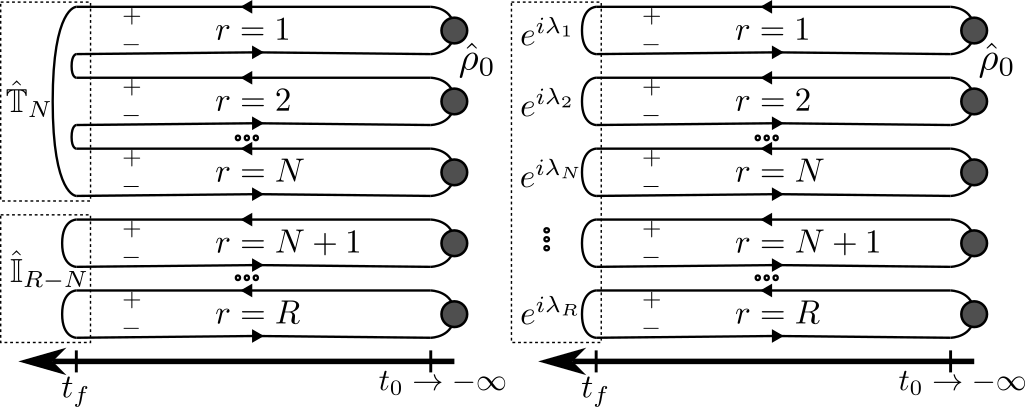}
    \caption{Replicated Keldysh contour and fermionic boundary conditions for the calculation of purity (left) and density generating functional (right). Lines with arrows correspond to the non-unitary evolution operator consisting of Kraus operators and Hamiltonian evolution for the ``$+$'' branches and conjugated evolution operator for the ``$-$'' branches within one replica; inserting $\hat{\rho}_0$ (denoted by grey circle) between them in each individual replica yields a non-normalized density matrix $\hat{D}$. Left: $N$ density matrices are multiplied, leading to cyclic closure of the Keldysh contours for adjacent replicas (and for $r=1,N$), as indicated by the twist matrix $\hat{\mathbb{T}}_N$. The remaining $(R-N)$ replicas are traced out, illustrated by independent closure of the contours indicated by $\hat{\mathbb{I}}_{R-N}$. Right: each matrix $\hat{D}$ is traced out independently, but density sources at $t = t_f$ produce additional phase shifts.}
    \label{fig:BC} 
\end{figure}

Fermionic boundary conditions for calculation of purity and density fluctuations are summarized and illustrated in Fig.~\ref{fig:BC}

\subsection{Gaussian models and symmetries}
\label{sec:gaussian-models-symmetries}

Let us temporarily restrict ourselves to the subclass of \emph{Gaussian} models, i.e., models that preserve the Gaussian property of the initial state upon time evolution. This implies that both unitary and measurement-induced contributions to the evolution should be Gaussian, i.e., the interaction is absent $\hat{H}_{\text{int}} = 0$, and the measurement operators are quadratic (with normalization factors ${\cal N}_{i,\alpha}$):
\begin{align}
\label{eq:GaussianMeasurementsDefinition}
{\cal M}_{i,\alpha}(\psi^{\ast},\psi)&=\psi^{\ast}\hat{M}_{i,\alpha}\psi+\ln{\cal N}_{i,\alpha} \,, \\{\cal M}_{i,\alpha}^{\dagger}(\psi^{\ast},\psi)&=\psi^{\ast}\hat{M}_{i,\alpha}^{\dagger}\psi+\ln{\cal N}_{i,\alpha}^{\ast} \,.
\end{align}
For such models, the Lagrangian that enters Eq.~\eqref{eq:DMatrixPathIntegral} is quadratic, albeit the corresponding matrix is non-Hermitian. It can, however, be made Hermitian by introducing \emph{chiral fermionic fields}, related to the Keldysh components $\psi_{\pm}$ as follows:
\begin{equation}
\psi\equiv\begin{pmatrix}\psi_{R}\\
\psi_{L}
\end{pmatrix}\equiv\begin{pmatrix}\psi_{+}\\
\psi_{-}
\end{pmatrix},\quad\psi^{\dagger}\equiv\begin{pmatrix}\psi_{R}^{\ast}, & \psi_{L}^{\ast}\end{pmatrix}\equiv\begin{pmatrix}-\psi_{-}^{\ast}, & \psi_{+}^{\ast}\end{pmatrix},
\end{equation}
such that the Lagrangian becomes
$$L(t,\psi^\ast, \psi, {\cal T}) = \psi^\dagger \hat{L}(t) \psi$$ 
with the matrix in right-left (RL) space
\begin{equation}
\hat{L}(t)=\begin{pmatrix}0 & i\partial_{t}-\hat{H}_{0}-i\hat{M}^{\dagger}(t) \\
i\partial_{t}-\hat{H}_{0}+i\hat{M}(t) & 0,
\end{pmatrix} \,,
\end{equation}
where
\begin{align}
 \hat{M}(t)&\equiv\sum_{m}\hat{M}_{i_{m},\alpha_{m}}\delta(t-t_{m}) \,.
\end{align}
Since $\hat{L}^\dagger(t) = \hat{L}(t)$, such Lagrangians can be thought of as describing a disordered (via operator $\hat{M}$) \emph{Hermitian} system in $(d+1)$ (space+time) dimensions. Furthermore, it
manifestly has a chiral symmetry in the RL space $\{\hat{L}(t),\hat{\tau}_z\} = 0$. For this reason, in the absence of additional symmetries, such a disordered system is expected to belong to the chiral unitary class AIII \cite{altland1997nonstandard, evers08} (see also the classification of random non-unitary circuits in Ref.~\cite{Jian2022}).

At this point, we would like to note that nearest-neighbor tight-binding models with density monitoring as studied in Refs.~\cite{Poboiko2023a,Poboiko2023b} possess an additional 
particle-hole (PH) symmetry, as was pointed out in Ref.~\cite{FavaNahum2024}.
 Generally, it can be shown to be present if (i) the Hamiltonian $\hat{H}_0$ itself has a PH symmetry, implying that in some basis it satisfies $\hat{H}_0 = -\hat{H}_0^T$, and (ii)  the measurement operators are real, $\hat{M}_{i,\alpha} = \hat{M}_{i,\alpha}^\ast$, in the same basis. When both conditions are fulfilled, one has $\hat{L}(t) = -\hat{L}^T(t)$, rendering the universality class to be chiral orthogonal, i.e., BDI (cf. Ref.~\cite{Jian2022}). Presence of both PH symmetry and chiral symmetry implies presence of a time-reversal symmetry (TRS) as well. Let us emphasize that the symmetry classification here corresponds to that of the Lagrangian $\hat{L}(t)$ and not of the Hamiltonian $\hat{H}_0$.
 In particular, the TRS of the Hamiltonian $\hat{H}_0$ does not imply the TRS of the operator $\hat{L}(t)$ because, upon transposition, the term $i \partial_t$ changes sign, while a time-reversal-symmetric Hamiltonian does not. As we discuss below, the physics for both classes AIII and BDI turns out to be very similar. 

\subsection{Projective density monitoring}
\label{sec:ProjectiveDensityMonitoring}

In the previous sections, we have discussed a generic situation of measurements described by arbitrary sets of Kraus operators $\hat{\mathbb{K}}_{i,\alpha}$. Let us now specify the model that will be studied in the rest of the present paper. We will consider projective measurements of local site occupation numbers $2\hat{\psi}^{\dagger}(\boldsymbol{x})\hat{\psi}(\boldsymbol{x})-1$, such that the index $i$ enumerates lattice sites $\boldsymbol{x}$. Each measurement can have two outcomes, either ``click'' ($\alpha = 1$) when the particle was found, or ``no-click'' ($\alpha = -1$) when it was not. The corresponding Kraus operators then become projectors onto the subspace with a given outcome:
\begin{equation}
\hat{\mathbb{K}}_{i,\alpha}\equiv\hat{\mathbb{P}}_{\alpha}(\boldsymbol{x})=\begin{cases}
\hat{\psi}^{\dagger}(\boldsymbol{x})\hat{\psi}(\boldsymbol{x}), & \alpha=+1 \,,\\
1-\hat{\psi}^{\dagger}(\boldsymbol{x})\hat{\psi}(\boldsymbol{x}), & \alpha=-1 \,.
\end{cases}
\end{equation}
The measurement rates $\gamma_i$ are assumed to be identical (independent of the site index) and will be denoted by $\gamma$.

The Keldysh coherent-state path-integral representation of these operators contains fields defined at precisely the same point in space and time, so that ill-defined objects such as time-ordered and anti-time-ordered Green's functions at coinciding times can appear in a perturbative expansion.
A standard prescription, which follows from the derivation of the path integral, is that the time (anti-)ordering at coinciding times has to be reduced to normal ordering.
This prescription, however, is inconvenient for the purposes of the present paper: we will adopt a ``principal value'' prescription, which postulates that these Green's functions should be understood as 
$$G(0)=\lim_{t\to 0}[G(t)+G(-t)]/2.$$
Such changes require introducing additional counter-terms in the action via the procedure that was thoroughly described in Appendix~A of Ref.~\cite{Poboiko2023a}; essentially, it requires first to perform symmetrization:
\begin{equation}
\hat{\mathbb{P}}_{\alpha}(\boldsymbol{x})=\frac{1}{2}+\frac{\alpha}{2}\left[\hat{\psi}^{\dagger}(\boldsymbol{x}),\hat{\psi}(\boldsymbol{x})\right],\quad \alpha = \pm 1,
\end{equation}
which then transforms to the coherent state representation:
\begin{equation}
\mathbb{P}_{\alpha}(\psi^{\ast},\psi)=\frac{1}{2}+\alpha\psi^{\ast}\psi=\frac{1}{2}\exp\left(2\alpha\psi^{\ast}\psi\right),
\end{equation}
implying
\begin{equation}
\hat{M}_{\alpha}=-2\alpha,\quad{\cal N}_{\alpha}=2 \,.
\end{equation}

\section{Non-interacting systems}
\label{Sec3:non-int}

In this Section, we employ the formalism developed in Sec.~\ref{sec:S2} to the model of free fermions subject to random projective measurements.
We will outline the main steps of the derivation of the NLSM for such monitored fermions, closely following Refs.~\cite{Poboiko2023a,Poboiko2023b} but introducing modifications necessary for the present work: (i) the derivation takes into account the particle-hole symmetric case corresponding to the symmetry class BDI (realized in the numerical simulations of Sec.~\ref{sec:numerics} below),  (ii) it is carried for both BDI and AIII symmetry classes in parallel, allowing us to discuss similarities and differences between these classes, (iii) it utilizes the original ($\pm$) Keldysh basis instead of Larkin-Ovchinnikov basis, clarifying the physical meaning of the ``replicon'' sector as the off-diagonal component of the full $Q$-matrix in the Keldysh basis, and (iv) it unifies calculations of both fluctuations of the charge and the entanglement entropy within the same field-theoretical framework. This will set the stage for the subsequent investigation of the interacting model, where exact relations \cite{KlichLevitov} between these quantities do not hold.

\subsection{Non-linear sigma model}
\label{sec:NLSM}

The starting point of our analysis is a $d$-dimensional \emph{non-interacting} system on the cubic lattice (lattice constant is set to unity), with the nearest-neighbor hopping described by a real hopping constant $J$. The single-particle hopping Hamiltonian in the Fourier space reads 
\begin{equation}
H_0(\boldsymbol{k}) = -2J\sum_{\alpha = 1}^{d} \cos k_\alpha.
\label{H0}
\end{equation}
This Hamiltonian does not correspond to a skew-symmetric matrix, i.e., $\hat{H}_0^T \neq -\hat{H}_0$. However, one can perform a local gauge transformation, $\exp(i \boldsymbol{K} \boldsymbol{x})$ with $\boldsymbol{K} = (\pi/2, \dots, \pi/2)$, which shifts all components of the  momentum by $\pi/2$. The transformed Hamiltonian
\begin{equation}
    H_0^\prime(\boldsymbol{k}) = 2J \sum_{\alpha=1}^{d} \sin k_\alpha = -H_0^\prime(-\boldsymbol{k}),
\end{equation}
is now indeed skew-symmetric: $\hat{H}_{0}^{\prime T}=-\hat{H}_{0}^{\prime}$, thus manifesting the PH symmetry. 
Such a transformation also works for any bi-partite lattice with real hoppings; however, e.g., adding real next-nearest-neighbor hoppings (or non-uniform on-site energy) leads to the breakdown of the PH symmetry. We will be working with the transformed Hamiltonian $H_0^\prime$, dropping for brevity the prime symbol from now on.

We choose to work in the original Keldysh basis (marked by index ``K''), although it will be convenient to introduce an extra minus sign for the $\psi^\ast_{-}$ components:
\begin{equation}
\psi=\begin{pmatrix}\psi_{+}\\
\psi_{-}
\end{pmatrix}_\text{K},\quad\psi^{\dagger}=\begin{pmatrix}\psi_{+}^{\dagger}, & -\psi_{-}^{\dagger}\end{pmatrix}_\text{K}.
\end{equation}
Furthermore, in order to incorporate the PH symmetry, we will introduce a PH space by effectively doubling the spinors:
\begin{equation}
\label{eq:PHSpinor}
\Psi\equiv\frac{1}{\sqrt{2}}\begin{pmatrix}\psi\\
\hat{\tau}_{x}\psi^{\ast}
\end{pmatrix}_\text{PH},\quad\bar{\Psi}\equiv\frac{1}{\sqrt{2}}\begin{pmatrix}\psi^{\dagger}, & \psi^{T}\hat{\tau}_{x}\end{pmatrix}_\text{PH}=\Psi^{T}\hat{{\cal C}},
\end{equation}
with the ``charge conjugation matrix''
\begin{equation}
\hat{{\cal C}}\equiv\hat{\tau}_{x}\hat{\sigma}_{x},
\end{equation}
where $\hat\sigma$ the Pauli matrices acting on the PH space and $\hat\tau$ acting on the Keldysh space.

Next, we derive the effective field-theory description---NLSM (see Appendix~\ref{sec:appendix:NLSM})---for monitored systems. For the symmetry class AIII, the NLSM is defined in terms of a $2R \times 2R$ matrix $Q_{ab}(\boldsymbol{r}) \sim 2 \psi_a (\boldsymbol{r}) \psi^\ast_b(\boldsymbol{r})$, 
which ``lives'' in the K$\times$R spaces. We have introduced here a short-hand notation $\boldsymbol{r}$
for the $(d+1)$-dimensional space-time argument. Due to enlarged space, for the symmetry class BDI, the theory is defined in terms of $4R \times 4R$ matrix $Q_{ab}(\boldsymbol{r}) \sim 4 \Psi_a(\boldsymbol{r})\bar{\Psi}_b(\boldsymbol{r})$, which, in addition to replica and Keldysh spaces, lives also in the PH space, and satisfies a constraint:
\begin{equation}
\bar{Q}\equiv{\cal C}Q^{T}{\cal C}=-Q.
\end{equation}
The target manifold for the NLSM, which incorporates all the relevant symmetries, is built in several steps, as detailed in  Appendix~\ref{sec:appendix:NLSM}.

First of all, there is a saddle point obtained within the self-consistent Born approximation (SCBA). It is trivial in the replica space and reads:
\begin{equation}
\label{eq:Lambda}
Q_{\text{SCBA}}\equiv\Lambda\equiv\begin{cases}
\Lambda_{0}, & \text{AIII},\\
\begin{pmatrix}\Lambda_{0} & 0\\
0 & -\tau_{x}\Lambda_{0}^{T}\tau_{x}
\end{pmatrix}_{\text{PH}}\!\!, & \text{BDI},
\end{cases}
\end{equation}
where matrix $\Lambda_0$ is defined in the Keldysh space:
\begin{equation}
\Lambda_{0}=\begin{pmatrix}1-2n_{0} & 2n_{0}\\
2(1-n_{0}) & -(1-2n_{0})
\end{pmatrix}_\text{K}.
\end{equation}
Here, the parameter $n_0 \in [0, 1]$ is the (conserved upon the evolution) filling factor of the band, as determined by the initial condition at $t_0 \to -\infty$. 

Secondly, the SCBA saddle point can be rotated by arbitrary unitary rotations in Keldysh space, which have a trivial structure in the replica and PH spaces, forming a \emph{replica-symmetric manifold} $\mathrm{S}^2$, the two-dimensional sphere:
\begin{equation}
Q_{\text{s}}=\begin{cases}
Q_{0}, & \text{AIII},\\
\begin{pmatrix}Q_{0} & 0\\
0 & -\tau_{x}Q_{0}^{T}\tau_{x}
\end{pmatrix}, & \text{BDI},
\end{cases}
\label{eq:Qs}
\end{equation}
with
\begin{equation}
Q_{0}={\cal R}_{0}\Lambda_{0}{\cal R}_{0}^{\dagger}=\begin{pmatrix}Q_{++} & Q_{+-}\\
Q_{-+} & Q_{--}
\end{pmatrix}_{\text{K}}
\end{equation}
and a unitary matrix ${\cal R}_0$.

Finally, the replica-symmetric matrix $Q_{\text{s}}$ can be further rotated by arbitrary special unitary matrices that are diagonal in the Keldysh space, but can have a non-trivial structure in the replica and PH spaces, yielding:
\begin{align}
\text{AIII:}\quad&Q=\begin{pmatrix}Q_{++} & Q_{+-}\hat{U}\\
Q_{-+}\hat{U}^{\dagger} & Q_{--}
\end{pmatrix}_{\text{K}},&U&={\cal V}_{+}{\cal V}_{-}^{\dagger},\nonumber\\
\text{BDI:}\quad&Q=\begin{pmatrix}Q_{++} & Q_{+-}\hat{U}\hat{\sigma}_{z}\\
Q_{-+}\hat{\sigma}_{z}\hat{U}^{\dagger} & Q_{--}
\end{pmatrix}_{\text{K}},&U&={\cal V}\sigma_{y}{\cal V}^{T}\sigma_{y},
\label{eq:SigmaModelQ}
\end{align}
with matrices ${\cal V}_\pm \in \mathrm{SU}(R)$ and ${\cal V} \in \mathrm{SU}(2R)$. Clearly, matrices ${\cal V}$ forming $\mathrm{SU}(R)$ group for AIII (those satisfying ${\cal V}_{+} = {\cal V}_{-}$), and $\mathrm{USp}(2R)=\mathrm{Sp}(2R,\mathbf{C})\cap \mathrm{U}(2R)$, i.e.,  unitary matrices with compact symplectic symmetry (satisfying $\mathcal{V} \sigma_y \mathcal{V}^T = \sigma_y$) for BDI, leave $U$ and $Q$ unchanged. Thus, the symmetric space for the replicon modes is:
\begin{equation}
\begin{split}
\text{AIII:}\quad & \mathrm{SU}(R)\times\mathrm{SU}(R)/\mathrm{SU}(R)=\mathrm{SU}(R),\\
\text{BDI:}\quad & \mathrm{SU}(2R)/\mathrm{USp}(2R).
\end{split}
\label{eq:SMManifold}
\end{equation}
The parametrization \eqref{eq:SigmaModelQ} includes all the Goldstone symmetries of the effective action and provides an explicit parametrization of the sigma-model manifold. 

The effective NLSM action can be obtained by means of the standard gradient expansion (see Appendix~\ref{sec:appendix:GradientExpansion}), and almost separates into the two sectors:
\begin{equation}
S[Q]=S_{\text{s}}[Q_{0}]+S_{R}[U,Q_{0}].
\end{equation}
The action $S_{\text{s}}$ for the replica-symmetric sector can be described on the level of matrix $Q_0$, and is identical to the one obtained earlier (cf. Ref.~\cite{Poboiko2023a}):
\begin{equation}
\label{eq:SymmetricLagrangian}
{\cal L}_{\text{s}}[Q_{0}]=\Tr\left[-\frac{1}{2}\hat{\Lambda}_{0}\hat{{\cal R}}_{0}^{\dagger}\partial_{t}\hat{{\cal R}}_{0}+\frac{D}{8}(\nabla\hat{Q}_{0})^{2}\right],
\end{equation}
with the trace taken only in the two-dimensional Keldysh space, the diffusion constant $D = v_0^2 / 2 \gamma$, and the root-mean-square velocity $v_0 = \sqrt{2} J$ (note that these expressions do not depend on the spatial dimension $d$). This effective theory describes the diffusive dynamics of the non-postselected averaged density matrix on top of an infinite-temperature heated state, on spatial scales larger than the mean-free path 
\begin{equation}
\ell_0 = v_0 / 2 \gamma.\label{eq:ell0}
\end{equation} 
At smaller scales, the effect of measurements is negligible and the behavior of the system is dominated by the unitary dynamics only (i.e., it is ``ballistic''; see the analysis of the ballistic regime in Ref.~\cite{Poboiko2023a}).

The diffusive form of the action in the replica-symmetric sector is guaranteed by the strict charge conservation (see Ref.~\cite{starchl2024generalized} for the case when the $\mathrm{U}(1)$ symmetry is preserved only on average).
Notably, there are no interference loop corrections to the replica-symmetric action~\cite{Poboiko2023a}, irrespective of the measurement rate $\gamma$.

The Lagrangian density for the replicon sector ${\mathcal L}_{R}$ can be conveniently written in the ``isotropic coordinates'' $r_\mu$ of the vector $\boldsymbol{r} = (v_0 t, \boldsymbol{x})$ with $\partial_\mu$ denoting the corresponding derivative:
\begin{equation}
\label{eq:RepliconLagrangian}
{\cal L}_{R}[U,Q_{0}]=\frac{\beta\, g(Q_{0})}{2}\Tr\left(\partial_{\mu}U^{\dagger}\partial_{\mu}U\right),
\end{equation}
where
\begin{equation}
\beta=\begin{cases}
1, & \text{AIII},\\
1/2, & \text{BDI},
\end{cases}
\end{equation}
the trace is taken in the replica and PH spaces, and the ``coupling constant'' $g(Q_{0})$ depends on the average (replica-symmetric) density 
\begin{equation}
\label{eq:RSDensity}
\rho\equiv1/2 - \Tr(Q_0 \tau_z) / 4
\end{equation}
via $g(Q_{0})=\rho(1-\rho)v_{0}/\gamma$. However, for the analysis in the present paper, the fluctuations in the replica-symmetric sector can be neglected, which is equivalent to the replacement $Q_0 \mapsto \Lambda_0$ and $\rho \mapsto n_0$ in $g(Q_0)$, yielding:
\begin{equation}
\label{eq:g0}
g(Q_0) \simeq g_0 \equiv \sqrt{2} n_0 (1-n_{0}) J/\gamma.
\end{equation}
The derivation of the sigma model is justified for $g_0 \gg 1$, i.e., for sufficiently rare measurements ($\gamma \ll J$). 

It is worth mentioning that the class-BDI symmetry of the NLSM in the replicon sector (in contrast to the NLSM in class AIII, as well as CI and DIII, where a manifold is a group) does not allow the introduction of an additional Wess-Zumino term in the action. Furthermore, contrary to the conventional $\text{U}(2R)/\text{USp}(2R)$ models of the symmetry class BDI, the $\text{SU}(2R)/\text{USp}(2R)$ symmetry does not support the Gade term \cite{Gade1991,Gade1993} in the NLSM action.

\subsection{Boundary conditions for NLSM}
\label{sec:boundary-NLSM}

The theory is defined on a semi-space $t < t_f$, where measurements occur, and has to be supplied with the boundary conditions at $t = t_f$. The form of boundary conditions for the NLSM originates from the boundary conditions for the fermionic fields, Eqs.~(\ref{eq:TFCS}) and (\ref{eq:TEnt}). For an arbitrary $\mathrm{U}(R)$ matrix $\hat{\mathbb{T}}(\boldsymbol{x})$, the boundary conditions acquire the following form:
\begin{equation}
Q(\boldsymbol{x},t=t_{f})=\begin{cases}
\Lambda_{b}(\boldsymbol{x}), & \text{AIII},\\
\begin{pmatrix}\Lambda_{b}(\boldsymbol{x}) & 0\\
0 & -\tau_{x}\Lambda_{b}^{T}(\boldsymbol{x})\tau_{x}
\end{pmatrix}_{\text{PH}}\!\!, & \text{BDI},
\end{cases}
\end{equation}
with the boundary matrix
\begin{equation}
\Lambda_{b}(\boldsymbol{x})=\begin{pmatrix}1-2n_{0} & 2n_{0}\,\hat{\mathbb{T}}(\boldsymbol{x})\\
2(1-n_{0})\,\hat{\mathbb{T}}^{\dagger}(\boldsymbol{x}) & -(1-2n_{0})
\end{pmatrix}_{\text{K}}.
\end{equation}
If, additionally, $\det \hat{\mathbb{T}}(\boldsymbol{x}) = 1$ (which is the case for all quantities of interest in the present paper), this boundary condition implies 
\begin{equation}
\hat{Q}_0(\boldsymbol{x}, t = t_f) = \Lambda_0,\quad \hat{U}(\boldsymbol{x}, t = t_f) = \hat{U}_b(\boldsymbol{x}),
\end{equation}
with
\begin{equation}
\begin{split}\text{AIII:}\quad & \hat{U}_b(\boldsymbol{x})=\hat{\mathbb{T}}(\boldsymbol{x}),\\
\text{BDI:}\quad & \hat{U}_b(\boldsymbol{x})=\begin{pmatrix}\hat{\mathbb{T}}(\boldsymbol{x}) & 0\\
0 & \hat{\mathbb{T}}^{T}(\boldsymbol{x})
\end{pmatrix}_{\text{PH}}\!\!.
\end{split}
\label{eq:UBoundaryConditions}
\end{equation}

We thus conclude that both the purity and the FCS generating functional can be expressed via the partition function with fixed boundary conditions defined in Eqs.~(\ref{eq:TFCS}), (\ref{eq:TEnt}):
\begin{equation}
\label{eq:GeneratingFunctional:TwistMatrix}
Z\left[\hat{\mathbb{T}}\right]=\lim_{R\to1}\int_{\hat{U}(\boldsymbol{x},t=t_{f})=\hat{U}_{b}(\boldsymbol{x})}
\!\!{\cal D}\hat{U}\,\exp\left(-S_{R}[\hat{U}]\right),
\end{equation}
with the action
\begin{equation}
S_{R}[\hat{U}]=\frac{\beta g_{0}}{2}\int_{-\infty}^{v_0 t_f}dr_{0}\int d^{d}\boldsymbol{r}\Tr\left(\partial_{\mu}\hat{U}^{\dagger}\partial_{\mu}\hat{U}\right).
\end{equation}
The FCS generating functional and the purity 
are then obtained from
\begin{equation}
\label{eq:ObservablesAsPartitionFunction}
Z_{A}^{(N)}(\lambda)=Z\left[\hat{\mathbb{T}}_{\text{FCS}}\right],\quad{\cal P}_{A}^{(N)}=Z[\hat{\mathbb{T}}_{\text{ent}}],
\end{equation}
respectively.
For the specific choice of $\hat{\mathbb{T}}$ corresponding to the calculation of the purity, the above boundary conditions can be compared with those introduced in Refs.~\cite{Fava2023,FavaNahum2024}.
The above NLSM boundary conditions can also be considered as a generalization of the ``twisted boundary conditions'' that were introduced for the calculation of the entanglement entropy in conformal field theories (cf. Refs.~\cite{Calabrese2004, Cardy2008, Calabrese2009}). The same boundary conditions will be used for interacting fermions in Sec.~\ref{Sec:Interaction} below.

\subsection{Relation between purity and full counting statistics for free fermions}
\label{Sec:KL}
Let us now discuss a direct consequence of the NLSM symmetry. The off-diagonal twist matrix that enters \eqref{eq:TFCS} can be diagonalized by means of a unitary transformation:
\begin{equation}
\hat{\mathbb{T}}_{N}=\hat{T}\,\exp\left[i\hat{\lambda}^{(\text{ent})}\right]\,\hat{T}^{\dagger},
\end{equation}
with the rotation matrix $\hat{T}$ operating in the replica space,
\begin{equation}
T_{rr^{\prime}}=\frac{1}{\sqrt{N}}
\exp\left[-\frac{i\pi r\,(2r^{\prime}-1)}{N}\right],
\label{Trotation}
\end{equation}
and the elements of the diagonal matrix determined by
\begin{equation}
\lambda_{r}^{(\text{ent})}=\pi\left(1-\frac{2r-1}{N}\right),
\label{lambda-ent}
\end{equation}
where $r,r^\prime = 1, \dots, N$. As a consequence, one can perform a gauge transformation in the path integral of the form:
\begin{equation}
\begin{split}\text{AIII:}\quad & \hat{U}=\hat{T}\hat{U}^{\prime}\hat{T}^{\dagger},\\
\text{BDI:}\quad & \hat{U}=\begin{pmatrix}\hat{T} & 0\\
0 & \hat{T}^{\ast}
\end{pmatrix}_{\text{PH}}\!\!\hat{U}^{\prime}\,\begin{pmatrix}\hat{T}^{\dagger} & 0\\
0 & \hat{T}^{T}
\end{pmatrix}_{\text{PH}}\!\!.
\end{split}
\label{eq:BCGaugeTransformation}
\end{equation}
After such transformation, the boundary conditions for matrix $\hat{U}^\prime$ retain the form of Eq.~\eqref{eq:UBoundaryConditions} but with the \emph{diagonal} matrix $\hat{\mathbb{T}}^\prime(\boldsymbol{x})$:
\begin{align}
\label{eq:TEnt:Diagonal}
\hat{\mathbb{T}}^{\prime}(\boldsymbol{x})&=\exp\left[i\hat{\lambda}^{(\text{ent})}(\boldsymbol{x})\right],\\
\lambda_{r}^{(\text{ent})}(\boldsymbol{x})&=\lambda_{r}^{(\text{ent})}\mathcal{I}_{A}(\boldsymbol{x}),\quad r=1,\dots,N,\\
\lambda_{r}^{(\text{ent})}(\boldsymbol{x})&=0,\quad r=N+1,\dots,R.
\end{align}
As a result, the purity can be exactly expressed via the same generating functional that defines the FCS of charge fluctuations:
\begin{equation}
\label{eq:PurityFCS}
{\cal P}_{A}^{(N)}=Z\left[\exp\left(i\hat{\lambda}^{(\text{ent})}\right)\right]
\end{equation}
This relation between the entanglement entropy and the fluctuations of charge is a direct manifestation of the Klich-Levitov identity \cite{KlichLevitov} that holds for Gaussian states:
\begin{equation}
\label{eq:KlichLevitov}
{\cal S}_{A}=\sum_{q=1}^{\infty}2\zeta(2q)\,{\cal C}_{A}^{(2q)},\qquad{\cal C}_{A}^{(2q)}\equiv\overline{\left\llangle \hat{N}_{A}^{2q}\right\rrangle}.
\end{equation}
We emphasize that our derivation of relation \eqref{eq:PurityFCS} was based on the fact that the action has a continuous NLSM symmetry, allowing us to perform the gauge transformation \eqref{eq:BCGaugeTransformation}. Below, we will see that interactions will violate this symmetry, breaking down the relation between the entanglement and the charge fluctuations.

\subsection{Semiclassical approximation}
\label{sec:semiclassical:non-int}
In the limit ${g_0\to\infty}$, the generating functional \eqref{eq:GeneratingFunctional:TwistMatrix} can be evaluated using the saddle-point approximation, which requires solving the \emph{classical} equations of motion for the NLSM action $S_R[\hat{U}]$ subject to boundary conditions:
\begin{equation}
\label{eq:SemiclassicalApproximation}
Z\left[\hat{\mathbb{T}}\right]\approx\exp\left(-S_R\left[\hat{U}_{\text{cl}}\right]\right).
\end{equation}
For the diagonal form of boundary conditions $$\hat{\mathbb{T}}(\boldsymbol{x})=\exp\left[i\hat{\lambda}(\boldsymbol{x})\right],$$ the classical solution can also be sought in the diagonal form 
\begin{equation}
\label{eq:SaddlePoint:DiagonalAnsatz}
\begin{split}\text{AIII:}\quad & \hat{U}_{\text{cl}}(\boldsymbol{r})=\exp\left[i\hat{\Phi}_{\text{cl}}(\boldsymbol{r})\right],\\
\text{BDI:}\quad & \hat{U}_{\text{cl}}(\boldsymbol{r})=\begin{pmatrix}\exp\left[i\hat{\Phi}_{\text{cl}}(\boldsymbol{r})\right] & 0\\
0 & \exp\left[i\hat{\Phi}_{\text{cl}}^{T}(\boldsymbol{r})\right]
\end{pmatrix}_{\text{PH}},
\end{split}
\end{equation}
with the diagonal matrix $\hat{\Phi}_{\text{cl}}(\boldsymbol{r})$ solving the Dirichlet problem for a $(d+1)$-dimensional Laplace equation:
\begin{equation}
\partial_{\mu}^{2}\hat{\Phi}_{\text{cl}}(\boldsymbol{r})=0,\quad\hat{\Phi}_{\text{cl}}(r_0 = 0,\boldsymbol{x})=\hat{\lambda}(\boldsymbol{x}).
\label{eq87}
\end{equation}
Here, we have set $t_f = 0$ and used the ``isotropic'' coordinates $r_\mu = (v_0 t, \boldsymbol{x})$.
The result for the action evaluated on the classical configuration $\hat{U}_{\text{cl}}(\boldsymbol{r})$ reads:
\begin{equation}
S_{R}[U_{\text{cl}}]=\frac{1}{2}\int d^{d}\boldsymbol{x}_{1}d^{d}\boldsymbol{x}_{2}\,C(\boldsymbol{x}_{1}-\boldsymbol{x}_{2})\tr_{R}\left[\hat{\lambda}(\boldsymbol{x}_{1})\hat{\lambda}(\boldsymbol{x}_{2})\right],
\end{equation}
with 
\begin{equation}
C(\boldsymbol{x})\simeq n_0(1-n_0) \delta(\boldsymbol{x})-\frac{2g_{0}}{\sigma_{d}|\boldsymbol{x}|^{d+1}},
\label{eq:Cx}
\end{equation}
where
\begin{equation}
\sigma_{d}=\frac{2\pi^{(d+1)/2}}{\Gamma\left(\frac{d+1}{2}\right)}
\end{equation} is the surface area of $d$-dimensional unit sphere.
The delta-function term ensures vanishing of the space integral: $\int d^d{\boldsymbol{x}}\,C(\boldsymbol{x}) = 0$. 
The Fourier transform of function $C(\boldsymbol{x})$ has a universal (independent of $d$) form:
\begin{equation}
    C(\boldsymbol{q})=g_{0}|q|. 
    \label{eq:tildeCq}
\end{equation}

Substituting the sources (\ref{eq:DensitySource:FCS}) and (\ref{eq:TEnt:Diagonal}) for the FCS generating function \eqref{eq:CumulantGeneratingFunction} and R{\'e}nyi entanglement entropy \eqref{eq:RenyiEntanglementEntropy}, respectively, we obtain:
\begin{align}
\chi_{A}^{(N)}(\lambda)&=\frac{i}{2}N\lambda^{2}{\cal C}_{A}^{(2)}, 
\label{eq:chi-C}\\
\mathcal{S}_{A}^{(N)}&=\left(\frac{1}{2}\frac{1}{N-1}\sum_{r=1}^{N}\lambda_{r}^{2}\right){\cal C}_{A}^{(2)}=\frac{\pi^{2}}{6}\left(1+\frac{1}{N}\right){\cal C}_{A}^{(2)},
\label{eq:KL-general}
\end{align}
with
\begin{multline}
{\cal C}_{A}^{(2)}\equiv\overline{\left\langle \hat{N}_{A}^{2}\right\rangle -\left\langle \hat{N}_{A}\right\rangle ^{2}}=\int_{A}d^{d}\boldsymbol{x}_{1}\int_{A}d^{d}\boldsymbol{x}_{2}C(\boldsymbol{x_{1}}-\boldsymbol{x}_{2})\\
=\frac{g_{0}}{\pi}\left\Vert \partial A\right\Vert \ln\frac{\ell_{A}}{\ell_{0}},
\label{eq:C2A}
\end{multline}
where $\ell_A$ is the linear size of subsystem $A$.

This result implies that, in the limit $g_0 \to \infty$, in the non-interacting system (i) fluctuations of charge become Gaussian, being fully determined by the second cumulant $\mathcal{C}_A^{(2)}$, (ii) the entanglement entropy and charge fluctuations follow identical scaling and differ only by a numerical prefactor, which is consistent with the Klich-Levitov relation \eqref{eq:KlichLevitov} for $N = 1$, (iii) both quantities follow the area$\times$log (i.e., neither volume-law nor area-law) scaling, and (iv) these predictions are identical for AIII and BDI symmetry classes.

\subsection{Renormalization group}
Relaxing the condition $g_0 \to \infty$, one needs to include loop corrections. The derivation of a one-loop RG equation for the dimensionless coupling constant 
\begin{equation}
    G(\ell) = g(\ell) \ell^{\epsilon},
\end{equation}
with $\epsilon \equiv d-1 \ll 1$ and $\ell$ being the RG ultraviolet length-scale, is outlined in Appendix~\ref{sec:appendix:BDIRG} (see also Ref.~\cite{koenig2012metal-insulator}). For the sake of comparison, we provide equations for both the AIII and BDI symmetry classes (the first one is adopted from Ref.~\cite{Poboiko2023b}; cf. Refs.~\cite{evers08,HikamiRG,WegnerRG}):
\begin{align}
\text{AIII:}\quad&\frac{dG}{d\ln\ell}=\epsilon\, G-\frac{R}{4\pi} + O(G^{-1}),
\label{RG-AIII}\\
\text{BDI:}\quad&\frac{dG}{d\ln\ell}=\epsilon\, G-\frac{R}{2\pi} + O(G^{-1}).
\label{RG-BDI}
\end{align}
Importantly, in the $R\to 1$ replica limit, these one-loop RG equations for the chiral classes describe the perturbative renormalization of $G$ for $\epsilon=0$ (in contrast to the conventional limit $R\to 0$ for disordered systems, where the perturbative beta-function is identically zero \cite{Gade1991,Gade1993,evers08,koenig2012metal-insulator}).

The one-loop RG equations (\ref{RG-BDI}) and (\ref{RG-AIII}) for classes BDI and AIII differ only by the factor of 2 in the second term of the beta-function. Therefore, qualitative predictions for the BDI symmetry class are nearly identical to those for AIII symmetry class studied analytically in Refs.~\cite{Poboiko2023a,Poboiko2023b} (see also Ref.~\cite{FavaNahum2024}, where analytical consideration was supplemented by numerics for class AIII): (i) the class-BDI RG equation in the relevant replica limit $R \to 1$ predicts ``localization''---i.e., area-law---behavior for one-dimensional systems ($\epsilon=0$) for arbitrary weak monitoring rates, with crossover between area$\times$log and area-law at exponentially large length scale \begin{equation}
\ell_{\text{loc}} \sim \ell_0 \exp(2 \pi g_0), \quad \text{BDI},
\label{lcorr-BDI}
\end{equation}
which differs only by a numerical factor in the exponential from the class-AIII correlation length:
\begin{equation}
    \ell_{\text{loc}}\sim \ell_0 \exp(4 \pi g_0), \quad \text{AIII};
    \label{lcorr-AIII}
\end{equation} 
(ii) it also predicts a ``metal-insulator'' transition---i.e., transition between area$\times$log and area-law scaling of the entanglement entropy and charge fluctuations---in dimensions $d > 1$. The difference between the two RG equations also explains the difference by a factor of 2 in the weak-localization correction observed in the numerical part of Ref.~\cite{Poboiko2023a}, where a class-BDI system was simulated, compared to the analytical prediction for the class AIII (see Fig.~3 of Ref.~\cite{Poboiko2023a}). As an additional check, we have performed numerical simulations of two similar models, one belonging to the AIII class and the other to the BDI class, and observed the weak-localization corrections consistent with the predictions given by Eqs.~(\ref{RG-AIII}) and (\ref{RG-BDI}), see Appendix~\ref{sec:appendix:WL-BDI-vs-AIII} for details.

Thus, the results for observables obtained for the non-interacting models of symmetry classes AIII and BDI are qualitatively the same. It is, therefore, natural to anticipate that the inclusion of interactions in the corresponding NLSM would also produce qualitatively similar results. While, as discussed above, the BDI model is easier to implement in numerical simulations, the AIII model turns out to be somewhat simpler in terms of analytical treatment (as it does not possess an extra symmetry). In order to understand the general phase diagram of the interacting models of monitored fermions, in what follows, we will focus our analytical consideration on the AIII model and add a weak short-range fermion-fermion interaction to its NLSM action.

\section{Interacting model: information-charge separation}
\label{Sec:Interaction}

In this section, we derive the additional term in the NLSM action, which is generated by two-body interaction of the form \eqref{eq:Hint}. 
We will then analyze the effect of the additional term on FCS and entanglement of monitored fermions. 
The effect of the two-body interaction  \eqref{eq:Hint} can be conveniently included in the fermionic path integral description \eqref{eq:DMatrixReplicatedAction} via an additional quartic Lagrangian term for each individual replica:
\begin{equation}
L_{\text{int}}
\!=\!-\frac{1}{2}\sum_{\boldsymbol{x}\boldsymbol{x}^{\prime}}V(\boldsymbol{x},\boldsymbol{x}^{\prime})\left[n_{+}(\boldsymbol{x})\,n_{+}(\boldsymbol{x}^{\prime})\!-\!n_{-}(\boldsymbol{x})\,n_{-}(\boldsymbol{x}^{\prime})\right],
\label{eq:Lint}
\end{equation}
where $n_{\pm}(\boldsymbol{x})=\psi_{\pm}^{\ast}(\boldsymbol{x})\psi_{\pm}(\boldsymbol{x})$.

The interaction-induced term in the Lagrangian has important implications from the point of view of symmetry.
In the absence of interactions, the replicated fermionic action \eqref{eq:DMatrixReplicatedAction} had a large continuous symmetry group $\mathrm{U}(R) \times \mathrm{U}(R)$, which, after averaging over trajectories, leads to the NLSM defined on the $\mathrm{SU}(R)$ symmetric space, see Eq.~\eqref{eq:SMManifold}.
However, continuous rotations, which mix different replicas, form a symmetry only for Gaussian models. Upon switching on the interactions in the unitary dynamics, two relevant symmetries of the fermionic action survive: (i) \emph{discrete} replica permutation symmetry $\mathbb{S}_R \times \mathbb{S}_R$, corresponding to independent permutation of fermionic fields on ``$+$'' and ``$-$'' branches of the Keldysh contour, and (ii) continuous $[\mathrm{U}(1)]^{2R}$ symmetry which is responsible for charge conservation on each individual replica. 
Below,  we will demonstrate in the NLSM language that interactions gap out some of the Goldstone modes on the  $\mathrm{SU}(R)$ replicon manifold, reducing the symmetry down to $\mathbb{S}_R \times [\mathrm{U}(1)]^{R-1}$.

Interactions also reduce the symmetry of the replica-symmetric sector down to $\mathrm{U}(1)$ corresponding to charge conservation. We anticipate the diffusive behavior of the replica-symmetric density to persist in the presence of interactions \footnote{Diffusive behavior 
of average density was shown to hold for quantum simple exclusion processes \cite{Bernard2022}, which correspond to a certain type of strong inter-particle interaction.}. In what follows, we focus on the replicon sector of the theory, which describes measurement-induced transitions in entanglement and in charge fluctuations.

\subsection{Non-linear sigma model for weakly interacting fermions}
\label{sec:4A}

We begin our analysis by assuming the interaction to be perturbatively weak, such that its symmetry-breaking effect is also weak. Formally, we will assume that the gap in the modes, which were Goldstone modes in the absence of interaction, is still smaller than the gap which describes fluctuations in the direction orthogonal to the NLSM manifold. This will allow us to study the effect of the interaction utilizing the NLSM formalism already developed in Sec.~\ref{Sec3:non-int} for non-interacting fermions. 

Our first goal is then to derive the relevant interaction-induced correction to the NLSM action. The full derivation and analysis of the interacting NLSM in the presence of PH symmetry will be presented elsewhere; here, we will restrict ourselves to the interaction-induced modification of the AIII NLSM. As discussed above, the phase diagrams for the two models (with and without the PH symmetry) are expected to be qualitatively the same.

The derivation follows the standard steps (cf.~Refs.~\cite{Finkelstein1990, Finkelstein2023scale-dependent, BelitzKirkpatrick1994, KamenevLevchenko}) and is outlined in Appendix~\ref{sec:appendix:Interaction}. An important difference to the derivation of the interacting NLSM in disordered systems is that the $Q$-matrices of the NLSM now remain local in space-time, as was already discussed above for the non-interacting model. The quartic term is decoupled via the Hubbard-Stratonovich transformation by introducing a set of ``plasma fields'' $\varphi_{\pm,r}(\boldsymbol{x},t)$ that can be combined into a diagonal matrix in the Keldysh space:
$
\hat{\varphi}=\text{diag}\left\{\varphi_{+},\,
\varphi_{-}\right\}_{K}.
$
Interactions then induce additional terms in the action, which can be derived by expanding the determinant, obtained after the integration over fermionic degrees of freedom, in powers of $\varphi$:
\begin{equation}
\label{eq:SintQ:intermediate}
S_{\text{int}}[\hat{Q}]=\!-\ln\!\! \int\!{\cal D}\varphi\,\exp\left\{\frac{i}{2}\varphi\hat{u}^{-1}\hat{\tau}_{z}\varphi+\Tr\ln(1-\hat{G}[\hat{Q}]\hat{\varphi})\right\},
\end{equation}
with the Green's function
\begin{equation}
\label{eq:GQ}
\left(\hat{G}[\hat{Q}]\right)^{-1}=i\partial_{t}-\hat{H}_{0}+i\gamma\hat{Q}.
\end{equation}

\begin{figure}
    \centering
    \includegraphics[width=\columnwidth]{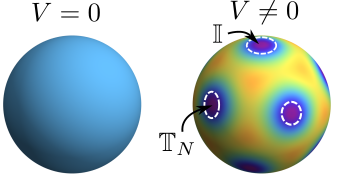}
    \caption{Cartoon of the NLSM symmetric manifold: a ``poor man's'' picture of the interaction-induced potential in the replicon sector breaking down the continuous $\mathrm{SU}(R)$ symmetry, with spheres used for illustrative purpose.  Left: in the absence of interaction, the full NLSM symmetry $\mathrm{SU}(R)$ is present and all points on the manifold are equivalent. Right: potential [as given by Eq.~\eqref{Lint-U}; colors mark the strength of the potential] leads to the appearance of a set of global minima forming a discrete replica permutation group $\mathbb{S}_R$ (which include, in particular, the identity matrix $\mathbb{I}$ and the cyclic replica permutation matrix $\mathbb{T}_N$ relevant for the boundary conditions). For each of the minima, some of Goldstone modes persist, forming a $[\mathrm{U}(1)]^{R-1}$ sector, as illustrated by dashed circles.}
    \label{fig:potential-cartoon}
\end{figure}

We will be interested in the effective action for the replicon sector described by matrix $\hat{U}$. 
Furthermore, we will focus on the terms
without gradients, which will create an effective ``potential”
on the NLSM manifold, leading to the breaking of the continuous $\mathrm{SU}(R)$ symmetry. For this reason, it is sufficient to
assume ${Q(x)=\text{const}}$ when deriving such terms. 
Performing integration over $\varphi$ fields, we see that the relevant terms in the NLSM action arise in the second-order perturbation theory in the interaction strength, or, equivalently, fourth order in $\varphi$. The resulting interaction-induced Lagrangian can be cast in the following form involving the matrix elements of the $\hat{U}$-matrices operative in the replica space:
\begin{equation}
{\cal L}_{\text{int}}[\hat{Q}]=Y_\text{HF}\,\rho^{2}(1-\rho)^{2}\, \left(R-\sum_{r_{1}r_{2}}|U_{r_{1}r_{2}}|^{4}\right).
\label{Lint-U}
\end{equation}
Here the prefactor $Y_\text{HF}$ is given by a sum of Hartree and Fock diagrams, whose analysis is performed in Appendix \ref{sec:appendix:HartreeFockDiagrams}.
The precise form of $Y_\text{HF}$ depends on the microscopic model; its parametric dependence 
for $\gamma \ll J$ and the short-range  interaction of strength $V$ is given by
\begin{equation}
Y_\text{HF}\sim\frac{V^{2}}{J}\times\begin{cases}
\displaystyle{\ln\frac{J}{\gamma}}, & d=1,\\
\text{const}, & d>1.
\end{cases}
\label{YHF}
\end{equation}
It is convenient to characterize the strength of the interaction-induced term in the NLSM by a ``mass'' $m$, 
\begin{equation}
    m^{2}\equiv2\rho(1-\rho)\frac{Y_\text{HF}}{D} \propto \frac{\gamma V^2}{J^3},
\end{equation}
with the associated (large) length scale 
\begin{equation}
\label{eq:l-int}
    \ell_{\text{int}} = m^{-1} \propto \frac{J}{|V|} \sqrt{\frac{J}{\gamma}},
\end{equation}
up to logarithmic corrections for $d=1$. 
The resulting $(d+1)$-dimensional Lagrangian density of the interacting NLSM takes the form:
\begin{equation}
\label{eq:InteractionLagrangian}
{\cal L}_R[U]=\frac{g}{2}\left[\Tr\left(\partial_\mu \hat{U}^{\dagger} \partial_\mu\hat{U}\right)+\frac{m^{2}}{2}\left(R-\sum_{r_{1}r_{2}}|U_{r_{1}r_{2}}|^{4}\right)\right].
\end{equation}
where $r_\mu$ is the coordinate in $(d+1)$-dimensional space-time.

The interaction-induced Lagrangian (\ref{Lint-U}) introduces a modulation of the $\mathrm{SU}(R)$ NLSM manifold, as illustrated in Fig.~\ref{fig:potential-cartoon}.  The minima of $\mathcal{L}_{\text{int}}$ are attained on the configurations, where each row and column of $U$ is ``maximally localized'', i.e., contain only a single non-zero element with an arbitrary phase: such configuration nullify $\mathcal{L}_{\text{int}}$. The manifold, on which this happens, is given by $\mathbb{S}_R \times [\mathrm{U}(1)]^{R-1}$, where $\mathbb{S}_R$ is the permutation group for $R$ replicas [cf. discussion of the symmetries below Eq.~(\ref{eq:Lint})]. For small fluctuations around each of those global minima, the interaction-induced term produces a (small) mass $m$ for the replica-off-diagonal Goldstone modes. 

Beyond the symmetry-breaking scale $\ell_{\text{int}}$, the replicon Goldstone modes responsible for the localization corrections in the free-fermion case are switched off. Therefore, for weak monitoring, the area-law phase of free fermions in 1D systems turns out to be destabilized by the inclusion of interactions if the interaction-induced length $\ell_{\text{int}}$ is shorter than the measurement-induced ``localization length'' $\ell_\text{loc}$. In the opposite limit of strong monitoring (large $\gamma/J$), the quantum Zeno effect is expected to overcome the effect of interaction (``localization'' occurs on the scale much smaller than $\ell_{\text{int}}$).
As a result, one anticipates that the phase diagram of monitored interacting fermions in 1D should reveal phase transition (or transitions), in contrast to free fermions. Below, we will analyze the behavior of FCS and entanglement in the presence of the interaction-induced term \eqref{Lint-U} in the NLSM.

\subsection{Semiclassical approximation and fluctuations of charge}
\label{sec:Gaussian_approx}

Following the recipe of Sec.~\ref{sec:semiclassical:non-int}, we start our analysis of the action \eqref{eq:InteractionLagrangian} with the help of the saddle-point approximation, valid in the limit $g \to \infty$. One then needs to solve the non-linear classical equations of motion, which read:
\begin{equation}
\label{eq:Sint:SaddlePointEquations}
\partial_{\mu}(\hat{U}^{\dagger}\partial_{\mu}\hat{U})_{rr^{\prime}}=\frac{m^{2}}{2}\sum_{r^{\prime\prime}}\left(U_{r^{\prime\prime}r}^{\ast}U_{r^{\prime\prime}r^{\prime}}|U_{r^{\prime\prime}r}|^{2}-\text{H.c.}\right),
\end{equation}
where the Hermitian conjugate stands for complex conjugation and swapping $r \leftrightarrow r^\prime$.
These equations are subject to boundary conditions $U(\boldsymbol{x},t=t_{f})=\mathbb{T}(\boldsymbol{x})$, with matrix $\mathbb{T}(\boldsymbol{x})$ determined by the observable of interest, as discussed earlier. Importantly, the structure of the solution depends drastically on the precise form of matrix $\mathbb{T}$.

For charge fluctuations, where matrix $\mathbb{T}(\boldsymbol{x})$ is diagonal, and one can easily check that the diagonal Ansatz \eqref{eq:SaddlePoint:DiagonalAnsatz} is consistent with Eq.~\eqref{eq:Sint:SaddlePointEquations}. Furthermore, this Ansatz completely nullifies the right-hand side of the equations of motion; for this reason, the remaining calculation and the results are identical to those of Sec.~\ref{sec:semiclassical:non-int}. Namely, the fluctuations of charge remain Gaussian, and the second cumulant follows the area$\times$log law scaling:
\begin{equation}
{\cal C}_{A}^{(2)}=\frac{g}{\pi}\left\Vert \partial A\right\Vert \ln\frac{\ell_{A}}{\ell_{0}},\quad m\neq 0, \quad g\to \infty,
\label{eq:second-cumulant}
\end{equation}
which is identical to Eq.~(\ref{eq:C2A}).
Thus, the first important prediction of our theory is that weak interactions have no qualitative effect on the behavior of charge fluctuations in the limit $g \to \infty$, and our predictions of Sec.~\ref{sec:semiclassical:non-int} for the charge sector also hold in the weakly interacting case.

\subsection{Entanglement entropy: Volume law}
\label{sec:EEInt}

The reasoning of Sec.~\ref{sec:Gaussian_approx} cannot be directly applied to calculate the entanglement entropy, since the corresponding boundary matrix $\mathbb{T}(\boldsymbol{x})$, Eq.~\eqref{eq:TN}, is not diagonal. Furthermore, the trick that was used in Sec.~\ref{Sec:KL} for non-interacting systems, which related the entanglement entropy to the fluctuations of charge, is also not applicable here. Indeed, the unitary transformation that diagonalizes the boundary conditions would also directly affect the interaction part of the action \eqref{eq:InteractionLagrangian} because this term in the action does not respect the full $\mathrm{SU}(R)$ symmetry. 

\begin{figure}
    \centering
    \includegraphics[width=\columnwidth]{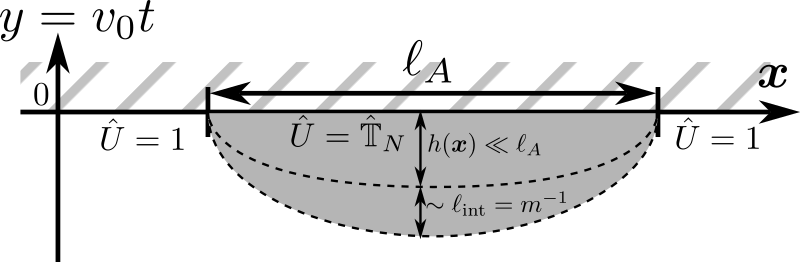}
    \caption{Schematic structure of the domain-wall solution in the $(d+1)$ space-time $(\boldsymbol{x}, y = v_0 t)$, which gives rise to the volume-law behavior of the $N$-th R{\'e}nyi entanglement entropy in the presence of weak interaction. Boundary conditions fix two separate global minima of the $\mathcal{L}_{\text{int}}$, circular permutation matrix $\hat{\mathbb{T}}_N$ and identity matrix $\hat{1}$, inside and outside the region $A$, correspondingly. A domain wall of width $\sim \ell_{\text{int}}$ is formed near the boundary $y = 0$, separated from the boundary by distance $h(\boldsymbol{x})$, such that $\ell_A \gg h(\boldsymbol{x}) \gg \ell_{\text{int}}$. The action on such configuration is proportional to the $d+1$-dimensional surface area of the domain wall, coinciding with the $d$-dimensional volume of the region $A$ in the leading order.}
    \label{fig:domain-wall}
\end{figure}

Nevertheless, some generic properties of the solution, which are based directly on symmetry considerations, can be inferred. 
We note that both matrices $\hat{\mathbb{I}}$ and $\hat{\mathbb{T}}_{N}$, which enter boundary conditions, provide local minima for the action. This implies that the solution will take the form of a $(d+1)$-dimensional \emph{domain wall} separating two regions, as illustrated in Fig.~\ref{fig:domain-wall}. The action calculated on such a solution will be proportional to the $(d+1)$-dimensional surface area of the domain wall. The width of the domain wall is determined by the interacting length scale $\ell_{\text{int}} = m^{-1}$, and its explicit shape (parametrized by $h(\boldsymbol{x})$ in Fig.~\ref{fig:domain-wall}) is determined by an interplay of two effects: (i) the system tries to minimize the surface area of the domain wall, and (ii) the domain wall experiences ``repulsion'' from the boundary where the boundary condition is set.

At large distances, $m h(\boldsymbol{x}) \gg 1$, the ``interaction'' falls off exponentially $\propto \exp[-m h(\boldsymbol{x})]$, implying that the typical distance of the domain wall from the boundary can be estimated as 
\begin{equation}
h(\boldsymbol{x}) \sim m^{-1} \ln (m \ell_A) \ll \ell_A,
\label{eq:DomainWallDistance}
\end{equation}
where $\ell_A$ is the linear size of the region $A$. For this reason, to leading order, the $(d+1)$-dimensional surface area of the domain wall in the leading order coincides with the $d$-dimensional \emph{volume} of the region $A$. The bending of the domain wall will contribute to the subleading terms, presumably having the form $\propto \Vert\partial A\Vert \ln \ell_A$, see Eq.~\eqref{eq:DomainWallDistance}.
As a result, we deduce that the entanglement entropy in the semiclassical approximation (without loop corrections) should follow the volume law:
\begin{equation}
\label{eq:VolumeLaw}
\mathcal{S}_{A}^{(N)}=s_{N}\, g\, m\,\left\Vert A\right\Vert,\quad m \ell_A \gg 1,\quad g \to \infty,
\end{equation}
where the prefactor $s_N = O(1)$ is determined by the energy per surface area of the domain wall. 

The above qualitative consideration of the domain-wall configurations in the context of entanglement entropy is similar to the one yielding the volume law in generic hybrid 
circuits, see, e.g., Ref.~\cite{Fisher2022} for review. 
We now proceed to a more specific analysis of the model. In Appendix~\ref{sec:appendix:RenyiGaussian}, we show that the saddle-point solution can still be found with the help of the unitary transformation \eqref{Trotation}, and has the following form:
\begin{equation}
\hat{U}(\boldsymbol{r})=\hat{T}\,\diag\left(\left\{ \exp\left[i\lambda_{r}\phi(\boldsymbol{r})/2\pi\right]\right\} _{r=1}^{N}\right)\,\hat{T}^{\dagger}\oplus\hat{\mathbb{I}}_{R-N},
\end{equation}
where function $\phi(\boldsymbol{r})$ satisfies the elliptic sine-Gordon equation:
\begin{equation}
\partial_{\mu}^{2}\phi=m^{2}\sin\phi,\quad\phi(\boldsymbol{x},y=0)=\begin{cases}
0, & \boldsymbol{x}\notin A,\\
2\pi, & \boldsymbol{x}\in A.
\end{cases}
\end{equation}

The transverse structure of the domain wall is then determined by the celebrated 1D kink solution of the sine-Gordon equation:
\begin{equation}
\phi(\boldsymbol{x},y)\approx4\arctan\exp\big(m[y-h(\boldsymbol{x})]\big).
\end{equation}
This solution determines the action per $(d+1)$-dimensional unit surface area of the domain wall, allowing us to derive the constant prefactor in Eq.~\eqref{eq:VolumeLaw}:
\begin{equation}
\label{eq:SN:prefactor}
s_N = \frac{2}{3}\left(1 + \frac{1}{N}\right).
\end{equation}

\subsection{Discussion: Information-charge separation}
\label{sec:4D}

The performed analysis leads to an important conclusion: while interactions do not change the behavior of the charge fluctuations (on the semiclassical level), they dramatically affect the behavior of the entanglement entropy---a phenomenon that we call ``information-charge separation''. It can be understood based on purely symmetry reasoning. 

As we discussed above (cf.~Fig.~\ref{fig:potential-cartoon}), the interaction breaks down the symmetry of the NLSM from $\mathrm{SU}(R)$ down to $[\mathrm{U}(1)]^{R-1} \times \mathbb{S}_R$. Furthermore, the semiclassical solution for the charge fluctuations lies entirely in the subgroup $[\mathrm{U}(1)]^{R-1}$, which remains Goldstone because of charge conservation. Thus, on the semiclassical level, one does not expect qualitative changes in the behavior of charge cumulants compared to the non-interacting case. In contrast, it is the discrete subgroup $\mathbb{S}_R$ that determines the behavior of the entanglement entropy. As discussed in Sec.~\ref{sec:EEInt}, this leads to the volume-law scaling of the entanglement entropy.

\section{Renormalization group and phase diagram for interacting fermions}
\label{sec:RG}

In Sec.~\ref{Sec:Interaction}, we have derived the interacting NLSM and analyzed the effect of interaction on FCS and entanglement at the semiclassical level. 
The results obtained for the charge fluctuations and entanglement entropy are directly applicable in the limit $g\to \infty$, provided the symmetry breaking has occurred, $\ell_A \gg \ell_\text{int}$. 
Now, we proceed with our analysis of the model defined by Eq.~\eqref{eq:InteractionLagrangian} using the RG approach. 
This will allow us to describe the system at arbitrary $g$, to determine the transient scaling of observables at intermediate length scales, and to construct the phase diagram of monitored interacting fermions.

\subsection{Perturbative renormalization group equations}
\label{sec:pert-RG}

\begin{figure}
    \centering    \includegraphics[width=0.7\columnwidth]{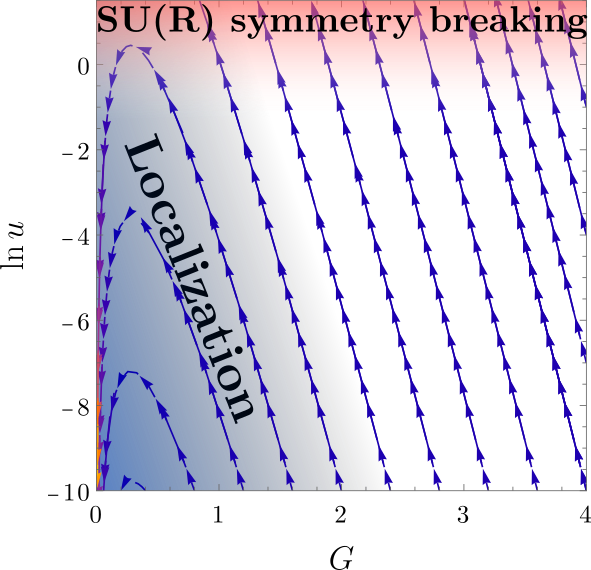}
    \caption{Renormalization-group flow according to Eqs.~\eqref{eq:RG} for the 1D model.   
    Blue region schematically marks area of initial parameters that flow to the weakly-interacting ``localized'' regime (i.e., $G \lesssim 1$ while $u \lesssim 1$, where the one-loop description is insufficient). Red region $u \gtrsim 1$ marks the area where RG equations become inapplicable because of $\mathrm{SU}(R)$ symmetry-breaking mass term becoming of the order of unity. If the system enters this regime (i.e., reaches $u \sim 1$) while $G$ is still large compared the unity, it is characterized by ``delocalized'' information and charge, i.e., volume law for the entanglement and logarithmic law for charge fluctuations.}
    \label{fig:RG}
\end{figure}

In $d$ dimensions, both couplings $g$ and $m$ in the interacting NLSM are dimensionful. Therefore, we will characterize the system by their dimensionless counterparts defined as:
\begin{equation}
G(\ell)=g(\ell)\ell^{d-1},\quad u(\ell)=m(\ell) \ell.
\label{eq:G-u-def}
\end{equation}
The one-loop RG equations for these two couplings are derived in Appendix~\ref{sec:appendix:RG}.
In the replica limit $R \to 1$, they read:
\begin{align}
\begin{split}
\frac{d G}{d\ln\ell}&=\beta_{G}(G,u) = (d-1)G-\frac{1}{4\pi},
\\
\frac{d\ln u}{d\ln\ell}&=\beta_{u}(G,u) = 1-\frac{7}{8\pi G}.
\end{split}
\label{eq:RG}
\end{align}
The first (constant) terms in the beta functions are ``natural'' engineering dimensions of corresponding operators, and the second terms represent quantum corrections. There also exists a one-loop correction term $O(u^4)$ in $\beta_G$, akin to the renormalization of the superfluid stiffness in the Berezinskii-Kosterlitz-Thouless (BKT) transition; however, it turns out that the prefactor in this term is proportional to $(R-1)$ and thus vanishes in the $R \to 1$ replica limit. The RG flow produced by Eqs.~(\ref{eq:RG}) for a 1D model is illustrated in Fig.~\ref{fig:RG}.

These perturbative RG equations are only valid provided $G(\ell) \gg 1$ (corresponding to the ``weak localization regime'') and $u(\ell) \ll 1$ (meaning that the interaction-induced mass is still small), and the RG flow has to be stopped once one of these two validity criteria violated. 

The first possibility is that, upon renormalization, the condition $G(\ell) \gg 1$ is violated first, i.e., the system reaches the regime $G(\ell) \sim 1$ when $u(\ell)$ is still small (blue region on Fig.~\ref{fig:RG}). At the corresponding length scale, the effect of interactions is still weak, so that the behavior of the system is qualitatively similar to that in the non-interacting case. This implies that the entanglement entropy and charge fluctuations remain related (like for Gaussian states). The flow into the strong-coupling regime signals localization, i.e., the area-law behavior for both charge and information.

The second possibility is that the system reaches region $u(\ell) \sim 1$ first. If we denote the corresponding scale by $\ell^\ast$, this means that $u(\ell^\ast) \sim 1$ and $G(\ell^\ast) \gg 1$. It follows from the second of RG equations \eqref{eq:RG} that $\ell^\ast \sim \ell_{\text{int}}$ to leading order, so that we will not make a distinction between $\ell^\ast$ and $\ell_{\text{int}}$ below.  As discussed earlier, at this scale, the $\mathrm{SU}(R)$ symmetry breaks down into two sectors with the symmetry $\mathbb{S}_R \times [\mathrm{U}(1)]^{R-1}$. 
The sector with the discrete symmetry group determines the behavior of the entropy. The effective dimensionless coupling constant, which governs the behavior in this sector, can be estimated as $K \sim G(\ell^\ast) u(\ell^\ast) \sim G(\ell^\ast)$. According to the reasoning in Sec.~\ref{sec:EEInt}, it determines the effective coupling between coarse-grained regions of the system corresponding to different permutations. This discrete symmetry can be spontaneously broken in  $d + 1$ dimensions for any spatial dimensionality $d > 0$, yielding a measurement-induced phase transition in the behavior of the entanglement entropy. The critical coupling, where such a transition happens, is naturally expected to be of the order of unity: $K_{c} \sim 1$.

In the large-$K$ region, the symmetry is spontaneously broken, meaning that there is a finite ``cost'' to create a domain wall, and, as a consequence, this phase exhibits the volume-law behavior of the entanglement entropy, see Eq.~\eqref{eq:VolumeLaw}.  The small-$K$ region 
of the $\mathbb{S}_R$ sector of the theory describes a ``paramagnetic'' phase 
with restored symmetry, implying the area-law scaling of the entanglement entropy. The correlation length, beyond which the area law is reached for entanglement, is expected, on general grounds, to exhibit a power-law scaling
\begin{equation}
\ell_{\text{E}}\sim\ell^*\,|K-K_{c}|^{-\nu_{\text{E}}},
\label{eq:lE}
\end{equation}
with a certain critical index $\nu_{\text{E}}$.

The remaining sector with continuous Goldstone symmetry $[\mathrm{U}(1)]^{R-1}$ governs the fluctuations of charge. 
The symmetry allows for vortex-like excitations, and their pairing in 1D systems can lead to the BKT transition. The field theory that incorporates vortex degrees of freedom in 1+1)-dimensional systems can be built in a standard manner, as detailed in Sec.~\ref{sec:BKT}. 

\subsection{Berezinskii-Kosterlitz-Thouless 
transition for charge fluctuations in the symmetry-broken regime}
\label{sec:BKT}

The effective field theory, which describes the $[\mathrm{U}(1)]^{R-1}$ sector, can be formulated in terms of $R$ phases $\Phi_r$, subject to constraint (which follows from $\det \hat{U} = 1$):
\begin{equation}
\label{eq:PhaseConstraint}
\sum_{r=1}^{R} \Phi_r = 0\quad (\mathrm{mod}\,2\pi),
\end{equation}
with the Lagrangian density
\begin{equation}
{\cal L}_R=\frac{g}{2}\sum_{r=1}^{R}(\partial_{\mu}\Phi_{r})^{2}.
\end{equation}
This Lagrangian is formally equivalent to the Lagrangian of (replicated) $(d+1)$-dimensional XY model. The bare value of the coupling constant $g$ at length scale $\ell_{\text{int}}$ is determined by the ``weak localization'' flow equations described in the previous section.

The main difference in comparison with the conventional XY model is that each individual vortex is described by $R$ quantized charges in each replica, and the constraint \eqref{eq:PhaseConstraint} requires that these charges add up to zero. This leads to an effective field theory for dual phases $\theta_r$ (cf.~Ref.~\cite{Barratt2022}):
\begin{equation}
\label{eq:BKTActionDual}
{\cal L}_{R}=\frac{1}{8 \pi^2 g}\sum_{r=1}^{R}(\partial_{\mu}\theta_{r})^{2}-\lambda\sum_{r^{\prime}>r}\cos\left(\theta_{r}-\theta_{r^{\prime}}\right),
\end{equation}
with coupling $\lambda$  playing the role of fugacity of the simplest replicated vortex with charges $(\pm 1, \mp 1)$ in replicas $(r, r^\prime)$, and subject to linear constraint $\sum_r \theta_r = 0$ also following from Eq.~\eqref{eq:PhaseConstraint}. 

Introducing the dimensionless fugacity $\kappa = \lambda \ell^2$ with the initial value $\kappa \sim \exp(-g)$ at $\ell\sim \ell_\text{int}$, we arrive at BKT-type flow equations (see Appendix~\ref{App:BKT}):
\begin{equation}
\label{eq:BKT:RG}
\begin{split}
\frac{d\kappa}{d\ln\ell} & =2\left(1-\pi g\right)\kappa,\\
\frac{dg}{d\ln\ell} & =-\kappa^{2}g^{4}.
\end{split}
\end{equation}
These equations possess a transition point separating the phase of ``screened Coulomb plasma'' and the power-law Goldstone phase.
The transition takes place at $g_c = 1/\pi$, which is twice smaller compared to the standard value $2 / \pi$ for the BKT transition, owing to the fact that the simplest vortex spans over two replicas \cite{Barratt2022}. This is a transition between area$\times$log scaling of the charge fluctuations at $g > g_c$, and area-law scaling at $g < g_c$. The critical scaling of the particle-number cumulant at the transition is logarithmic in a 1D system:
\begin{equation}
\label{eq:BKTScaling}
{\cal C}_{\ell}^{(2)}=\frac{2g_{c}}{\pi}\ln\frac{\ell}{\ell_\text{int}}=\frac{2}{\pi^{2}}\ln\frac{\ell}{\ell_{\text{int}}}\quad(\text{1D, transition}).
\end{equation}
On the localized side of the transition, the logarithmic behavior of charge cumulants becomes an intermediate asymptotic, saturating at  
distances $\ell > \ell_{\text{C}}$, where
$\ell_{\text{C}}$ is the BKT correlation length,
\begin{equation}
\ell_{\text{C}}\sim\ell_\text{int}\exp\left(1/\sqrt{|g-g_{c}|}\right).
\label{eq:lBKT}
\end{equation}

A question to be asked at this point concerns the mutual location of the two transitions (for information and charge). It is natural to expect that, if the discrete permutation symmetry  $\mathbb{S}_R$ is restored, then the Goldstone $[\mathrm{U}(1)]^{R-1}$ symmetry describing phase fluctuations around a replica permutation is restored as well. 
In other words, if entanglement is localized, then charge is localized as well. 
This suggests that the charge-fluctuation transitions takes place within the volume-law phase for the entanglement (cf.~Ref.~\cite{Barratt2022}). 
If this is indeed the case, the two transitions are controlled by fixed points 
of two different theories ($\mathbb{S}_R$ and 
$[\mathrm{U}(1)]^{R-1}$), as discussed above.

We speculate, however, that there might be an alternative possibility. Note that, according to the above discussion, both transitions take place when $\ell_\text{int}\sim \ell_\text{loc}$. The system, that has initially (at a scale $\ell_0$) small couplings $G(\ell_0), u(\ell_0) \ll 1$ then flows under RG \eqref{eq:RG} to a point where simultaneously $G(\ell) \sim 1$ and  $u(\ell) \sim 1$. We cannot exclude a possibility that there is a fixed point in this strong-coupling regime that controls the transition in both, entanglement and charge fluctuations. The difficulty in giving a definite answer on whether this may indeed happen stems from the fact that this requires understanding the RG flow at strong coupling, $G(\ell) \sim 1$ and  $u(\ell) \sim 1$, which is a highly challenging task. We will return
to this issue below in Sec.~\ref{sec:phase-dia} where we discuss the phase diagram.

\subsection{Evolution of observables with increasing subsystem size for 1D systems}

Having performed the RG analysis of the interacting NLSM, 
we proceed with the predictions for the behavior of observables for finite subsystem size $\ell$, assuming thermodynamic limit for the size of the whole system $L \to \infty$. However, we note that the scaling with $\ell$ will be the same for the case $\ell = L/2$ (as realized in the numerical simulations) by virtue of continuity, although numerical prefactors may differ.
For definiteness, we will focus our discussion here on the case of 1D systems. A generalization of our analysis on higher dimensions is straightforward.

\begin{figure}
    \centering
    \includegraphics[width=\columnwidth]{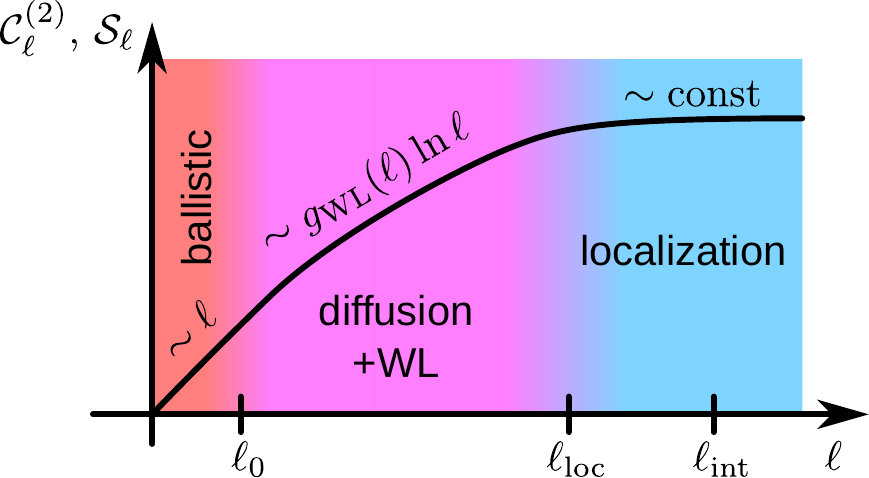}
    \caption{Characteristic scale dependence of entropy $\mathcal{S}_{\ell}$ and charge fluctuations $\mathcal{C}_{\ell}^{(2)}$ for 1D systems in the parameter region $\ell_{\text{loc}} \ll \ell_{\text{int}}$. Interactions are essentially unimportant in this case, and the length scale $\ell_\text{int}$ does not appear in the observables. For strong monitoring, when $\ell_\text{loc}\sim \ell_0$, the intermediate diffusion region disappears, and the ballistic ``volume law'' crosses over directly to the area-law regime, which might create an impression of a volume-to-area transition in small systems. For weaker monitoring, the system size can be smaller than the exponentially large $\ell_\text{loc}$. Then the finite-size results may suggest an area-to-log transition, which is absent in the truly thermodynamic limit.}
    \label{fig:behavior:localization}
\end{figure}

We start with the smallest subsystem sizes. Our predictions imply that both the charge fluctuations and entanglement in subsystems whose size is smaller or comparable to the ``mean free path'', Eq.~\eqref{eq:ell0}, $\ell\lesssim \ell_0$ (this condition is easily realized in not-too-large systems at weak monitoring), are governed by the ballistic dynamics. As was demonstrated in Ref.~\cite{Poboiko2023a}, the scaling of the particle-number cumulant and the entanglement entropy in a 1D free-fermion system in the ballistic regime is linear in $\ell$ (measured in units of the lattice constant):
\begin{equation}
\begin{split}
 &\mathcal{S}_{\ell} \simeq -\left[n_0\ln n_0 +(1-n_0)\ln(1-n_0)\right]\ell, \quad \ell\lesssim \ell_0,\\  &\mathcal{C}^{(2)}_{\ell}\simeq  n_0(1-n_0)\ell, \quad \ell\lesssim \ell_0.
 \end{split}  
    \label{eq120}
\end{equation}
For half-filling, the prefactors in front of $\ell$ are equal to $\ln2$ and $1/4$ for $\mathcal{S}_{\ell}$ and $\mathcal{C}^{(2)}_{\ell}$, respectively.

Interactions with $\ell_\text{int}>\ell_0$ (which is the assumption used throughout the paper) do not essentially affect the ballistic scaling.
As a result, numerical simulations at weak monitoring can deliver a deceptive identification of a ``volume-law-phase'' behavior for charge fluctuations and entanglement. This behavior will, in fact,  transform into the sub-extensive scaling~\eqref{eq:second-cumulant} with increasing the system size. 

\begin{figure*}
    \centering    \includegraphics[width=\textwidth]{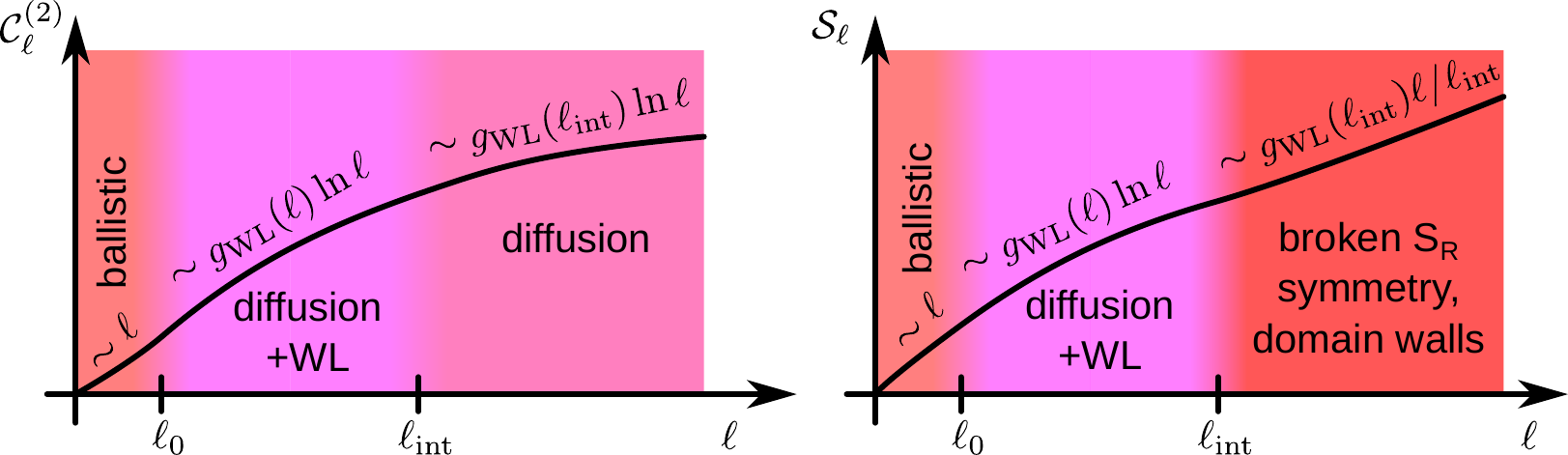}
    \caption{
    Characteristic scale dependence of charge fluctuations $\mathcal{C}_{\ell}^{(2)}$ (left) and entanglement entropy $\mathcal{S}_{\ell}$ (right) in 1D systems when $\ell_{\text{loc}} \gg \ell_{\text{int}}$. 
    Beyond the length scale $\ell_{\text{int}}$, the breaking of $\mathrm{SU}(R)$ symmetry stops the localization, leading to stabilization of the logarithmic phase for charge fluctuations (left panel) and emergence of volume-law behavior for entanglement (right panel).}
    \label{fig:behavior:int}
\end{figure*}

\begin{figure*}
    \centering    \includegraphics[width=\textwidth]{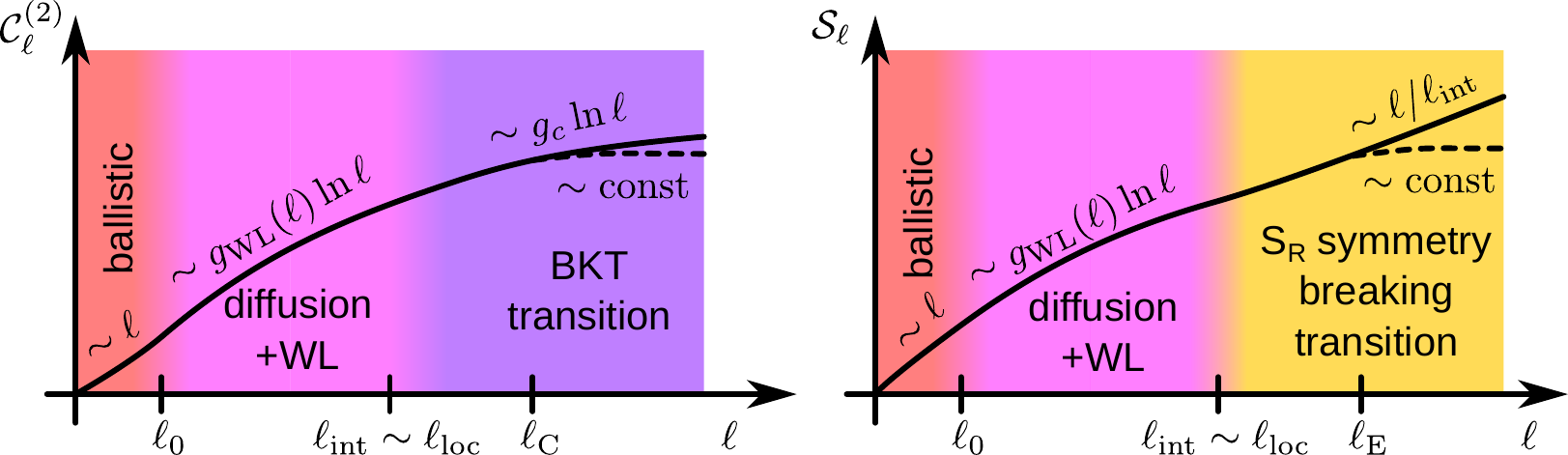}
    \caption{
    Characteristic scale dependence of charge fluctuations $\mathcal{C}_{\ell}^{(2)}$ (left) and entanglement entropy $\mathcal{S}_{\ell}$ (right) in 1D systems when $\ell_{\text{loc}} \sim \ell_{\text{int}}$. Left panel: beyond the length scale $\ell_{\text{int}}$, the BKT physics develops in the $[\mathrm{U}(1)]^{R-1}$ sector of the theory, yielding a BKT transition between the logarithmic (solid) and area-law (dashed) behavior of the cumulant in the thermodynamic limit, which is reached at scales larger than $\ell_{\text{C}}$. The transition is marked by the critical value $g_c = 1 / \pi$ [see Eq.~\eqref{eq:BKTScaling}]. Right panel: the transition between the volume-law (solid) and the area-law (dashed) behavior of entanglement takes place in the symmetry-broken regime and becomes evident at scales larger than $\ell_{\text{E}}$.}
    \label{fig:behavior:crit}
\end{figure*}

As discussed in Sec.~\ref{sec:pert-RG}, the behavior at larger scales depends crucially on the relation between the measurement-induced localization scale $\ell_\text{loc}$ [Eqs.~\eqref{lcorr-BDI} and \eqref{lcorr-AIII}] and $\ell_\text{int}$ [Eq.~\eqref{eq:l-int}].
Using Eqs.~\eqref{lcorr-AIII}, \eqref{eq:l-int} and setting $J=1$, the condition 
$\ell_\text{int}\sim \ell_\text{loc}$ translates for $\gamma\ll 1$ and half-filling into
\begin{equation}
V\sim \sqrt{\gamma}\exp\left(-\sqrt{2}\,\pi/\gamma\right). 
\label{eq:V-crit}
\end{equation}
If the interaction $V$ is much smaller (larger) than the right-hand side of Eq.~\eqref{eq:V-crit}, we have $\ell_\text{loc}\ll \ell_\text{int}$
(respectively, $\ell_\text{loc}\gg \ell_\text{int}$).

Let us first consider the case $\ell_\text{loc}\ll \ell_\text{int}$.
For a very high measurement rate, $\ell_\text{loc}\sim \ell_0$, the system gets immediately localized (the interaction does not play a role), showing the area-law behavior. 
For less frequent measurements, a window opens for the diffusive dynamics, $\ell_0\ll \ell \ll \ell_\text{loc}\ll \ell_\text{int}$, before localization sets in. The interaction-induced mass in replica-off-diagonal Goldstone modes can be neglected. The behavior of the charge fluctuations and entanglement follows the prediction~\cite{Poboiko2023a} for non-interacting systems: the semiclassical logarithmic scaling is affected by the weak-localization correction (cf. Appendix~\ref{sec:appendix:WL-BDI-vs-AIII}):
\begin{equation}
 \frac{3}{\pi^2}\,    \mathcal{S}_{\ell} \simeq \mathcal{C}^{(2)}_{\ell}
     \simeq \frac{2}{\pi}\,
     g_\text{WL}(\ell)\ln \ell, \quad \ell_0\ll \ell\ll \ell_\text{loc},\ell_\text{int} \,,
\label{eq122}   
\end{equation} 
where $g_\text{WL}(\ell)$ is governed by Eq.~(\ref{eq:RG}). When $\ell$ reaches $\ell_\text{loc}$, a crossover to the area-law regime [Eq.~(\ref{eq121})] occurs:
both $\mathcal{C}^{(2)}_{\ell}$ and $\mathcal{S}_{\ell}$ do not depend on the subsystem size (area law) when it exceeds $\ell_\text{loc}$:
\begin{equation}
   \mathcal{S}_{\ell}, \, \mathcal{C}^{(2)}_{\ell}\sim g_0^2\propto \text{const}(\ell), \quad \ell\gtrsim \ell_\text{loc} \ \ (\ell_\text{loc}\ll \ell_\text{int}). 
\label{eq121}   
\end{equation}
These results for $\ell_\text{int}\gg \ell_\text{loc}$ are illustrated in Fig.~\ref{fig:behavior:localization}.

In the opposite limit, $\ell_\text{int}\ll \ell_\text{loc}$,
the behavior of observables at scales between $\ell_0$ and $\ell_\text{int}$ is still barely affected by the interaction and is again described by Eq.~(\ref{eq122}).
For $\ell\gtrsim \ell_\text{int}$, however, the situation changes drastically: the interaction now gaps some of the Goldstone modes and gives rise to charge-information separation.
The coupling $g$ at $\ell\sim \ell_\text{int}$ is still large.
The BKT RG, Eq.~(\ref{eq:BKT:RG}), then predicts that the renormalization of $g$ is negligible. 
The particle-number cumulant is described by the semiclassical result, Eq.~(\ref{eq:second-cumulant}), where $g$ is given by $g_\text{WL}(\ell_\text{int})$, corresponding to the purely logarithmic scaling of charge fluctuations in the thermodynamic limit:
\begin{equation}
    \mathcal{C}^{(2)}_{\ell}\simeq \frac{2}{\pi}\, g_\text{WL}(\ell_\text{int})\ln \ell, \quad \ell\gtrsim \ell_\text{int} \ \ (\ell_\text{int}\ll \ell_\text{loc}). 
    \label{C-log-law}
\end{equation}
Turning to the behavior of the entanglement entropy for $\ell\gtrsim \ell_\text{int}$ for the interaction-dominated regime $\ell_\text{int}\ll \ell_\text{loc}$, we use Eq.~\eqref{eq:VolumeLaw} with the renormalized value of $g$ at the length scale $\ell_\text{int}$. 
The asymptotic behavior of the entropy is then given by 
\begin{equation}
    \mathcal{S}_{\ell}\simeq \frac{4}{3} g_\text{WL}(\ell_\text{int})\frac{\ell}{\ell_\text{int}}, \quad \ell\gg \ell_\text{int} \ \ (\ell_\text{int}\ll \ell_\text{loc}), 
    \label{eq125}
\end{equation}
which corresponds to the volume-law phase. Thus, for $\ell_\text{int}\ll \ell_\text{loc}$, both observables demonstrate the growth with increasing system size---``metallic'' behavior. This is illustrated in Fig.~\ref{fig:behavior:int}.

Finally, we address the scale dependence of observables around $\ell_\text{loc}\sim\ell_\text{int}$,
where the transition from metallic to insulating behavior takes place.
In the semiclassical formula for the particle-number cumulant, Eq.~(\ref{eq:second-cumulant}), $g$ is now
replaced with the decreasing coupling $g_\text{BKT}(\ell)$ governed by the BKT RG, Eq.~(\ref{eq:BKT:RG}). Exactly at the BKT separatrix, one has $g_\text{BKT}(\infty) = g_c=1/\pi$, yielding the critical scaling of the cumulant, Eq.~(\ref{eq:BKTScaling}).
On the delocalized side of the charge-fluctuation transition, the coupling $g_\text{BKT}(\ell)$ reaches a constant value  $g_\text{BKT}(\infty) > 1/\pi$ (BKT line of fixed points) at the characteristic length $\ell_\text{C}$ given by Eq.~\eqref{eq:lBKT}. The scaling of the cumulant is described by Eq.~(\ref{C-log-law}), with a replacement
$g_\text{WL}(\ell_\text{int}) \mapsto g_\text{BKT}(\infty)$. 
When the initial condition for the BKT RG is on the opposite (``localized'') side of the separatrix, the charge fluctuations first follow the critical logarithmic scaling and then saturate  at the scale  $\ell_\text{C}$,
\begin{equation}
 \mathcal{C}^{(2)}_{\ell}\simeq \frac{2}{\pi^2}\ln\ell_\text{C}, \quad g<g_c,\ \ell\gg \ell_\text{C} \,.
\end{equation}
which means that, in the thermodynamic limit, the system is in the area-law phase. 

As discussed in Sec.~\ref{sec:EEInt} above Eq.~\eqref{eq:VolumeLaw}, the leading correction to the volume-law growth of $\mathcal{S}$ in the ``metallic'' regime, Eq.~\eqref{eq125}, is logarithmic in $\ell$. One can then speculate that the following behavior might persist near the transition:
$\mathcal{S}_{\ell}\sim s_\ell\, \ell + s_\text{log} \ln \ell$,     
where $s_\ell$ and $s_\text{log}$ are coefficients depending on $\gamma$ and $V$, with $s_\ell$ vanishing at the transition $K=K_c$. 
Assuming that $s_\text{log}$ stays finite at the transition, the critical behavior of the entropy at the entanglement transition will be logarithmic, as in $d=1$ hybrid quantum circuits described by conformal field theories
(see, e.g., Refs.~\cite{Jian2020a,Li2021}).
Near the transition for $K>K_c$,
the volume-law asymptotics in the thermodynamic limit is established at $\ell\gtrsim \ell_\text{E}\gg \ell_\text{int}$.
For stronger monitoring, $K<K_c$, the entropy growth is cut off at $\ell_\text{E}$ [Eq.~\eqref{eq:lE}], leading to the crossover to the area law, 
\begin{equation}
\mathcal{S}_{\ell}=\text{const}(\ell),\quad K<K_c, \  \ell\gg \ell_\text{E}.   
\label{eq131}
\end{equation}
The analysis of the critical behavior of entanglement at the transition in our model is a challenging task that requires a separate study. However, the precise form of the critical scaling does not affect qualitatively the picture of volume-to-area-law transition, with Eqs.~(\ref{eq125}) and (\ref{eq131}) describing the entropy scaling in the corresponding phases. 

The above results for the charge fluctuations and entanglement at $\ell_\text{int}\sim \ell_\text{loc}$  (i.e., around the transitions for charge and information) are illustrated in Fig.~\ref{fig:behavior:crit}.

\subsection{Phase diagrams for 1D and 2D monitored interacting systems}
\label{sec:phase-dia}

\begin{figure*}
    \centering
    \includegraphics[width=\textwidth]
    {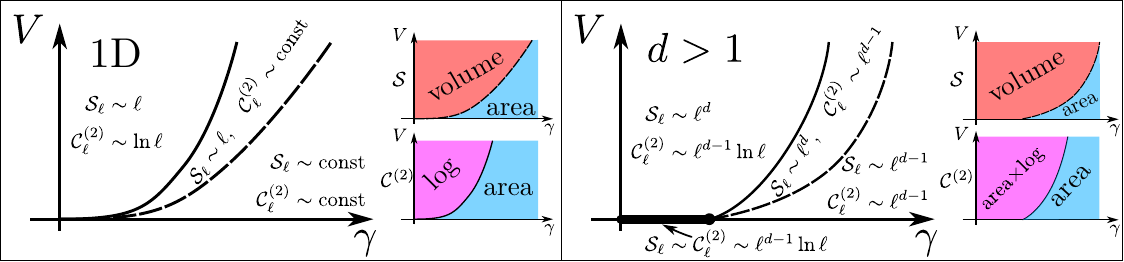}
    \caption{Schematic phase diagrams for 1D  (left) and higher-$d$ (right) systems. Insets show separately the phase diagrams for entanglement entropy $\mathcal{S}$ and for charge fluctuations characterized by $\mathcal{C}^{(2)}$. Dashed lines mark the area-to-volume-law transition for $\mathcal{S}$, associated with the spontaneous breaking of discrete replica permutation symmetry. Solid lines mark the transition between the area-law and  area$\times$log phases for charge fluctuations, associated with the BKT transition (1D) or spontaneous breaking ($d>1$) of the $[\mathrm{U}(1)]^{R-1}$ symmetry.  Both lines are given by $\ell_{\text{int}} \sim \ell_{\text{loc}}$,
    yielding Eq.~\eqref{eq:V-crit},
    so that the intermediate phase is expected to be rather narrow. It is also possible that, in the range of sufficiently small $V$, this phase shrinks so that there is a single transition for both charge and information.}
    \label{fig:phase-dia}
\end{figure*}

Finally, we present the phase diagrams for monitored interacting fermions in dimensions $d=1$ and $d>1$ (exemplified by the $d=2$ case). In the absence of interactions, the phases for entanglement and particle-number fluctuations coincide \cite{Poboiko2023a, Poboiko2023b} in view of the Klich-Levitov identity. In 1D non-interacting system, there is only a single phase---the area-law phase, where both the entanglement entropy and the particle-number cumulant saturate in the thermodynamic limit because of ``Anderson localization'', irrespective of the measurement rate.
For non-interacting in dimension $d>1$, an Anderson-like ``metal-insulator transition'' takes place at a critical value of $\gamma$,
separating the area-law phase at $\gamma>\gamma_c$ and the area$\times$log phase at $\gamma<\gamma_c$. The resulting phase diagrams contain only one axis. 
The inclusion of interactions (parametrized by $V>0$) adds another dimension to phase diagrams that are now defined in the  
$\gamma$-$V$ plane, with the $V=0$ axis corresponding to the non-interacting case.
In addition, the behavior of charge fluctuations and entanglement decouples as a manifestation of charge-information separation.

The phase diagrams for monitored interacting fermions are shown in Fig.~\ref{fig:phase-dia}. The left panel refers to the case $d=1$. The main plot shows phase diagrams for entanglement and charge fluctuations (also shown separately in the insets). The transition lines are determined by the condition $\ell_\text{int}\sim\ell_\text{loc}$,
which translates (for $\gamma\ll 1$ and a half-filled band) into 
Eq.~\eqref{eq:V-crit}.
The phase above both lines is characterized by the following behavior of observables: 
$\mathcal{S}\propto \ell$ and $\mathcal{C}^{(2)}\propto \ln\ell$ (volume law for entanglement and logarithmic law for the particle-number fluctuations), while in the phase below both lines we have $\mathcal{S}\propto  \text{const}$ and $\mathcal{C}^{(2)}\propto \text{const}$ (area law for both entanglement and particle-number fluctuations). Finally, an intermediate phase (between the two lines) is characterized by $\mathcal{S}\propto \ell$ and $\mathcal{C}^{(2)}\propto \text{const}$ (volume law for entanglement and area law for the particle-number fluctuations). 
We emphasize that both transition lines are governed by the same condition $\ell_\text{int}\sim\ell_\text{loc}$, i.e., may differ by a numerical coefficient only. The intermediate phase is thus {\it not} parametrically wide. 

As was discussed in the end of Sec.~\ref{sec:BKT}, we cannot exclude a possibility that the two phase transition lines actually coincide for small $V$ and $\gamma$ (where our derivation of the NLSM field theory holds). On the other hand, for $V\sim 1$ and $\gamma \sim 1$, the entanglement and charge fluctuations are decoupled already at the microscopic scale, making it very likely that the two transitions are distinct in this part of the phase diagram. There is thus a possibility that there is a single phase boundary for information and charge at small $V$ and $\gamma$, which then splits into two phase boundaries at a certain point where $V\sim 1$ and $\gamma \sim 1$.

The phase diagram is presented for sufficiently weak interactions; 
in this case, the critical measurement strength for each of the transition lines monotonically increases as a function of $V$, as we have derived analytically for small $V$, Eq.~\eqref{eq:V-crit}. We speculate that, for stronger interactions,  the transition lines bend toward smaller values of $\gamma$, since a very strong interaction is expected to block dynamics, thus favoring the quantum Zeno effect and correspondingly the area-law phase.

The phase diagram for the 2D case (and, more generally, for $d>1$) is shown in the right panel of Fig.~\ref{fig:phase-dia}. The transition lines now start at the point $\gamma=\gamma_c,\, V=0$, which is the transition point separating the area-law phase from the area$\times$log phase in non-interacting systems. With finite interaction included, we again have three phases, as in the 1D case, with the logarithmic phase for the charge fluctuations becoming the area$\times$log phase. Noteworthy, an arbitrarily weak interaction destabilizes the area$\times$log phase for entanglement, which was found for free fermions, driving the entanglement entropy for the same values of $\gamma$ to grow according to the volume law in the limit of large system size. 
The condition $\ell_{\text{int}} \sim \ell_{\text{loc}}$, which determines both transition lines, translates for $d>1$ into
\begin{equation}
V\sim (\gamma-\gamma_{c})^{\nu},
\end{equation}
with $\nu$ being the correlation-length critical index for the associated transition in the non-interacting system. For $d=2$, the value of $\nu$ was found to be ${\nu\approx1.4}$ by numerical modeling in Ref.~\cite{Poboiko2023b}.

\section{Numerical analysis}
\label{sec:numerics}

To verify and complement the analytical predictions, we have performed a numerical analysis of the problem. Specifically, we studied numerically the model defined by the Hamiltonian  
$\hat{H}=\hat{H}_{0}+\hat{H}_{\text{int}}$ given by Eqs.~\eqref{eq:H0} and \eqref{eq:Hint}, with the nearest-neighbor hopping $J$ and nearest-neighbor interaction $V$, subjected to projective measurements of on-site densities with a rate $\gamma$. This can be viewed as a ``minimal model'' of monitored interacting fermions (with a time-independent Hamiltonian). In the non-interacting limit ($V=0$), a detailed numerical study of this model (which belongs to the symmetry class BDI, see above) was carried out in Ref.~\cite{Poboiko2023a} where system sizes up to $L=2000$ were reached. Clearly, in the presence of interaction, the range of $L$ accessible to a numerical study is substantially reduced. 

For all numerical simulations, the nearest-neighbor hopping strength is fixed at \(J\equiv 1\), i.e., all energy parameters are given in units of $J$. We choose the open boundary conditions. The system is initiated in a half-filled product state (\(n_0 = 1 / 2\)), with half of the sites (chosen randomly) occupied and the remaining sites empty. This state is then time-evolved according to our protocol for \(5 \Delta t_{\rm obs}= 25 / \gamma\) units of time, which means that during this stage every site is measured (on average) 25 times. 
We have verified that the time offset of \(5\Delta t_{\rm obs}\) is indeed sufficient to reach the dynamical steady state. 
Starting from this offset, observables are calculated in time intervals of \(\Delta t_{\rm obs}\) until \(t_{\max} = 100 / \gamma\) is reached. To obtain a numerical result for the sought average~\eqref{eq:TrajectoryAverage},
observables are averaged over these time points and, in addition, over multiple quantum trajectories.

In full correspondence with the analytical part of our investigation, we consider numerically the observables that characterize entanglement and the particle-number fluctuations. First, we calculate the standard (von Neumann, \(N \rightarrow 1\)) entanglement entropy \(\mathcal{S}_A\), Eq.~\eqref{eq:EntanglementEntropy}, at the half-cut \(A = \{1, 2, \ldots, L / 2\}\). 
Second, we determine the second cumulant of the particle number 
$\mathcal{C}^{(2)}$, Eq.~\eqref{eq:second-cumulant}, for the same region \(A\). 
In the plots below, we rescale the cumulant according to
\begin{equation}
\tilde{\mathcal{C}}^{(2)} = (\pi^2 / 3)\, \mathcal{C}^{(2)},
\end{equation}
to match the leading term of the expansion \eqref{eq:KlichLevitov} that holds in the free-fermion case (Gaussian states). 
As was shown in Ref.~\cite{Poboiko2023a}, \(\tilde{\mathcal{C}}^{(2)}\) approximates \(\mathcal{S}\) extremely well in the non-interacting case in the whole range of the considered measurement strengths, so that substantial deviations between \(\tilde{\mathcal{C}}^{(2)}\) and \(\mathcal{S}\) reveal the effect of interactions. In addition, we calculate the (minus) particle-number covariance \(\mathcal{G}\) 
\begin{align}
    \mathcal{G}_\text{BC} \equiv -\overline{\langle \hat{N}_B \hat{N}_C \rangle - \langle \hat{N}_B \rangle \langle \hat{N}_C \rangle}
\end{align}
between the first and last third of the system (\(B=\{1, \ldots, L / 3\}\) and \(C = \{L - L / 3, \ldots, L\}\)). This quantity was shown in Ref.~\cite{Poboiko2023b} to be the proper scaling quantity characterizing the ``Anderson-localization'' physics for monitored free fermions, where it plays a role similar to the conductance in the RG analysis of Anderson transitions. For Gaussian states, $\mathcal{G}$ is directly related to the mutual information $I(B:C)$ via the Klich-Levitov identity~\ref{eq:KlichLevitov},
\begin{equation}
I(B:C)={\cal S}(B)+{\cal S}(C)-{\cal S}(B\cup C)\simeq\frac{2\pi^{2}}{3}\mathcal{G}_\text{BC};
\label{mutual}
\end{equation}
this relation is broken by interactions, as they destroy Gaussianity.

We started numerical studies of our interacting model by performing exact simulations (``exact diagonalization'', ED) on the full Hilbert space for relatively small systems of length $L \le 24$. 
The ED simulations were performed using the \texttt{QuSpin} package~\cite{quspin1,quspin2}.
It turns out, however, that, for the considered model, the system sizes $L \le 24$ are too small, so that resolving the  
regimes between the ballistic one (at small $\gamma$) and fully localized one (at large $\gamma$) is very problematic. We used the ED to assess the accuracy of other numerical methods discussed below. 

The main computational approach used in this work is a variant of the time-dependent variational principle for matrix product states~\cite{tdvp_mps, tdvp_for_ql}, see Sec.~\ref{sec:numerics:tdvp},
which has allowed us to proceed reliably up to system sizes $L=72$ in the relevant range of $\gamma$. As an auxiliary approach for studying density fluctuations, we also used time-dependent Hartree-Fock simulations~\cite{tdhf_eq_ref}, see Sec.~\ref{sec:numerics:tdhf}.

All these methods provide ways to calculate the unitary time evolution of the many-body state, either exactly (ED) or within some approximation (MPS-TDVP and TDHF). The effect of projective measurements is taken into account exactly in all the approaches. Crucially, our numerical methods give access to individual quantum trajectories, with an outcome of each measurement chosen at random, according to the Born rule~\eqref{eq:BornRule}.

\subsection{Time-dependent variational principle with matrix product states}
\label{sec:numerics:tdvp}

\begin{figure*}
    \centering
    \includegraphics[width=\textwidth]{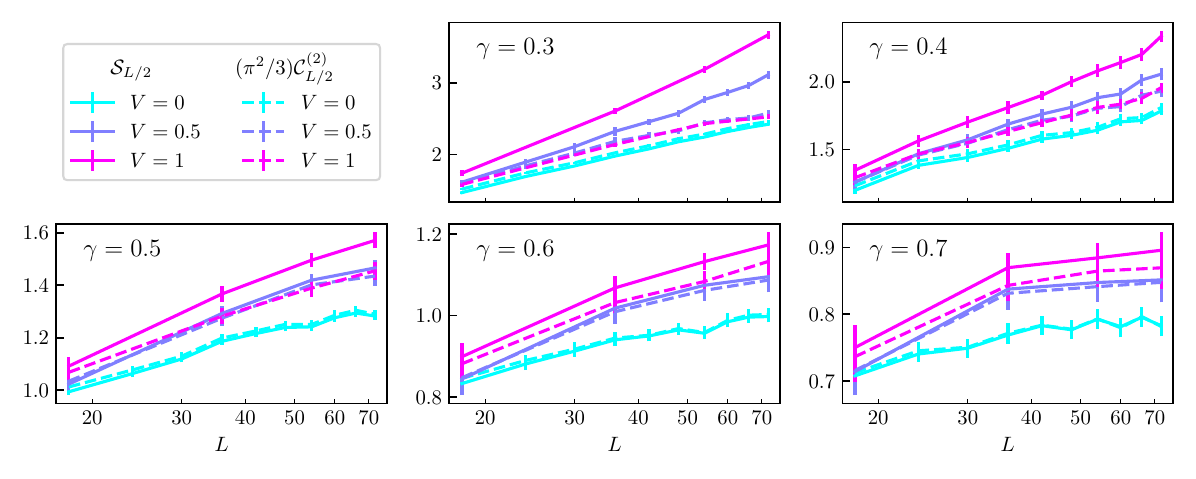}
    \caption{MPS-TDVP results for the entanglement entropy $\mathcal{S}$ (solid) and the rescaled second particle-number cumulant $\mathcal{C}^{(2)}$ (dashed), for measurement rate $\gamma \in [0.3, 0.7]$ and interaction strength $V = \{0.5, 1\}$, together with exact results for non-interacting system ($V = 0$) for comparison.
    Both quantities are evaluated for a cut dividing the system in two parts of equal length $L/2$ and are shown as functions of $L$. 
    For \(\mathcal{S} \leq 2.25\), the maximum bond dimension \(\chi_{\rm max} = 128\) was imposed. 
    For \(\mathcal{S} > 2.25\), the simulations were performed with  \(\chi_{\rm max} = 256\), except for the last two points ($L=54$ and 72) in the upper left plot ($V=1$, $\gamma=0.3$) where \(\chi_{\rm max} = 512\) was used.
    The dashed lines for $V=0.5$ and $V=1$ are almost indistinguishable from each other for $\gamma=0.3,\,0.4,$ and 0.5.}
    \label{fig:tdvp-entropy-cumulant}
\end{figure*}

The TDVP approximates the time evolution of a state in the space of MPS with a limited bond dimension~\cite{tdvp_for_ql}. This is achieved by projecting the time evolution of the state on the MPS subspace and numerically integrating the resulting equations with a small time step~\cite{tdvp_mps}. An MPS is characterized by a set of bond dimensions that characterize the bonds between the system sites by the maximum number of non-zero singular values for a Schmidt decomposition of the state on that bond. Any state in a 1D system can be accurately described by an MPS with a maximum bond dimension exponential in the number of system sites. However, the computational complexity of an MPS algorithm scales with the bond dimensions, so that, for large $L$, the method is applicable in practice only for states that can be well approximated by MPS with a not-too-large bond dimension, i.e., for states with relatively low entanglement.

The MPS-TDVP has been used successfully to describe the time evolution of measured interacting systems~\cite{Goto2020a,VanRegemortel2021a,Tang2020a,Doggen2022a,Doggen2023,doggen2023ancilla,Cecile2024}. Applicability of this approach is particularly clear in the case of an area-law phase, for which R{\'e}nyi entropies of the system are bounded by $L$-independent values. Consequently, the state can be represented as an MPS with a relatively small bond dimension, and its time evolution can thus be simulated efficiently with MPS-based methods~\cite{mps_simulability}. However, we are also interested in studying systems with the half-chain entropy growing as a function of the system size $L$. In particular, in the volume-law phase, the bond dimension required for an accurate simulation grows exponentially as a function of system size~\cite{mps_simulability}, so that describing a system with an MPS approach (with a decent accuracy) eventually becomes computationally unrealistic when $L$ grows. 
Nonetheless, there is a range of parameters for which the MPS approach permits one to proceed controllably to  system sizes $L$ considerably larger than those accessible to ED. This is, in particular, expected to be the case for the vicinity of the measurement-induced entanglement transition, where the growth of the entanglement entropy with $L$ is relatively slow. We use this idea in the present work.

The entropy of an MPS, cut off at a bond of dimension \(\chi\), cannot be larger than \(\mathcal{S}_{\rm max}(\chi) = \ln \chi\). For our simulations, we have found that the bond dimension \(\chi_{\rm max}=128\), with $\mathcal{S}_{\rm max}(2^7) \approx 4.85$,  is sufficient to simulate states with a half-chain entanglement entropy up to \(\mathcal{S} \approx 2.25\). This result, which is in agreement with observations in Ref.~\cite{doggen2023ancilla}, follows from the comparison the values of observables obtained with \(\chi_{\rm max}=128\) and with a doubled bond dimension, \(\chi_{\rm max}=256\). 

Raising the bond dimension by a factor of two increases the maximum entropy by $\ln 2 \approx 0.69$. This implies that the MPS calculation with \(\chi_{\rm max}=256\), with $\mathcal{S}_{\rm max}(2^8) \approx 5.54$, describes the system well up to $\mathcal{S}\approx 2.94$. 
In view of this, we limit the bond dimension to \(\chi_{\max} = 128\) for entropy values below \(\mathcal{S} = 2.25\) and use \(\chi_{\max} = 256\) in the range of parameters $(V, \gamma, L)$ where $\mathcal{S} > 2.25$. Almost all our data satisfied the condition $\mathcal{S} < 3$. For two points in the parameter space, where this condition was not fulfilled ($V=1$, $\gamma=0.3$, $L=54$ and 72), we carried out simulations with \(\chi_{\max} = 512\), which should give a good accuracy in the range $\mathcal{S}\le 3.63$. This was well satisfied for $L=54$ but still slightly violated for $L=72$. We thus argue that all our MPS-TDVP results have good accuracy, with a somewhat higher numerical error expected for the latter point, see a detailed discussion below. 

\begin{figure*}
    \centering    \includegraphics[width=0.75\textwidth]{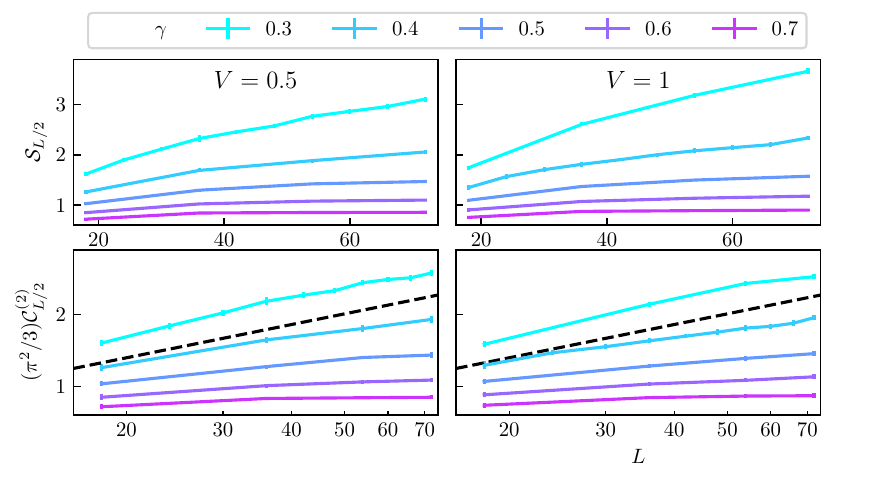}
    \caption{Entanglement entropy (top) and rescaled charge cumulant (bottom) as obtained by MPS-TDVP simulations at different measurement rates $\gamma$, as functions of system size $L$, for interaction strengths $V = 0.5$ (left) and $V = 1$ (right).
    This is an alternative presentation of the data shown in Fig.~\ref{fig:tdvp-entropy-cumulant}.
    Note that the scale on the $L$ axis is linear in the entanglement plots and logarithmic in the cumulant plots.
    The data visualize a transition from area law to volume law for the entanglement entropy and from area law to logarithmic law for the charge cumulant. Note that the largest-$L$ points on upper ($\gamma=0.3$) curves in the $V=1$ graphs are probably prone to a larger error in view of not fully sufficient MPS bond dimension, so that the actual values of the entropy and the cumulant for this point are expected to be slightly larger, see a discussion in the text. 
    The dashed straight lines in the right panels are  $(2/3)\ln L+\const$, corresponding to the critical slope predicted by Eq.~\eqref{eq:BKTScaling}.}
            \label{fig:tdvp-summary}
\end{figure*}

To control the accuracy of our MPS-TDVP simulations, we used several complementary approaches. First, as mentioned above, we benchmarked the simulations for relatively small systems by comparing them to the ED results. Second, we performed a comparison to exact results obtained by means of time-evolved Green's functions in the absence of interaction (cf. Appendix~\ref{sec:appendix:WL-BDI-vs-AIII}), in which case the exact calculation can be carried out in the whole range of length used in our simulations \cite{Poboiko2023a}. Third, as also explained above, we verified the accuracy by performing simulations with different bond dimensions \(\chi_{\max}\). The saturation with respect to the  bond dimension is illustrated in Fig.~\ref{fig:entE:bonddim} of
Appendix~\ref{sec:appendix:numerical-details} for the case of the largest entanglement in our simulations  (largest system size, $L=72$, strongest interaction, $V=1$, and smallest $\gamma=0.3$). Finally, to optimize our simulations, we used a variant of the TDVP algorithm that allows one to keep all bond dimensions variable throughout the simulation in an efficient way~\cite{cbe_dmrg,cbe_tdvp,comment_on_cbe}, see Appendix~\ref{sec:appendix:numerical-details} for details.

While our analytical theory was developed for a weak interaction (small $V$), the interaction-induced length $\ell_{\rm int}$ is parametrically large in this case, see Eq.~\eqref{eq:l-int}, which makes it very difficult to study interaction-induced effects numerically. For this reason, we perform simulations for quite strong interaction, choosing two values, $V=0.5$ and $V=1$.

In Fig.~\ref{fig:tdvp-entropy-cumulant}, numerical results for $\mathcal{S}$  (solid lines) and \(\tilde{\mathcal{C}}^{(2)}\) (dashed lines) are presented for \(V=0.5\) 
and \(V= 1\). 
The measurement rate $\gamma$ is chosen to be
$\gamma = 0.3$, 0.4, 0.5, 0.6, and 0.7,
increasing from the first to the last panel. For comparison, we also show the corresponding data for a non-interacting system obtained by exact Green's function simulations. As discussed above, the maximum bond dimension in the MPS-TDVP simulations for interacting systems was set to be \(\chi_{\rm max}=128\), 256, or 512, depending on the value of $\mathcal{S}$, see also the figure caption. 
For each set of parameters ($V, \gamma, L$), we simulated from 20 (for \(\gamma = 0.3\)) to 40 (for \(\gamma =  0.7\)) different quantum trajectories. The vertical bars represent a statistical error from finite averaging, which grows with increasing measurement strength because of increasing sample-to-sample fluctuations. Statistical error bars are calculated using a bootstrap procedure~\cite{bootstrap}.

Let us first discuss the results for $V=0.5$ (dark blue curves).
The highest considered measurement rate \(\gamma=0.7\) is chosen such that the entropy and the particle-number cumulant can be seen to approach a plateau. The saturation of the entanglement entropy and the cumulant as a function of system size indicates that this point in the parameter space ($V=0.5$, $\gamma=0.7$) corresponds to the area-law phase. Furthermore, the (rescaled) cumulant $\tilde{\mathcal{C}}^{(2)}$ is almost identical to the entropy \(\mathcal{S}\), as in the non-interacting case. This is exactly what is expected when the system is deeply in the area-law phase, with the ``localization length'' $\ell_{\text{loc}}$ being considerably smaller than the interaction-induced length $\ell_{\text{int}}$. The only visible effect of the interaction here is a relatively small ($\sim 8\%$) renormalization of the saturation value.

As the measurement strength is gradually reduced to $\gamma = 0.6$ and then $\gamma =0.5$, the full saturation is not reached anymore. At the same time, the data show a clear tendency to saturation. This is a manifestation of the fact that, with decreasing $\gamma$, the ``localization length'' $\ell_{\text{loc}}$ gradually increases, becoming comparable with, and eventually larger than, our largest system size $L = 72$. The results for $\tilde{\mathcal{C}}^{(2)}$ and \(\mathcal{S}\) remain almost identical for $\gamma = 0.6$ and 0.5, with the overall behavior very similar to that of the non-interacting model. This indicates that the interacting system (with $V=0.5$) is still in the area-law phase with respect to both $\tilde{\mathcal{C}}^{(2)}$ and \(\mathcal{S}\) at these values of $\gamma$.

With further decrease of the measurement rate, $\gamma = 0.4$ and 0.3, the behavior changes qualitatively. First, a clear difference between $\tilde{\mathcal{C}}^{(2)}$ and 
\(\mathcal{S}\) emerges, which grows with increasing $L$. Second, as is seen in the $\gamma=0.3$ panel, while the cumulant $\tilde{\mathcal{C}}^{(2)}$ grows logarithmically with $L$, the entropy $\mathcal{S}$  exhibits a faster-than-logarithmic growth, indicating the onset of the volume law. This is in full consistency with the behavior that is predicted analytically, see Secs. \ref{sec:Gaussian_approx} and \ref{sec:RG}.

For $V=1$ (magenta curves), the overall picture of the evolution from $\gamma=0.7$ to $\gamma=0.3$ is similar but the effect of interaction is somewhat stronger. This particularly manifests itself in a larger difference between $\tilde{\mathcal{C}}^{(2)}$ and 
\(\mathcal{S}\). Here, the faster-than-logarithmic increase of \(\mathcal{S}\) starts to show up already for $\gamma=0.4$ and is particularly well pronounced at $\gamma=0.3$. We note that, for $\gamma=0.3$ and the largest system size, $L=72$, the obtained value of \(\mathcal{S}\) is slightly above the estimated threshold for accurate results at our largest bond dimension, $\chi_{\rm max}=512$, see a discussion above. We thus expect that the actual values of $\tilde{\mathcal{C}}^{(2)}$ and 
\(\mathcal{S}\) in this point ($V=1$, $\gamma=0.3$, $L=72$) are somewhat larger than obtained numerically. 

We also observe that the dashed lines for $V=0.5$ and $V=1$ overlap for $\gamma=0.3,\,0.4,$ and 0.5, signalizing that the second particle-number cumulant is essentially unaffected by the interaction in this range of parameters. This is again consistent with the analytical predictions (see Sec.~\ref{sec:Gaussian_approx}). The weak residual dependence on the interaction strength can be attributed to subleading, interaction-induced corrections to the gradient term in the NLSM action, which were neglected in our analysis, see a discussion preceding Eq.~\eqref{Lint-U}.

In Fig.~\ref{fig:tdvp-summary}, the same TDVP data are presented in a complementary way. Here, the top panels show the results for the entropy \(\mathcal{S}\), with a linear scale of the $L$ axis, for different values of $\gamma$. The plots support our above conclusion that the entanglement transition between area-law and volume-law phases predicted by the analytic theory likely happens for $\gamma$ between 0.4 and 0.5, both for $V=0.5$ and $V=1$. 
The bottom panels 
show the (rescaled) cumulant $\tilde{\mathcal{C}}^{(2)}$ in a similar way but with a logarithmic scale of the $L$ axis. For larger $\gamma$, the behavior is clearly of the area-law type, as we have already seen in Fig.~\ref{fig:tdvp-entropy-cumulant}. With reducing $\gamma$, the scaling of $\tilde{\mathcal{C}}^{(2)}$ becomes logarithmic. To better estimate the position of the transition, we plot straight lines $(2/3) \ln L + {\rm const}$ corresponding to the analytical prediction \eqref{eq:BKTScaling} for the critical slope of $\mathcal{C}^{(2)}$. Comparing the numerical data with this analytical prediction, we estimate that the charge-fluctuation transition between the phases with area-law and logarithmic scaling is located in the interval $0.3 <\gamma < 0.4$.

Thus, the estimated value of critical measurement rate $\gamma_C$ for the charge-fluctuation transition is close to that for the entanglement transition ($\gamma_E$). This supports our analytical results that positions of the two transitions are determined by the same condition $\ell_{\text{int}} \sim \ell_{\text{loc}}$, and correspondingly by the same Eq.~\eqref{eq:V-crit}, i.e., can differ by a numerical factor only. 
Further, the estimated value of $\gamma_C$ is slightly below that for $\gamma_E$, in consistency with the analytical expectation that $\gamma_C \le \gamma_E$ and providing (weak) evidence that the intermediate phase does exist, at least for these values of interaction. However, our data do not permit us to exclude the possibility that both transitions take place at the same value of $\gamma$ in our model. A further improvement in the precision of numerical investigation would be needed to unambiguously resolve the two transitions.

\begin{figure*}
    \centering    \includegraphics[width=0.75\textwidth]{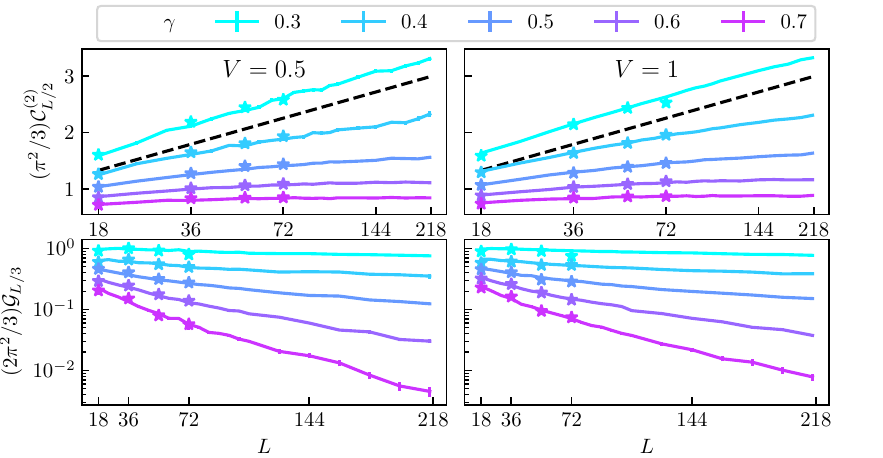}
    \caption{TDHF results for the rescaled charge cumulant (top) and covariance (bottom) for system sizes up to $L=220$ at interaction strengths $V=0.5$ (left) and $V=1$ (right). 
    Stars show TDVP results when available (for system sizes up to $L=72$). 
    The dashed straight lines in the left panels are described by $(2/3)\ln L+\const$, corresponding to the critical slope predicted by Eq.~\eqref{eq:BKTScaling}
}
    \label{fig:tdhf-c2-cov}
\end{figure*}

\subsection{Time dependent Hartree-Fock simulations}
\label{sec:numerics:tdhf}

In addition to our main computational tool, the MPS-TDVP, we have also performed simulations by using the TDHF approximation. This is a self-consistent approximation for the equation of motion for the lesser Green's function, which is obtained by discarding all but the lowest-order (Hartree and Fock) skeleton diagrams in the derivation of the self-consistent self-energy~\cite{tdhf_eq_ref}. At zero interaction, the self-energy vanishes and the method becomes exact. Another important feature of this method is that it allows one to proceed to rather large system sizes, at the same time retaining partly (but not fully) important aspects of the interaction-induced physics. Recently, the TDHF method was applied in the context of many-body localization (MBL) \cite{tdhf_wgk,tdhf2,tdhf3}. 
While it was found that this approximation destabilizes the MBL transition in 1D systems, replacing it by a crossover \cite{tdhf2,tdhf3}, it was also shown that the method describes remarkably well the character of the transport on the ergodic side of the transition \cite{tdhf_wgk,tdhf2}.

To avoid confusion, let us emphasize right away that the TDHF approximation certainly does not represent a good approximation for all observables in the present problem. Indeed, it preserves the Gaussian character of the many-body states and thus a relation between the entropy and the cumulant. As a consequence, the TDHF fails completely to describe the volume-law scaling of the entropy. At the same time, we observe that TDHF results are in very good agreement with the results for the charge fluctuations---the second cumulant $\tilde{\mathcal{C}}^{(2)}$ and the (minus) covariance $\mathcal{G}$---for all parameters and all lengths $L$ for which MPS-TDVP calculation is accurate. This shows that TDHF certainly captures (some of) interaction-induced physics with respect to charge fluctuations. In particular, it nicely extrapolates the logarithmic behavior of the cumulant $\tilde{\mathcal{C}}^{(2)}$ that we observed with MPS-TDVP to considerably larger system sizes. Specifically, since TDHF is computationally more efficient than MPS-TDVP, we used it to proceed up to $L=220$. It is not clear to us at this stage whether there is a true area-law-to-logarithmic-law transition in the TDHF approximation, or else, the logarithmic behavior that we observe is only transient as in non-interacting systems. This is an interesting open question that remains to be resolved. However, in any case, the TDHF appears to be useful for studying the charge fluctuations in large interacting systems, even if it does not describe correctly the $L\to \infty $ limit. 

Figure \ref{fig:tdhf-c2-cov} shows TDHF results at different measurement strengths (colorful lines) for the rescaled second cumulant (top panels) and the correspondingly rescaled [see Eq.~\eqref{mutual}] covariance (bottom panels) 
at interaction strengths \(V= 0.5\) (left panels) and \(V=1\) (right panels). The star symbols show the TDVP results for $L \le 72$ which were presented in Sec.~\ref{sec:numerics:tdvp}.  It is seen that the TDHF results agree very well with those of MPS-TDVP in the range of $L$ where the latter are controllable. The only notable differences are for \(V=1\), the lowest measurement strength \(\gamma = 0.3\), and the largest system size \(L=72\), which is exactly the point in the parameter space for which we expect the TDVP to be slightly inaccurate due to bond-dimension limitation, see a discussion above.

The top panels presenting the cumulant provide full support to our earlier conclusions. We observe a transition from the area-law to the logarithmic behavior of the charge cumulant. Comparing to the analytical prediction for the critical slope \eqref{eq:BKTScaling} (shown by dashed lines), we
conclude again that the critical point of the charge-fluctuation transition is in the range $[0.3,\, 0.4]$, in agreement with what we found above based on TDVP data. 

The (minus) covariance $\mathcal{G}$ shown in the bottom panels of Fig.~\ref{fig:tdhf-c2-cov} is a counterpart of the dimensionless conductance at the Anderson transition \cite{Poboiko2023b}. In the area-law phase, $\gamma = 0.7$, 0.6, and 0.5, an exponential decay of $\mathcal{G}$ at large $L$ is observed. This decay can be used to determine the localization length. In the ``metallic'' phase, $\mathcal{G}$ should have a finite limit at $L\to \infty$. Indeed, we observe at $\gamma = 0.4$ and 0.3 a behavior that is very close to saturation. However, a close inspection of the data shows a very slow decay of $\mathcal{G}$ with $L$ even for $\gamma=0.3$. This might be either a manifestation of finite-size corrections to the asymptotic $L\to \infty$ value, or else, an indicator of the fact that TDHF approximation does not rigorously describe the metallic phase. As discussed above, the exact status of the TDHF approximation in this respect remains to be understood. 

\section{Summary and discussion}
\label{sec:Summary}

\subsection{Summary of main results}

In this article, we have explored the effect of measurements on interacting fermionic systems whose unitary dynamics is governed by a time-independent Hamiltonian. Starting with a generic model of interacting fermions subjected to generalized measurements, we have developed a Keldysh field-theoretical framework  that serves as a starting point for a systematic study of key physical observables, providing a unified approach to entanglement entropies and full counting statistics of charge fluctuations (Sec.~\ref{sec:S2}).  

In the absence of interaction (and for measurements preserving Gaussianity of the theory), this formalism allows one to identify the symmetry class of the problem 
in terms of the symmetry of the Lagrangian,
Sec.~\ref{sec:gaussian-models-symmetries}. For complex fermions, this places the problem in the symmetry class AIII or BDI, depending 
on the presence of additional particle-hole symmetry of Lagrangian, in agreement with earlier results. We have derived the replicated Keldysh non-linear sigma model for both symmetry classes, Sec.~\ref{sec:NLSM}. The model incorporates boundary conditions (Fig.~\ref{fig:BC}) specifically designed to produce generating functions for charge cumulants and R{\'e}nyi entropies directly in the language of the sigma model (Sec.~\ref{sec:boundary-NLSM}).

For free fermions, owing to the Gaussian nature of the many-body states, the charge fluctuations and entanglement are intrinsically related. Furthermore, the behavior in classes AIII and BDI is essentially the same, with the only factor-of-two difference in the ``weak-localization correction''. For 1D geometry, systems of both these classes exhibit a crossover from logarithmic to area-law scaling at a scale of ``localization length'' $\ell_{\rm loc}$ that becomes exponentially large for rare measurements. We have verified these predictions by numerically modeling non-interacting systems of classes AIII and BDI with all parameters being identical, up to the particle-hole symmetry breaking, see Appendix~\ref{sec:appendix:WL-BDI-vs-AIII}.

We have included a weak interaction in the derivation of the NLSM (Sec.~\ref{Sec:Interaction}). The interaction-induced terms in the action \eqref{eq:InteractionLagrangian} modifies the symmetry of the theory, which has a dramatic effect on the behavior of physical observables. Specifically, the symmetry is broken down to $\mathbb{S}_R \times [\mathrm{U}(1)]^{R-1}$, involving the discrete replica permutation group $\mathbb{S}_R$ and phase symmetries $[\mathrm{U}(1)]^{R-1}$ that do not mix replicas, as illustrated in the cartoon of Fig.~\ref{fig:potential-cartoon}. The boundary conditions for the calculation of the entanglement entropy involve two distinct global minima within the $\mathbb{S}_R$ subgroup. As a consequence, the classical space-time configuration of the sigma-model field acquires the form of the domain wall separating them (Fig.~\ref{fig:domain-wall}), which leads to the emergence of the volume-law scaling of the entanglement entropy [Eq.~\eqref{eq:VolumeLaw}]. This is in contrast with the particle-number statistics, where the boundary conditions are diagonal and belong to the continuous $[\mathrm{U}(1)]^{R-1}$ subgroup, with the Goldstone modes producing the area$\times$log scaling of the charge fluctuations [Eq.~\eqref{eq:second-cumulant}].

In Sec.\ref{sec:RG}, we have analyzed the effect of interactions on monitored fermions by means of the renormalization-group approach [Eq.~\eqref{eq:RG} and Fig.~\ref{fig:RG}] and determined the scale dependence of observables (Figs.~\ref{fig:behavior:localization}--\ref{fig:behavior:crit}). This analysis has allowed us to construct the phase diagrams for monitored interacting fermions in arbitrary dimension $d$, see  Fig.~\ref{fig:phase-dia}. 

The inclusion of interactions gives rise to several major modifications of the theory. First, it breaks down the Gaussianity of the states, thus inducing the ``information-charge separation’’:  the charge cumulants and entanglement entropies become decoupled. Second, the drastic effect of the interaction on the symmetry of the model stabilizes the volume-law phase for entanglement. 
Third, in 1D systems, the interaction also stabilizes the phase with logarithmic growth of charge cumulants (in the thermodynamic limit).  The interaction thus leads to the existence of the measurement-induced transitions for interacting fermions in arbitrary dimension $d$, as illustrated in Fig.~\ref{fig:phase-dia}. This contrasts with monitored free fermions, where no such transition occurs in one dimension, and the area-law phase remains stable regardless of measurement rate. One of two distinct transitions in the interacting model is associated with the change of scaling of particle number fluctuations with the subsystem size (which in 1D systems belongs to the BKT universality class, see Sec.~\ref{sec:BKT}). The other transition takes place for the entanglement entropy and is related to the spontaneous breaking of the replica permutation symmetry $\mathbb{S}_R$.
Our analysis shows that, for weak interaction, both transitions take place at $\ell_{\text{int}} \sim \ell_{\text{loc}}$ (where $\ell_{\text{int}}$ is the interaction-induced length scale), which translates into Eq.~\eqref{eq:V-crit} in terms of microscopic parameters of the model. The only difference is (or, more accurately, may be) in the value of a numerical coefficient. It is likely that the two transitions are distinct, in which case the charge-fluctuation transition should happen (for given $V$) at smaller $\gamma$ than the entanglement transition. However, we cannot exclude a possibility that the locations of two transitions coincide in a certain range of $V$. 

We have corroborated these analytical results with a numerical study of an interacting 1D model (Sec.~\ref{sec:numerics}). Since exact diagonalization proved insufficient in view of system-size limitations, we have employed the time-dependent variational principle for matrix product states, Sec.~\ref{sec:numerics:tdvp}) enabling us to investigate significantly larger systems. The numerically observed behavior of both the entanglement entropy and charge cumulants (Figs.~\ref{fig:tdvp-entropy-cumulant} and \ref{fig:tdvp-summary}) aligns very well with our analytical predictions. In particular, our MPS-TDVP numerics demonstrate the separation between information and charge, characteristic of the phase with ``delocalized'' information and charge fluctuations. Furthermore, the behavior of the corresponding behavior of entanglement and charge cumulant is consistent with the predicted area-law and logarithmic-law scaling, respectively. Our estimated values for critical $\gamma$ for both transitions show that two transitions are located quite close to each other. In addition, we have conducted time-dependent Hartree-Fock simulations (Sec.~\ref{sec:numerics:tdhf} and Fig.~\ref{fig:tdhf-c2-cov}) as a complementary method to estimate the transition point between the area-law phase and the logarithmic phase for charge fluctuations. While the numerical data suggest that the entanglement transition takes place at a somewhat larger value of $\gamma$ than the charge-fluctuation transition (as in the phase diagram in Fig.~\ref{fig:phase-dia}), the accuracy is not sufficient to deduce this with certainty.

\subsection{Comparison with related works}

At this point, it is worth comparing our findings with the results of related studies in the literature. 

In papers \cite{Agrawal2022, Barratt2022}, a ``charge-sharpening'' transition was predicted in certain hybrid quantum circuits involving Haar-random gates and qudits. The charge sharpening means that repeated measurements on a state initially lacking a definite ``charge'' lead to a well-defined charge value. The number of measurements required for this sharpening was predicted to scale differently with system size in the ``charge-fuzzy'' and ``charge-sharp'' phases, thus implying a phase transition. The analytical description of charge sharpening in Refs.~\cite{Barratt2022} was developed in the limit of the number of qudit states (i.e., Hilbert space dimensionality at each site) going to infinity. While the charge sharpening is a dynamical process, it was argued in Ref.~\cite{Barratt2022} that equal-time density correlation functions, similar to those analyzed in our work, can be used to experimentally probe this phenomenon~\cite{Agrawal2023}. This suggests that the charge-sharpening transition and the charge-fluctuation transition that we address are two manifestations of the same phase transition. 

Although the approach of Refs.~\cite{Agrawal2022, Barratt2022} was very different, the symmetry of the effective theory was analogous to that in the symmetry-broken regime in our work, with two sectors corresponding to entanglement and charge.
Consequently, the BKT transition for charge fluctuations in our work is by and large identical to that found for qudit hybrid circuits \cite{Barratt2022}. This is quite remarkable, taking into account that the random hybrid quantum circuit model is very different from that of interacting fermions.
Our model and analysis are complementary to that of Refs.~\cite{Agrawal2022,Barratt2022} in two respects. First, our model deals with fermions (i.e., two states for each site as for a qubit) rather than qudits.
Second, quantum circuits as studied 
in Refs.~\cite{Agrawal2022, Barratt2022} can be viewed as strongly interacting, while we start from the model of non-interacting fermions and then ``switch on'' interaction, exploring the corresponding phase diagram and emerging regimes. We expect that the NLSM framework developed in our work can be extended to quantum circuits by generalizing onto the case of time-dependent Hamiltonians and Floquet dynamics.

The effect of interactions on entanglement in 1D monitored Hamiltonian systems was addressed in Ref.~\cite{Xing2023}. In that work, several variants of spin-$1/2$ chains (which can be mapped to fermionic models by the Jordan-Wigner transformation) were studied by means of exact numerical simulations of dynamics (the approach that we abbreviated as ED, see Sec.~\ref{sec:numerics}) for system sizes $12 \le L\le 28$. This included the XXZ model with measurement of $\sigma_z$, which directly maps on the interacting fermionic model with measurement of the density that we studied, as well the XX model that is equivalent to our non-interacting ($V=0$) model. The measurement protocol in Ref.~\cite{Xing2023} involved weak measurements, at variance with projective measurements in our work. We expect, however, that this difference does not essentially affect the physics addressed. The parameter $\gamma$ characterizing the measurement strength in Ref.~\cite{Xing2023} is a direct counterpart of our $\gamma$ (measurement rate). The simulations in Ref.~\cite{Xing2023} were performed for a single value of this parameter, $\gamma=0.1$. Comparing the values of the entanglement entropy, we find that this roughly corresponds to $\gamma \simeq 0.1$ in our model. 

The results of Ref.~\cite{Xing2023} suggested a volume-law scaling of the entanglement entropy in the interacting model (XXZ), which is in agreement with our results, since $\gamma =0.1$ is well on the volume-law side of the entanglement transition. Furthermore, the analysis of different models in Ref.~\cite{Xing2023} suggested that the extensive (volume-law) scaling of the entanglement entropy in the weakly-monitored phase is a universal feature of interacting models, irrespective of the additional symmetries or integrability. This is also fully consistent with our findings in the present paper.
For the non-interacting (XX) model, the entropy data in Ref.~\cite{Xing2023} showed a (weak) bending down on the linear scale, which was interpreted as a manifestation of the logarithmic scaling. This is again in consistency with our results: the entropy behaves logarithmically in the absence of interaction up to a scale $\xi_{\rm loc}$, which is exponentially large for small $\gamma$ (and, in particular, is very large numerically for $\gamma = 0.1$). There is, however, a caveat here. For $\gamma = 0.1$, the measurement-induced mean free path $\ell_0 \sim 1/\gamma$ is comparable to system sizes studied in Ref.~\cite{Xing2023}, which suggests that the results may be substantially affected by ballistic dynamics at scales $\ell \lesssim \ell_0$. In particular, the results for the non-interacting (XX) model may be in the crossover from the ballistic (volume law) to the diffusive (logarithmic law) regimes.  
We have mentioned difficulties with exploring different regimes by ED in Sec.~\ref{sec:numerics}. Also, as we demonstrated in Sec.~\ref{sec:numerics}, a comparison of the second particle-number cumulant with the entanglement entropy is a very efficient tool to pinpoint the combined effect of measurements and interactions that leads to information-charge separation. However, Ref.~\cite{Xing2023}
did not study correlation functions related to the particle-number statistics considered in our work. 

In Ref.~\cite{Lumia2023}, the 1D interacting fermion model with projective measurements of the local density (equivalent to our model with $J=1/2$) was studied numerically 
by means of the MPS time-evolving block-decimation for the chain length $L=30$. The measurement-induced volume-to-area-law entanglement transition, with the logarithmic scaling of $\mathcal{S}$ at criticality, was identified in Ref.~\cite{Lumia2023} at $\gamma=0.21$, which translates into $\gamma=0.42$ in the units ($J=1$) used in our simulations, Sec.~\ref{sec:numerics}. This is perfectly consistent with our analytical predictions and with the results of our MPS-TDVP simulations for larger systems, yielding $0.4<\gamma_E<0.5$ for the entanglement transition. The entanglement transition in Ref.~\cite{Lumia2023} was accompanied by a transition in the spectral properties of the one-body reduced density operator, which was used to characterize the Gaussianity of the states. It was further found in Ref.~\cite{Lumia2023} that a fully analogous entanglement transition emerges for $d=1$  due to non-Gaussian measurements, which thus have essentially the same effect as the interaction.

In a number of works \cite{Tang2020a, Goto2020a, Fuji2020a, Fuji2021, Doggen2022a, Doggen2023, doggen2023ancilla}, the entanglement in 1D models of interacting bosons subjected to generalized measurements was studied numerically. These works reported evidence of a transition from area-law to volume-law scaling of the entanglement entropy. While some of these models were different from our interacting-fermion model, and the measurement protocols were also different, one can expect that the physics that we address has a large degree of universality. In this sense, the conclusions of the above works on the area-law-to-volume-law entanglement transition are in agreement with our results. However, the scaling of particle-number fluctuations for interacting bosons, as well as the associated charge-sharpening transition, had not been explicitly addressed in these past papers.

Very recently, the problem of phase transitions in monitored interacting spin chains was addressed in Ref.~\cite{Cecile2024} by utilizing the MPS-TDVP numerical machinery (also employed in our work, see Sec.~\ref{sec:numerics:tdvp}). In this paper, an approach based on the dependence of errors in the MPS method on the bond dimension was put forward to identify the measurement-induced transitions. One of two models studied in Ref.~\cite{Cecile2024} was the XXX model equivalent to our interacting fermionic model. While Ref.~\cite{Cecile2024} used continuous monitoring, one can make an identification of the parameter $\gamma$ governing the measurement strength by comparing the value of the second cumulant of charge (which is the spin cumulant in Ref.~\cite{Cecile2024}). This leads to the conclusion that $\gamma$ in Ref.~\cite{Cecile2024}) corresponds to our $\gamma$ without any substantial rescaling. 
The authors of Ref.~\cite{Cecile2024} reported
two transitions for this model: the entanglement transition and the charge-fluctuation transition, with the critical values of $\gamma$ being approximately 0.2 and 0.1, respectively. This is qualitatively consistent with our findings. At the same time, these critical values of $\gamma$ are substantially smaller than those that we find. Furthermore, Ref.~\cite{Cecile2024} reported a volume-law scaling of the spin cumulant (particle-number cumulant, in our terminology) in the weakly-monitored phase (i.e., for $\gamma$ below the charge transition). 
Our findings exclude the possibility of the volume-law scaling of charge fluctuations in the thermodynamic limit. At the same time, such behavior of the particle-number cumulants is characteristic of the ballistic regime [cf. Eq.~\eqref{eq120}] realized in not-too-large systems at weak monitoring, see discussion of system-size limitations in Sec.~\ref{sec:numerics} above. Indeed, for $\gamma = 0.1$, we do observe the ballistic effects in the range of system sizes used in Ref.~\cite{Cecile2024}. In view of the above, we argue that the value $\gamma = 0.1$ identified as the point of the charge transition in Ref.~\cite{Cecile2024} is probably located well on the delocalized side of this transition, and the actual value 
is close to our estimate, $0.3 < \gamma_C < 0.4$. Indeed the scaling of the cumulant for such $\gamma$ in Ref.~\cite{Cecile2024}  appears to be consistent with the logarithmic scaling predicted by our theory. It is also worth noting that the values of bond dimension used in Ref.~\cite{Cecile2024} are in the range $16 \le \chi \le 100$, i.e., substantially lower than $\chi= 128$, 256, and 512 that we used.

\subsection{Outlook}

Before closing the paper, we briefly discuss some prospective research directions motivated by our work:

\begin{itemize}

\item[(i)]
The field-theoretical (NLSM) approach to interacting monitored systems developed in our work can be extended to systems of other symmetry classes and to other types of measurements (including, in particular, non-Gaussian measurements and measurements of non-commuting observables).

\item[(ii)]
The approach can also be directly generalized to describe a monitored open system in contact with environment, whose evolution includes measurements and non-unitary Lindbladian dynamics (see, e.g., Refs.~\cite{Sieberer2016, Sieberer2023Review} for review). For such systems, however, the entanglement entropy and particle number cumulant cannot serve as good observables to define the transition as their behavior is  expected to follow generically a volume law due to effect of the environment. On the other hand, the volume-law contribution of ``classical correlations'' is expected to cancel~\cite{HorodeckiRevModPhys09} in the scaling of the mutual information and particle-number covariance, making them suitable observables to study possible MIPTs.

\item[(iii)] Including quenched disorder in the theory (in addition to monitoring and interaction) is a further challenging goal. 

\item[(iv)] Another important direction is to extend the theory to study dynamics (i.e., correlation functions with different time arguments. This will allow one, in particular, to directly explore the dynamics of purification and charge sharpening. 

\item[(v)] Finally, there is a clear need for further numerical studies of measurement-induced physics in interacting models. Important goals include, in particular, verifying whether the charge and information transitions are indeed distinct and exploring the corresponding critical properties. 
Clearly, analytical prospects mentioned in previous items in this list also call for corresponding computational studies.

\item[(vi)] Last but not least, we hope that our work will also stimulate further experimental studies in the field of measurement-induced phase transitions, which remains largely theoretical at present. 

\end{itemize}

\acknowledgments

We are very grateful to  Elmer Doggen, Adam Nahum, Pavel Ostrovsky, Marcin Szyniszewski, and Romain Vasseur for insightful discussions.

\textit{Note added:} While preparing the manuscript for publication, we became aware of related work~\cite{guo2024} on monitored interacting fermions. We thank Matthew Foster for informing us about this work. 

\appendix

\section{Derivation of the NLSM}
\label{sec:appendix:NLSM}

In this Appendix, we provide details on the derivation of the NLSM for monitored fermions.

\subsection{Summation over quantum trajectories}
\label{sec:appendix:Summation}

The averaging over quantum trajectories (i.e., summation over them with the corresponding weights) can be performed on the level of the action \eqref{eq:DMatrixReplicatedAction}, akin to averaging over strong Poissonian impurities in the context of disordered systems. This can be most conveniently done starting from Eq.~\eqref{eq:DMatrixKeldyshContour}:
\begin{multline}
\label{eq:app:DRK}
\hat{D}^{\otimes R}=\int{\cal D}\psi^{\ast}{\cal D}\psi\exp\left(i\sum_{r=1}^{R} S_{0}[\psi^{\ast}_{r},\psi_r]\right)\\
\times\prod_{r=1}^{R}\prod_{m}\mathbb{K}_{i_{m},\alpha_{m}}[\psi_{r,+}(t_{m})]\mathbb{K}_{i_m,\alpha_m}^{\dagger}[\psi_{r,-}(t_{m})],
\end{multline}
where the action $S_0$ describes the unitary part of the evolution. 
The summation over trajectories $\sum_{{\cal T}}$ introduced in
Eq.~\eqref{eq:TrajectorySummation} of the main text includes: (i) averaging over the total number of measurements $M_i$ of each type $i$, which has a Poissonian distribution with average $\overline{M_i} = \gamma_i T$ where $T = t_f - t_0$, (ii) averaging over times $t_m$ of each individual measurement, and (iii) a \emph{summation} over outcomes of each measurement. Explicitly, we have
\begin{equation}
\label{eq:sum-t-notation}
\sum_{{\cal T}}\equiv\prod_{i}\left[\sum_{M_{i}=0}^{\infty}\frac{\left(\gamma_{i}T\right)^{M_{i}}}{M_{i}!}e^{-\gamma_{i}T}\left(\prod_{m_{i}=1}^{M_{i}}\int_{t_{0}}^{t_{f}}\frac{dt_{m_{i}}}{T}\sum_{\alpha_{m_{i}}}\right)\right].
\end{equation}
Combining Eq.~\eqref{eq:sum-t-notation} with Eq.~\eqref{eq:app:DRK}, we observe that individual integrals over $t_{m_i}$ factorize, so that the product over $m_i$ simply reduces to a power. The summation over $M_i$ then yields an exponential function, and we obtain:
\begin{multline}
\label{eq:sum-K-T}
\sum_{{\cal T}}\mathbb{K}({\cal T})=\prod_{i}\Bigg[\sum_{M_{i}=0}^{\infty}\frac{e^{-\gamma_{i}T}}{M_{i}!}\Bigg(\gamma_{i}\sum_{\alpha_{m}}\int_{t_{0}}^{t_{f}}dt_{m}\\\times\prod_{r=1}^{R}\mathbb{K}_{i,\alpha_{m}}[\psi_{r,+}(t_{m})]\mathbb{K}_{i,\alpha_{m}}^{\dagger}[\psi_{r,-}(t_{m})]\Bigg)^{M_{i}}\Bigg]\\
=\exp\left(i \sum_{i}\gamma_{i} \int_{t_0}^{t_f} dt\,L_{M,i}[\psi^{\ast}(t),\psi(t)]\right).
\end{multline}
In the last line of Eq.~\eqref{eq:sum-K-T}, we have introduced 
\begin{equation}
i L_{M,i}[\psi^\ast, \psi] \equiv \sum_{\alpha}\prod_{r=1}^{R}\mathbb{K}_{i,\alpha}[\psi_{r,+}]\mathbb{K}_{i,\alpha}^{\dagger}[\psi_{r,-}]-1 \,,
\end{equation}
which is the contribution of each individual measured observable to the Lagrangian.
For Gaussian measurements, defined via Eqs.~(\ref{eq:KraussOperatorCoherent}) and (\ref{eq:GaussianMeasurementsDefinition}), this further simplifies to
\begin{equation}
iL_{M,i}[\psi^{\ast},\psi]=\sum_{\alpha}\frac{\exp\left(-\psi_{+}^{\dagger}\hat{M}_{i,\alpha}\psi_{+}-\psi_{-}^{\dagger}\hat{M}_{i,\alpha}^{\dagger}\psi_{-}\right)}{\left|{\cal N}_{i,\alpha}\right|^{2R}}-1,
\end{equation}
where $\psi_{\pm}$ are spinors in the replica space, and matrices $\hat{M}_{i,\alpha}$ act trivially on this space.

For the case of density monitoring, Sec.~\ref{sec:ProjectiveDensityMonitoring}, which represents the main focus of this paper, the index $i$ enumerates lattice sites, so that $L_{M,i}$ translates to the Lagrangian density ${\cal L}_M$. Possible measurement outcomes in this case are $\alpha = \pm 1$, and corresponding matrices $\hat{M}_{\alpha}$ are $\hat{M}_{\alpha} = -2\alpha$. The action then further simplifies to
\begin{equation}
i{\cal L}_{M}[\psi^{\ast},\psi]=\frac{2}{4^{R}}\cosh(2\psi^{\dagger}\psi)-1 \,.
\end{equation}

\subsection{Particle-hole symmetry}

For the purposes of derivation of the NLSM, it will be convenient to introduce an extra minus sign for $\psi^\ast_{-}$, i.e., to perform a redefinition $\psi^{\dagger}\mapsto\psi^{\dagger}\hat{\tau}_{z}$ (whereas $\psi\mapsto\psi$;  this can be done because $\psi^\ast$ and $\psi$ are formally independent Grassmann variables).  The matrices $\hat{\tau}_{j}$ act in Keldysh (K) space. The action then reads $S = S_0 + \gamma S_M$, with the corresponding Lagrangian densities given by
\begin{equation}
\label{eq:app:L:AIII}
{\cal L}_{0}=\psi^{\dagger}\left(\hat{\varepsilon}-\hat{H}_{0}\right)\psi,\quad i{\cal L}_{M}=\frac{2}{4^{R}}\cosh(2\psi^{\dagger}\hat{\tau}_{z}\psi).
\end{equation}
To incorporate the PH symmetry, we double the size of spinors, introducing spinors $\Psi$ and $\bar{\Psi}$ according to Eq.~\eqref{eq:PHSpinor} of the main text, which are related to each other via the ``charge conjugation matrix'' $\mathcal{C}$:
\begin{equation}
\label{eq:app:BDIChargeConjugation}
\bar{\Psi} = \Psi^T \mathcal{C},\quad \mathcal{C} = \tau_x \sigma_x
\end{equation}

For a PH-symmetric system, one has $(\varepsilon-\hat{H}_0)^T = -(\varepsilon - \hat{H}_0)$, and, as a consequence, the Lagrangian acquires the equivalent form in terms of doubled spinors:
\begin{equation}
\label{eq:app:L:BDI}
{\cal L}_{0}=\bar{\Psi}\left(\hat{\varepsilon}-\hat{H}_{0}\right)\Psi,\quad i{\cal L}_{M}=\frac{2}{4^{R}}\cosh(2\bar{\Psi}\hat{\tau}_{z}\Psi).
\end{equation}

Note that due to formal equivalence between Eqs.~(\ref{eq:app:L:AIII},\ref{eq:app:L:BDI}), for the derivation of the NLSM for symmetry class AIII (which does not require spinor doubling) it is convenient to \emph{define} $\Psi \equiv \psi$ and $\bar{\Psi} \equiv \psi^\dagger$ --- bearing in mind that for symmetry class AIII they become independent variables. Such definition allows us to carry on the derivation using the same formulae for both AIII and BDI symmetry classes, emphasizing the difference where it arises.

\subsection{Generalized Hubbard-Stratonovich transformation}

Following Ref.~\cite{Poboiko2023a}, we perform a generalized Hubbard-Stratonovich transformation by introducing two matrices, ${\cal G} = -i \Psi \bar{\Psi}$ (including only space components that are slow in the momentum) and an auxiliary matrix $\Sigma$ that fixes this relation via an identity
\begin{equation}
1=\int{\cal D}{\cal G}\exp\left[-i\Tr\left(\hat{\Sigma}\hat{{\cal G}}\right)-\bar{\Psi}\hat{\Sigma}\Psi\right].
\end{equation}

For symmetry class BDI, it is convenient to define the ``bar'' operation on an arbitrary $4R \times 4R$ matrix $\mathcal{O}$ defined in PH$\times$K$\times$R space via
\begin{equation}
\label{eq:app:BarOperation}
\bar{\mathcal{O}}\equiv{\cal C}\mathcal{O}^{T}{\cal C},
\end{equation}
and note that, by construction, one has $\bar{{\cal G}} = -{\cal G}$ and $\bar{\Sigma} = -\Sigma$. 

Using the definition of matrix ${\cal G}$, we perform decoupling of the measurement-induced ``interaction'' in all channels, which, as was shown in Ref.~\cite{Poboiko2023a}, accounts to performing a Gaussian average of the form:
\begin{multline}
\label{eq:app:LM}
\!\!\!\!\! -{\cal L}_{M}[{\cal G}]= \!\!\int \!\! \frac{{\cal D}\chi^{\ast}{\cal D}\chi}{\det^{\beta}\left(-i\hat{{\cal G}}^{-1}\right)}\exp\left(i\bar{\chi}\hat{{\cal G}}^{-1}\chi\right)i{\cal L}_{M}[\chi^{\ast},\chi]-1\\
=\sum_{\alpha=\pm1}\det\nolimits^{\beta}\left(\frac{1}{2}+i\alpha\hat{{\cal G}}\hat{\tau}_{z}\right)-1,
\end{multline}
with
\begin{equation}
\beta=\begin{cases}
1, & \text{AIII}\\
1/2, & \text{BDI}
\end{cases}
\end{equation}
The remaining integral over Grassmann fields $\Psi$, $\bar{\Psi}$ is Gaussian and thus can be straightforwardly evaluated.
As a result, we obtain the following effective action
\begin{eqnarray}
\label{eq:app:SGSigma}
S[{\cal G},\Sigma] &=& -\Tr\left[\beta\ln\left(\hat{\varepsilon}-\hat{H}_{0}+i\Sigma\right)-i\Sigma{\cal G}\right]
\nonumber \\
&+& \gamma\int d^{d+1}\boldsymbol{x}{\cal L}_{M}[{\cal G}],
\end{eqnarray}
with  ${\cal L}_{M}[{\cal G}]$ given by Eq.~\eqref{eq:app:LM}.
Note that the matrix action (\ref{eq:app:SGSigma}) is defined as Euclidian, i.e., the averaging takes place with the weight $\exp(-S[{\cal G},\Sigma])$. 

\subsection{Replica-symmetric manifold}

The saddle-point equations for the action \eqref{eq:app:SGSigma} yield the SCBA solution that is consistent with the initial condition on the Keldysh contour and reads:
\begin{equation}
{\cal G}=-\frac{i\beta}{2} \Lambda,\quad\Sigma=\gamma \Lambda,
\end{equation}
with $\Lambda$ given by Eq. \eqref{eq:Lambda} from the main text.

Further analysis of the Goldstone modes of the action reveals (cf. Ref.~\cite{Poboiko2023a}) that the theory splits into the replica-symmetric and replicon sectors.
The symmetry of the replica-symmetric sector can be understood by directly setting $R=1$.
Then, for an arbitrary matrix ${\cal G}$ (satisfying ${\bar{{\cal G}}=-{\cal G}}$ for symmetry class BDI), there is an exact identity for the measurements part of the action, which leads to enlargement of the symmetry,
\begin{equation}
-{\cal L}_{M}^{(R=1)}[{\cal G}]=2\det\nolimits^{\beta}\hat{{\cal G}}-\frac{1}{2} \,.
\end{equation}
This implies that, on the replica-symmetric manifold, an \emph{arbitrary} unitary rotation is a symmetry:
\begin{equation}
{\cal G}=-\frac{i \beta}{2}Q_{\text{s}},\quad\Sigma=\gamma Q_{\text{s}},\quad Q_{\text{s}}={\cal R}_{\text{s}}\Lambda {\cal R}_{\text{s}}^{-1},
\end{equation}
with rotation matrices satisfying
\begin{equation}
\label{eq:app:RotationInverse}
{\cal R}_{\text{s}}^{-1}=\begin{cases}
{\cal R}_{\text{s}}^{\dagger}, & \text{AIII},\\
\bar{{\cal R}}_{\text{s}}, & \text{BDI},
\end{cases}
\end{equation}
which define group $\mathrm{U}(2)$ for AIII and $\mathrm{O}(4)$ for BDI. The subgroup of this group that does not rotate $\Lambda$ satisfies an additional constraint, 
\begin{equation}
{\cal R}_{\text{s}} \Lambda {\cal R}_{\text{s}}^{-1} = \Lambda,
\end{equation}
which defines the group $\mathrm{U}(1) \times \mathrm{U}(1)$ for AIII and $\mathrm{U}(2)$ for BDI. Interestingly, as a result, the replica-symmetric manifolds for both cases coincide with a two-dimensional sphere:
\begin{align}
\text{AIII:}\quad&\mathrm{U}(2)/\mathrm{U}(1)\times\mathrm{U}(1)\simeq\mathrm{S}^{2}\\
\text{BDI:}\quad&\mathrm{O}(4)/\mathrm{U}(2)\simeq\mathrm{S}^{2}
\end{align}
Thus, presence or absence of the particle-hole symmetry plays no role on the level of the Lindbladian dynamics, i.e., dynamics of the average density matrix.

For symmetry class BDI, the replica-symmetric rotation matrix has the following explicit form in the PH space:
\begin{equation}
{\cal R}_{\text{s}}\equiv\begin{pmatrix}{\cal R}_{0} & 0\\
0 & \tau_{x}{\cal R}_{0}^{\ast}\tau_{x}
\end{pmatrix}_\text{PH}\!\!,\quad\bar{{\cal R}}_{s}=\begin{pmatrix}{\cal R}_{0}^{\dagger} & 0\\
0 & \tau_{x}{\cal R}_{0}^{T}\tau_{x}
\end{pmatrix}_\text{PH}\!\!,
\end{equation}
which leads to Eq.~\eqref{eq:Qs} of the main text.

\subsection{Replicon manifold}

Since we are interested in observables that are non-linear functions of the density matrix, our main focus is on the replicon manifold.  The full action \eqref{eq:app:SGSigma} possesses a 
``replicon'' symmetry 
which holds for arbitrary $R$:
\begin{equation}
{\cal G}\mapsto{\cal R}_{R}{\cal G}{\cal R}_{R}^{-1},\quad\Sigma\mapsto{\cal R}_{R}\Sigma{\cal R}_{R}^{-1},
\end{equation}
with matrices ${\cal R}_{R}^{-1}$, in addition to constraint analogous to Eq.~\eqref{eq:app:RotationInverse}, preserving the  $\tau_z$ matrix:
\begin{equation}
\label{eq:app:RCond}
\mathcal{R}_{R}^{-1}\tau_{z}{\cal R}_{R}=\tau_{z}\,.
\end{equation}
This leads to the replicon NLSM manifold of solutions:
\begin{equation}
\label{eq:app:manifold}
{\cal G}=-\frac{i \beta}{2}Q,\quad\Sigma=\gamma Q,\quad Q={\cal R}_{R}Q_{s}{\cal R}_{R}^{-1} \,,
\end{equation}
where $Q_s$ belongs to the replica-symmetric manifold \eqref{eq:Qs}.

The most general form of the matrix ${\cal R}_{R}$ that satisfies the constraints (\ref{eq:app:RotationInverse},\ref{eq:app:RCond}) is:
\begin{align}
\text{AIII:}\quad&{\cal R}_{R}=\begin{pmatrix}{\cal V}_{+} & 0\\
0 & {\cal V}_{-}
\end{pmatrix}_{\text{K}},&{\cal V}_{\pm}\in\mathrm{SU}(R),\\
\text{BDI:}\quad&{\cal R}_{R}=\begin{pmatrix}{\cal V} & 0\\
0 & \sigma_{x}{\cal V}^{\ast}\sigma_{x}
\end{pmatrix}_{\text{K}},&{\cal V}\in\mathrm{SU}(2R).
\end{align}
[We note that $\mathrm{U}(1)$ phase factors $\mathcal{V} = \exp(i \phi)$ are already included in the replica-symmetric manifold.] The full rotated $Q$-matrix then acquires the form given by Eq.~\eqref{eq:SigmaModelQ} of the main text.

\subsection{Gradient expansion}
\label{sec:appendix:GradientExpansion}

Having established the relevant symmetries and the manifold, we proceed to the derivation of an effective action for the Goldstone modes associated with these symmetries. We substitute $Q = {\cal R} \Lambda {\cal R}^{-1}$ in the action \eqref{eq:app:SGSigma}. 
The second term in Eq.~\eqref{eq:app:SGSigma} is constant on the NLSM manifold; furthermore, the constant should be zero since the Keldysh partition function is equal to unity in the replica limit $R \to 1$. Thus, the only term that remains is the fermionic determinant, which can be written as follows:
\begin{equation}
S[\hat{Q}]=-\beta\Tr\ln(1+i\hat{G}\hat{W}),\quad\hat{G}^{-1}=i\partial_{t}-\hat{H}_{0}+i\gamma\hat{\Lambda} \,.
\end{equation}
Here $\hat{G}$ is the SCBA Green's function, and the matrix $\hat{W}$ contains only gradients:
\begin{equation}
\hat{W}=-i{\cal R}^{-1}\left[\hat{\varepsilon},{\cal R}\right]+i{\cal R}^{-1}\left[\hat{H}_{0},{\cal R}\right]\approx{\cal R}^{-1}\partial_{t}{\cal R}+\hat{\boldsymbol{v}}{\cal R}^{-1}\nabla{\cal R},
\end{equation}
with the  single-particle operator of group velocity
$$\hat{\boldsymbol{v}}=i[\hat{H}_{0},\hat{\boldsymbol{x}}].$$

In the first-order expansion in $\hat{W}$, only the term with the temporal derivative survives, as the second term is off-diagonal in the coordinate space. Furthermore, substituting ${\cal R}={\cal R}_{R}{\cal R}_{s}$, we see that
\begin{equation}
\hat{W}_{t}={\cal R}_{s}^{-1}\partial_{t}{\cal R}_{s}+{\cal R}_{s}^{-1}({\cal R}_{R}^{-1}\partial_{t}{\cal R}_{R}){\cal R}_{s}.
\end{equation}
Here, only the replica-symmetric term survives, as, in the replica space, the second term is proportional to $\tr_{R}\left({\cal V}^{\dagger}\partial_{t}{\cal V}\right)=0$, since $\det {\cal V} = 1$. This produces the following contribution to the replica-symmetric action:
\begin{equation}
\label{eq:L-WZ}
{\cal L}^{(1)}[Q_{0}]=-\frac{1}{2}\tr_{\text{K}}\left(\Lambda_{0}{\cal R}_{0}^{\dagger}\partial_{t}{\cal R}_{0}\right),
\end{equation}
which has a 1D Wess-Zumino-type form (it cannot be represented through $Q_{0}$ itself and requires explicit parametrization). 

To treat the term of second order in $\hat{W}$, we note that the SCBA Green's function can be rewritten identically as:
\begin{equation}
\hat{G}=\frac{1}{2}\hat{G}^{R}(1+\Lambda)+\frac{1}{2}\hat{G}^{A}(1-\Lambda),
\end{equation}
with $\hat{G}^{R/A} = (\hat{\varepsilon} - \hat{H}_0 \pm i \gamma)^{-1}$. The second-order term then yields:
\begin{equation}
\label{eq:S2}
S^{(2)}=-\frac{\beta}{4}\Tr\left[\hat{G}^{R}(1+\Lambda)\hat{W}\hat{G}^{A}(1-\Lambda)\hat{W}\right].
\end{equation}
Equation \eqref{eq:S2} gives rise to all terms with two derivatives in the NLSM, both 
in the replica-symmetric and replicon sectors.

First of all, Eq.~\eqref{eq:S2} produces a term with the two spatial gradients acting on the replica-symmetric manifold:
\begin{equation}
\label{eq:L-21}
{\cal L}^{(2,1)}[\hat{Q}_0]=\frac{D}{8}\tr_{\text{K}}(\nabla\hat{Q}_{0})^{2},
\end{equation}
with the diffusion coefficient defined via the integral over the Brillouin zone:
\begin{equation}
D\equiv v_{0}^{2}/2\gamma,\quad v_{0}^{2}=\frac{1}{d}\int_{BZ}\frac{d^{d}\boldsymbol{k}}{(2\pi)^{d}}\left[\nabla_{\boldsymbol{k}}H_{0}(\boldsymbol{k})\right]^{2}.
\end{equation}
For the nearest-neighbor hopping, Eq.~\eqref{H0},
we have $v_{0}^2 = 2 J^2$ and $D = J^2 / \gamma$. 

Secondly, Eq.~\eqref{eq:S2} produces a term with two spatial gradients for the replicon manifold.  Taking the trace over the Keldysh space explicitly, using the fact that ${Q_0^2=1}$ and $\Tr Q_0 = 0$, and introducing the replica-symmetric density via Eq.~\eqref{eq:RSDensity} of the main text, we obtain this term in the form
\begin{equation}
\label{eq:L-22}
{\cal L}^{(2,2)}[Q_{0},\hat{U}]=\beta D\rho(1-\rho)\tr_\text{R,PH}\left(\nabla \hat{U}^{\dagger}\nabla \hat{U}\right).
\end{equation}
Finally, an analogous term with two time derivatives of the replicon field emerges.  It is found to have a similar structure, up to a replacement $D \mapsto 1/2\gamma$:
\begin{equation}
\label{eq:L-23}
{\cal L}^{(2,3)}[Q_{0},\hat{U}]=\frac{\beta \rho(1-\rho)}{2\gamma}\tr_\text{R,PH}\left(\partial_{t}\hat{U}^{\dagger}\partial_{t}\hat{U}\right). 
\end{equation}
The sum of the terms \eqref{eq:L-WZ}
and \eqref{eq:L-21} gives the theory in the replica-symmetric sector, Eq.~\eqref{eq:SymmetricLagrangian}, while the sum  of the terms 
\eqref{eq:L-22} and \eqref{eq:L-23} yields the NLSM in the replicon sector, Eq.~\eqref{eq:RepliconLagrangian}.

\section{One-loop renormalization group for non-interacting systems of BDI symmetry class}
\label{sec:appendix:BDIRG}

In this Appendix, we discuss the one-loop renormalization group correction for the NLSM in the BDI symmetry class. The theory is described by the Lagrangian density \eqref{eq:RepliconLagrangian} with $g(Q_0)$ replaced by the running coupling constant $g(\ell)$, whose bare value at the ultraviolet scale is given by $g(\ell_0) \equiv g_0$, see Eq.~\eqref{eq:g0}. It will be convenient to re-define the ``bar'' operation acting on matrices defined in the PH$\times$R space as follows:
\begin{equation}
\label{eq:app:BarDef}
\bar{\mathcal{V}} \equiv \sigma_y \mathcal{V}^T \sigma_y,
\end{equation}
and thus $U = \mathcal{V} \bar{\mathcal{V}}$. We perform standard splitting of the modes into ``slow'' and ``fast'' components as $\mathcal{V} = \mathcal{V}_s \mathcal{V}_f$, thus $U = \mathcal{V}_s U_f \bar{\mathcal{V}}_s$. The action then acquires the following form: 
\begin{align}
{\cal L}_{R}[U]&={\cal L}_{R}[U_{f}]+{\cal L}_{R}[U_{s}]+{\cal L}_{\text{fs}}[U_{f},U_{s}],\\
{\cal L}_{R}[U_{f}]&=\frac{g}{4}\tr\left(-iU_{f}^{\dagger}\partial_{\mu}U_{f}\right)^{2},\\
{\cal L}_{R}[U_{s}]&=\frac{g}{4}\tr\left(A_{s}+\bar{A}_{s}\right)^{2},\\
{\cal L}_{\text{fs}}^{(1)}[U_{f},U_{s}]&=-\frac{ig}{2}\tr\left(\partial_{\mu}U_{f}U_{f}^{\dagger}A_{\mu}+U_{f}^{\dagger}\partial_{\mu}U_{f}\bar{A}_{\mu}\right),\\
{\cal L}_{\text{fs}}^{(2)}[U_{f},U_{s}]&=\frac{g}{2}\tr\left(A_{\mu}U_{f}\bar{A}_{\mu}U_{f}^{\dagger}-A_{\mu}\bar{A}_{\mu}\right).
\end{align}
Here, we have defined 
\begin{equation}
A_{\mu}=-i\mathcal{V}_{s}^{\dagger}\partial_{\mu}\mathcal{V}_{s},\quad\bar{A}_{\mu}=-i\partial_{\mu}\bar{\mathcal{V}}_{s}\bar{\mathcal{V}}_{s}^{\dagger},
\end{equation}
which are related to each other via Eq.~\eqref{eq:app:BarDef}. 

We proceed with the exponential parametrization of the fast modes, keeping slow modes arbitrary:
\begin{align}
U_{f}&=e^{i\Phi_{f}}\approx1+i\Phi_{f}-\frac{1}{2}\Phi_{f}^{2},\\
-iU_{f}^{\dagger}\partial_{\mu}U_{f}&\approx\partial_{\mu}\Phi_{f}-\frac{i}{2}\left[\Phi_{f},\partial_{\mu}\Phi_{f}\right],
\\-i\partial_{\mu}U_{f}U_{f}^{\dagger}&\approx\partial_{\mu}\Phi_{f}+\frac{i}{2}\left[\Phi_{f},\partial_{\mu}\Phi_{f}\right],
\end{align}
with traceless Hermitian generators obeying the symmetry constraint $\bar{\Phi}_f = \Phi_f$. This implies that their Gaussian correlation function reads:
\begin{multline}
\Big\langle \big(\Phi_{f}(\boldsymbol{r})\big)_{ij}\big(\Phi_{f}(\boldsymbol{r}^{\prime})\big)_{kl}\Big\rangle_f =G_f(\boldsymbol{r}-\boldsymbol{r}^{\prime})\\
\times\left(\delta_{il}\delta_{jk}+\sigma_{ik}^{y}\sigma_{lj}^{y}-\frac{1}{R}\delta_{ij}\delta_{kl}\right),
\end{multline}
with the bare Green's function $G_f(\boldsymbol{r})$ whose Fourier transform is
\begin{equation}
G_{f}(\boldsymbol{q})=1/gq^{2}.
\label{eq:Gf}
\end{equation}

In the one-loop approximation, several terms are generated. The first term originates from the first order of perturbation theory from quadratic in $\Phi_f$ term in the expansion of $\mathcal{L}_{\text{int}}^{(2)}$, which reads as
\begin{equation}
    {\cal L}_{\text{fs}}^{(2,2)}[\Phi_{f},U_{s}]=\frac{g}{2}\tr\left(-\frac{1}{2}\Phi_{f}^{2}\left\{ A_{\mu},\bar{A}_{\mu}\right\} +A_{\mu}\Phi_{f}\bar{A}_{\mu}\Phi_{f}\right),
\end{equation}
leading to
\begin{equation}
\delta{\cal L}_{R}^{(1)}[U_s]=\left\langle {\cal L}_{\text{fs}}^{(2,2)}\right\rangle\!=\!-{\cal I}_{1}\!\left[\left(2R-1\right)\tr\left(A_{\mu}\bar{A}_{\mu}\right)\!+\!\tr\left(A_{\mu}^{2}\right)\right].
\end{equation}
Here, the factor ${\cal I}_{1}$ is given by the integral (in dimensions $d = 1+\epsilon$ with $\epsilon \ll 1$):
\begin{equation}
{\cal I}_{1}=\frac{g}{2}G_{f}(0)=\frac{1}{2}\int\frac{d^{d+1}\boldsymbol{q}}{(2\pi)^{d+1}}\frac{1}{q^{2}}=\frac{\ln(\ell^\prime / \ell)}{4\pi\ell^{\epsilon}}.
\end{equation}
The second term originates as the second-order  perturbation from quadratic terms in $\mathcal{L}_{\text{fs}}^{(1)}$:
\begin{equation}
{\cal L}_{\text{fs}}^{(1,2)}[\Phi_{f},U_{s}]=\frac{ig}{2}\tr\left(\left[\Phi_{f},\partial_{\mu}\Phi_{f}\right]A_{\mu}\right).
\end{equation}
Using standard ``center-of-mass'' coordinates, the renormalization of the Lagrangian then reads:
\begin{multline}
\delta{\cal L}_{R}^{(2)}\!=-\frac{1}{2}\!\int\!\! d^{d+1}\boldsymbol{r}\left\langle\left\langle {\cal L}_{\text{fs}}^{(1,2)}(\boldsymbol{R}\!+\!\boldsymbol{r}/2){\cal L}_{\text{fs}}^{(1,2)}(\boldsymbol{R}\!-\!\boldsymbol{r}/2)\right\rangle\right\rangle _{f}\\={\cal I}_{2}^{\mu\nu}\,(R-1)\Big[\tr(A_{\mu}\bar{A_{\nu}})-\tr(A_{\mu}A_{\nu})\Big],
\end{multline}
with the integral
\begin{equation}
{\cal I}_{2}^{\mu\nu}=g^{2}\!\int\!
d^{d+1}\boldsymbol{r}\partial_{\mu}G_{f}(\boldsymbol{r})\partial_{\nu}G_{f}(\boldsymbol{r})=\frac{2}{d+1}\delta_{\mu \nu} \mathcal{I}_1.
\end{equation}
As a result, the overall renormalization of the Lagrangian acquires the following form:
\begin{equation}
\delta{\cal L}_{R}[U_s]=-\frac{R}{8\pi \ell^\epsilon}\tr(A_{\mu}+\bar{A}_{\mu})^{2}\,\ln \frac{\ell^\prime}{\ell},
\end{equation}
which is identical to the one-loop RG equation for the dimensionless coupling constant $G(\ell) = g(\ell) \ell^\epsilon$:
\begin{equation}
\frac{dG}{d\ln\ell}=\epsilon\, G-\frac{R}{2\pi}.
\end{equation}

\section{Numerical comparison of weak-localization corrections in AIII and BDI classes}
\label{sec:appendix:WL-BDI-vs-AIII}

In order to illustrate the effect of the particle-hole symmetry on ``weak localization'', we have performed numerical simulations of projectively monitored free fermions for two models described by the one-dimensional Hamiltonian that, in addition to nearest-neighbor hopping ($J_1$), includes next-nearest-neighbor hopping ($J_2$):
\begin{align}
&\hat{H}=-\sum_{n}\left(J_{1}\hat{\psi}_{n}^{\dagger}\hat{\psi}_{n+1}+J_{2}\hat{\psi}_{n}^{\dagger}\hat{\psi}_{n+2}+\text{H.c.}\right),\\
\text{AIII:}\quad&J_{1}=0.6,\quad J_{2}=0.4,\\
\text{BDI:}\quad&J_{1}=0.6,\quad J_{2}=0.4i.
\end{align}
One can easily check that the parameter choice for the BDI model indeed corresponds to the PH-symmetric system, whereas the choice for AIII does not; furthermore, the numbers are chosen such that the root-mean-square velocities $v_0$ for both systems coincide, implying that values of the coupling constant $g_0$ also coincide (for the same measurement rate $\gamma$). 

We utilized the Green's function method, as described in Ref.~\cite{Poboiko2023a}, and calculated the equal-time density-density correlation functions in the steady state:
\begin{equation}
C(x)=\overline{\left\langle \hat{n}(x)\hat{n}(0)\right\rangle -\left\langle \hat{n}(x)\right\rangle \left\langle \hat{n}(0)\right\rangle },
\end{equation}
whose Fourier transform can be related to the running coupling constant at lengthscale $\ell = q^{-1}$ via 
\begin{equation}
\label{eq:app:gq}
    g(q) = C(q) / q.
\end{equation}
The simulations were carried out on systems of size $L = 1000$ sites, with periodic boundary conditions; averaging involved over 100 independent quantum trajectories for each value of $\gamma$.

Subtracting the Gaussian result, which interpolates between ballistic and diffusive regimes at $q \sim \ell_0^{-1}$, and whose precise form is of no importance here (apart from the fact that it saturates at value $g_0$ at $q \to 0$), we have extracted the ``weak-localization'' correction $\delta g(q)$. It follows from Eqs.~\eqref{RG-AIII} and \ref{RG-BDI} that, for the system sizes smaller than the ``localization length'', the weak-localization correction should have a universal form (i.e., the slope of the correction on the linear-log plot should not depend on any parameters):
\begin{align}
\text{AIII:}\quad&\delta g(q)=-\frac{1}{4\pi}\ln\frac{1}{q\ell_{0}},
\label{eq:app:WLPrediction:AIII}
\\
\text{BDI:}\quad&\delta g(q)=-\frac{1}{2\pi}\ln\frac{1}{q\ell_{0}}.
\label{eq:app:WLPrediction:BDI}
\end{align}

\begin{figure}
    \centering
    \includegraphics[width=\columnwidth]{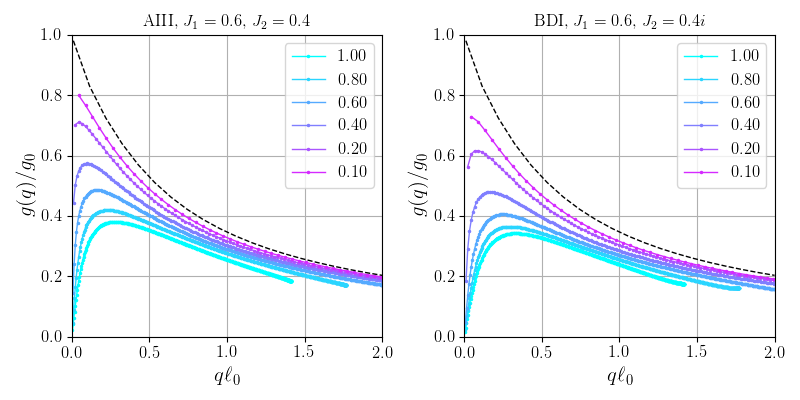}
    \caption{Effective ``coupling constant'' $g(q)/g_0$, Eq.~(\ref{eq:app:gq}), extracted from the Fourier transform of the density correlation function via Eq.~\eqref{eq:app:gq}, as a function of $q \ell_0$ for AIII (left) and BDI (right) models. Dashed line marks a universal curve interpolating between ballistic and diffusive behavior.}
    \label{fig:app:Cq}
\end{figure}
\begin{figure}
    \centering
    \includegraphics[width=\columnwidth]{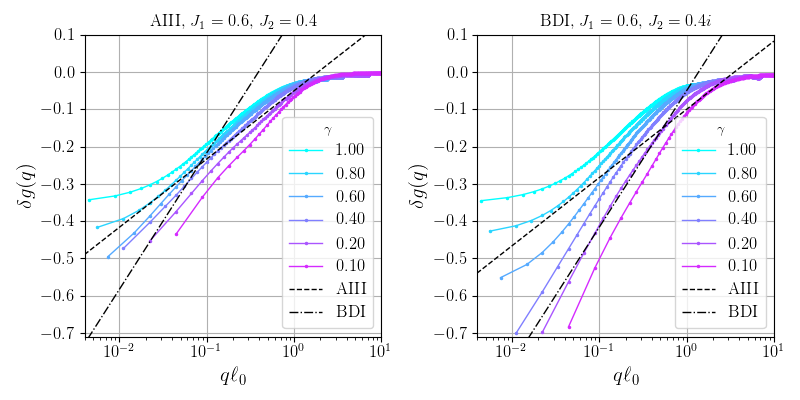}
    \caption{``Weak-localization'' correction $\delta g(q)$ for the AIII (left) and BDI (right) models as a function of $q \ell_0$ in linear-log axes, for several values of $\gamma$. Dashed and dash-dotted lines: analytical predictions for class AIII and BDI, respectively, Eqs.~ (\ref{eq:app:WLPrediction:AIII}) and (\ref{eq:app:WLPrediction:BDI}). }
    \label{fig:app:WL}
\end{figure}

In Fig.~\ref{fig:app:Cq}, we plot the behavior of the running coupling constant for both symmetry classes (left panel: AIII, right panel: BDI), as obtained from the density correlation function. The dependence of $g(q)$ at large momenta corresponds to the ballistic behavior: $C(q) \propto \const \Rightarrow g(q) \propto 1/q$; the dashed line marks a universal interpolating function between ballistic ($q \ell_0 \gtrsim 1$) and diffusion ($q \ell_0 \ll 1$) regions. Evident bending down of each curve is a manifestation of localization. It is evident that the qualitative behavior of the curves in the two panels of Fig.~\ref{fig:app:Cq} is identical. At the same, a closer inspection reveals that the localization effects are stronger in the BDI model than in the AIII model, in consistency with analytical predictions

The difference between the curves and dashed line allows us to extract the ``weak-localization correction'', which is plotted in Fig.~\ref{fig:app:WL}. One clearly sees that the slope of the weak-localization correction indeed changes noticeably between two models considered here: it is larger for the BDI model than for the AIII model, in agreement with the analytical predictions, Eqs.~(\ref{eq:app:WLPrediction:AIII}) and (\ref{eq:app:WLPrediction:BDI}). The slope for the class-BDI system (right panel) is perfectly consistent with the analytically predicted  value $1/2\pi$, Eq.~(\ref{eq:app:WLPrediction:BDI}).
The slope for the AIII model (left panel) is clearly smaller than the one for BDI, although still appears to be somewhat larger than the predicted value $1/4\pi$, Eq.~(\ref{eq:app:WLPrediction:AIII}). This deviation can be presumably attributed to finite-size effects.

\section{Interaction term in NLSM}
\label{sec:appendix:Interaction}

In this Appendix, we derive the effect of a short-range interaction on the replicon sector of the NLSM, assuming the interaction to be sufficiently weak, so that it can be treated perturbatively. As described in the main text, for simplicity, we focus on the symmetry class AIII; a detailed derivation for the symmetry class BDI will be presented elsewhere.

We start from the fermionic path integral, where the effect of the interaction is described via the Lagrangian term \eqref{eq:Lint}. We introduce a set of Hubbard-Stratonovich ``plasma fields'' for each individual replica $\varphi_{\pm,r}(\boldsymbol{r})$ via the relation:
\begin{equation}
e^{iS_{\text{int}}}=\int{\cal D}\varphi\exp\left(\frac{i}{2}\varphi\hat{\tau}_{z}\hat{u}^{-1}\varphi-i\psi^{\dagger}\hat{\varphi}\psi\right)
\label{Sint-ferm}
\end{equation}
with the matrix $\hat{\varphi}$ having the following structure (diagonal in replica and Keldysh spaces):
\begin{equation}
\hat{\varphi}=\begin{pmatrix}\varphi_{+} & 0\\
0 & \varphi_{-}
\end{pmatrix}_{K}=\frac{1}{\sqrt{2}}\left(\hat{\varphi}_{\text{cl}}+\hat{\tau}_{z}\hat{\varphi}_{\text{q}}\right).
\end{equation}
Here, we utilized the standard notation of ``classical'' and ``quantum'' field components for the Keldysh contour:
\begin{equation}
\varphi_{\text{cl},\text{q}}=\frac{1}{\sqrt{2}}\left(\varphi_{+}\pm\varphi_{-}\right)
\end{equation}

We then follow the derivation of the NLSM from the fermionic action, as described in Refs.~\cite{Poboiko2023a,Poboiko2023b} (cf. Sec.~\ref{Sec3:non-int} and Appendix~\ref{sec:appendix:NLSM}). The Gaussian integration over the fermionic degrees of freedom then yields an additional term in the action, describing the interaction of the collective field $Q(\boldsymbol{r})$ and plasma fields $\varphi$ of the following form:
\begin{equation}
\label{eq:app:Sint:TrLog}
S_{\text{int}}[Q,\varphi]\approx-\Tr\ln\left(1-\hat{G}[\hat{Q}]\hat{\varphi}\right),
\end{equation}
with the Green's function given by Eq.~\eqref{eq:GQ} of the main text.

For spatially homogeneous $\hat{Q}$-matrices, we utilize the expansion:
\begin{equation}
\label{eq:app:GQGRGA}
\hat{G}[\hat{Q}]=\frac{1}{2}\left[\hat{G}_{R}(1+\hat{Q})+\hat{G}_{A}(1-\hat{Q})\right],
\end{equation}
where $G_{R,A}$ are retarded and advanced SCBA Greens functions that have the following form in the momentum representation:
\begin{equation}
G_{R,A}(\varepsilon,\boldsymbol{k})=(\varepsilon-\xi_{\boldsymbol{k}}\pm i\gamma)^{-1}.
\end{equation}
The effective action for the $Q$-field is then obtained by integrating out the auxiliary plasma fields:
\begin{equation}
S_{\text{int}}[Q]=-\ln\left\langle \exp\left(-S_{\text{int}}[Q,\varphi]\right)\right\rangle _{\varphi},
\end{equation}
[cf. Eq.~\eqref{eq:SintQ:intermediate} of the main text]. Here, the angular brackets denote Gaussian averaging over $\varphi$ fields, whose correlation function follows from the quadratic part of Eq.~\eqref{Sint-ferm} and, in terms of ``classical'' and ``quantum'' components, reads:
\begin{equation}
\left\langle \varphi_{\text{cl},a}(\boldsymbol{r})\varphi_{\text{q},b}(\boldsymbol{r}^{\prime})\right\rangle =iV(\boldsymbol{x},\boldsymbol{x}^{\prime})\delta(t-t^{\prime})\delta_{ab}.
\end{equation}

We expand the trace-log term in Eq.~\eqref{eq:app:Sint:TrLog} up to the fourth order and substitute Eq.~\eqref{eq:app:GQGRGA} into this expansion; because of instantaneous interaction and retardicity, only the following terms are of interest:
\begin{align}
S_{\text{int}}^{(2)}[Q,\varphi]&=\frac{1}{4}\Tr\left(\hat{\varphi}\hat{G}_{R}(1+\hat{Q})\hat{\varphi}\hat{G}_{A}(1-\hat{Q})\right),
\\
S_{\text{int}}^{(4)}[Q,\varphi]&=\frac{1}{32}\Tr\left(\left[\hat{\varphi}\hat{G}_{R}(1+\hat{Q})\hat{\varphi}\hat{G}_{A}(1-\hat{Q})\right]^{2}\right),
\end{align}
and the effective action then reads:
\begin{equation}
\label{eq:app:Sint}
S_{\text{int}}[\hat{Q}]\simeq-\frac{1}{2}\left\langle (S_{\text{int}}^{(2)})^{2}\right\rangle_\varphi + \left\langle S_{\text{int}}^{(4)}\right\rangle_\varphi .
\end{equation}

Next, we perform Wick's contraction of $\varphi$-fields. Each contraction has an associated replica index, over which the summation is performed. Green's functions and interaction lines fix the spatial separation between different $Q$-matrices entering this expression to be much smaller than the mean-free path, allowing us to effectively take them at coinciding points. The spatial integration then naturally produces ``Hartree'' and ``Fock'' matrix elements as prefactors:
\begin{align}
Y_{\text{H}}&=\int d^{d+1}\boldsymbol{r}_{1,2,3}\,\,V_{13}V_{24}G_{12}^{R}G_{21}^{A}G_{34}^{R}G_{43}^{A},\\
Y_{\text{F}}&=-\int d^{d+1}\boldsymbol{r}_{1,2,3}\,\,V_{13}V_{24}G_{12}^{R}G_{23}^{A}G_{34}^{R}G_{41}^{A},
\end{align}
with the short-hand notation $G_{ij}^{R/A}\equiv G^{R/A}(\boldsymbol{r}_{i}-\boldsymbol{r}_{j})$, and $V_{ij}\equiv V(\boldsymbol{x}_{i}-\boldsymbol{x}_{j})\delta(t_{i}-t_{j})$. These matrix elements are represented diagrammatically in Fig.~\ref{fig:app:HartreeFock}, and will be estimated in Appendix~\ref{sec:appendix:HartreeFockDiagrams}. 

Finally, we substitute the $Q$-matrix given by Eq.~\eqref{eq:SigmaModelQ} into Eq.~(\ref{eq:app:Sint}) and calculate the trace over the Keldysh space. All the replica-symmetric matrix elements $Q_{\pm,\pm}$ then form a combination $[\rho(1-\rho)]^2$, with the replica-symmetric density $\rho$ given by Eq.~\eqref{eq:RSDensity}. Furthermore, the Keldysh and replica structures of both terms in Eq.~\eqref{eq:app:Sint} turn out to be identical, yielding the final result given by Eq.~\eqref{Lint-U} of the main text, with $Y_{\text{HF}} = Y_{\text{H}} + Y_{\text{F}}$.

\section{Calculation of the Hartree and Fock diagrams for the interaction Lagrangian}
\label{sec:appendix:HartreeFockDiagrams}

\begin{figure}
    \centering
    \includegraphics[width=\columnwidth]{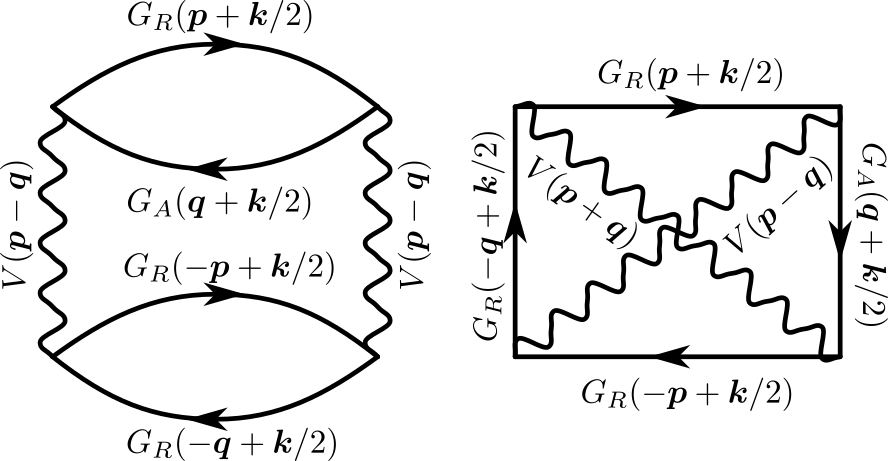}
    \caption{Hartree (left) and Fock (right) diagrams for the matrix element $Y_\text{HF}$ given by Eq.~\eqref{eq:HartreeFockMatrixElement}, contributing to the effective action for the $\mathrm{SU}(R)$ NLSM to leading order in interaction strength.}
    \label{fig:app:HartreeFock}
\end{figure}

This Appendix is devoted to the calculation of the prefactor $Y_\text{HF}$ in the interaction-induced Lagrangian (\ref{Lint-U}). This prefactor is given by the sum of the Hartree and Fock diagrams shown in Fig.~\ref{fig:app:HartreeFock}.
In what follows, we assume that the interaction $V(\boldsymbol{x},\boldsymbol{x}')$ in Eq.~\eqref{eq:Hint} depends only on the distance between $\boldsymbol{x}$ and $\boldsymbol{x}'$.
In the mixed time-momentum representation, the Hartree and Fock diagrams are combined to give:
\begin{multline}
Y_\text{HF}=\frac{1}{2}\int_{0}^{\infty}\!dt\int(d^d\boldsymbol{k})(d^d\boldsymbol{p})(d^d\boldsymbol{q}) \left(V_{\boldsymbol{p}-\boldsymbol{q}}-V_{\boldsymbol{p}+\boldsymbol{q}}\right)^{2}\\
\times G_{\boldsymbol{p}+\boldsymbol{k}/2}^{R}(t)\,G_{\boldsymbol{q}+\boldsymbol{k}/2}^{A}(-t)\,G_{-\boldsymbol{p}+\boldsymbol{k}/2}^{R}(t)\,G_{-\boldsymbol{q}+\boldsymbol{k}/2}^{A}(-t),
\label{eq:HartreeFockMatrixElement}
\end{multline}
with the retarded (advanced) Green's functions given by 
\begin{equation}
G_{\boldsymbol{k}}^{R,A}(t)=\pm i\exp\left[\left(\mp\gamma - i\xi_{\boldsymbol{k}}\right)t\right],
\end{equation}
$V_{\boldsymbol{p}}$ the Fourier transform of 
$V(\boldsymbol{x}-\boldsymbol{x}^\prime)$,
and the integration over each of the momenta $\boldsymbol{k}$, $\boldsymbol{p}$, $\boldsymbol{q}$ performed over $d$-dimensional Brillouin zone $[-\pi; \pi]^d$.

It is evident that for the point-like interaction $V_{\boldsymbol{p}} \equiv V = \const$, the two diagrams cancel each other, which is a manifestation of the Pauli exclusion principle. For the sake of transparency, we will focus on the simplest model with nearest-neighbor interaction and nearest-neighbor hopping (although this formally corresponds to a system of class BDI), such that the momentum dependence of both $\xi_{\boldsymbol{k}}$ and $V_{\boldsymbol{k}}$ can be expressed via the cubic lattice form-factor $f_{\boldsymbol{k}}$ as follows:
\begin{equation}
\xi_{\boldsymbol{k}}=-2Jf_{\boldsymbol{k}},\quad V_{\boldsymbol{q}}=2V f_{\boldsymbol{q}},\quad f_{\boldsymbol{k}}\equiv\sum_{\alpha=1}^{d}\cos k_{\alpha}.
\label{eq:fk}
\end{equation}
For other microscopic models, the form-factors will be different, leading to different numerical prefactors, but the parametric dependence of $Y_\text{HF}$ will remain the same.

We introduce a dimensionless time $\tau = 4 J t$ and denote the ratio $z \equiv \gamma / J \ll 1$, arriving at
\begin{equation}
Y_\text{HF}=\frac{V^{2}}{J}\int_{0}^{\infty}d\tau\,e^{-z\tau}I(\tau),
\label{Y-int}
\end{equation}
where
\begin{multline}
    I(\tau)=\frac{1}{2}\int(d^{d}\boldsymbol{k})\int(d^{d}\boldsymbol{p})(d^{d}\boldsymbol{q})\left(f_{\boldsymbol{p}-\boldsymbol{q}}-f_{\boldsymbol{p}+\boldsymbol{q}}\right)^{2}\\
\times\exp\left(\frac{i\tau}{2}\left[f_{\boldsymbol{p}+\boldsymbol{k}/2}+f_{\boldsymbol{p}-\boldsymbol{k}/2}-f_{\boldsymbol{q}+\boldsymbol{k}/2}-f_{\boldsymbol{q}-\boldsymbol{k}/2}\right]\right).
\end{multline}
As the next step, we substitute the explicit expression for the form-factor $f_{\boldsymbol{k}}$, Eq.~\eqref{eq:fk}, which yields
\begin{multline}
I(\tau)=2\int(d^{d}\boldsymbol{k})
\left(\prod_{\alpha=1}^{d}\int_{-\pi}^{\pi}\frac{dp_{\alpha}}{2\pi}e^{i\tau\cos(k_{\alpha}/2)\cos p_{\alpha}}\right)\\
\times\left(\prod_{\alpha=1}^{d}\int_{-\pi}^{\pi}\frac{dq_{\alpha}}{2\pi}e^{-i\tau\cos(k_{\alpha}/2)\cos q_{\alpha}}\right)
\left(\sum_{\alpha=1}^{d}\sin p_{\alpha}\sin q_{\alpha}\right)^{2}.
\label{eq:I-tau}
\end{multline}
We expand now the square of the sum in the last factor and observe that the cross-terms, which are odd functions of $p_\alpha$ and $q_\alpha$, vanish after the integration over the corresponding momentum components (the rest of the integrand is an even function of these components). 
To calculate the diagonal terms (there are $d$ identical terms), we first take integrals over components of momenta $\boldsymbol{p}$ and $\boldsymbol{q}$ utilizing the following integrals yielding the Bessel functions $J_n(t)$ with $n=0$ and $n=1$:
\begin{equation}
\int_{-\pi}^{\pi}\frac{dp_{\alpha}}{2\pi}\exp\left(\pm i\tau\cos\frac{k_{\alpha}}{2}\cos p_{\alpha}\right)=J_{0}\left(\tau\cos\frac{k_{\alpha}}{2}\right),
\end{equation}
\begin{equation}
\int_{-\pi}^{\pi}\frac{dp_{\alpha}}{2\pi}\exp\left(\pm i\tau\cos\frac{k_{\alpha}}{2}\cos p_{\alpha}\right)\sin^{2}p_{\alpha}=\frac{J_{1}\left(\tau\cos\displaystyle\frac{k_{\alpha}}{2}\right)}{\tau\cos\displaystyle\frac{k_{\alpha}}{2}}.
\end{equation}

\begin{figure}
    \centering
    \includegraphics[width=\columnwidth]{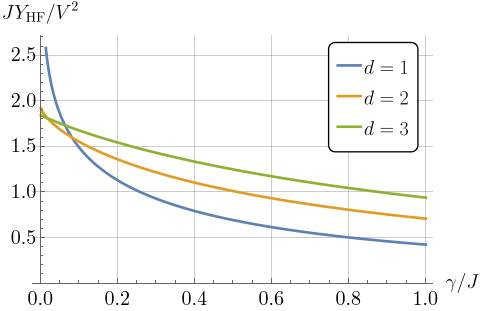}
    \caption{Rescaled Hartree-Fock matrix element $J Y_\text{HF} / V^2$ as a function of $\gamma / J$ in spatial dimensions $d = 1,2,3$.}
    \label{fig:app:YHF}
\end{figure}

Next, we calculate the integrals over the components of momentum $\boldsymbol{k}$ utilizing further two integrals, which can be expressed via the generalized hypergeometric function $\,_2F_3$:
\begin{equation}
I_{0}(\tau)\equiv\int_{-\pi}^{\pi}\frac{dk_{\alpha}}{2\pi}J_{0}^{2}\left(\tau\cos\frac{k_{\alpha}}{2}\right)=\,_{2}F_{3}\left(\frac{1}{2},\frac{1}{2};1,1,1;-\tau^{2}\right),
\end{equation}
\begin{multline}
I_{1}(\tau)\equiv\int_{-\pi}^{\pi}\frac{dk_{\alpha}}{2\pi}\left[\frac{J_{1}\left(\tau\cos\frac{k_{\alpha}}{2}\right)}{\tau\cos\frac{k_{\alpha}}{2}}\right]^{2}\\
=\frac{1}{4}\,_{2}F_{3}\left(\frac{1}{2},\frac{3}{2};1,2,3;-\tau^{2}\right).
\end{multline}
As a result, we obtain for $I(\tau)$, Eq.~\eqref{eq:I-tau},
\begin{equation}
I(\tau)=2d\,I_{0}^{d-1}(\tau)I_{1}(\tau).
\end{equation}

The asymptotic behavior of the integrals at large times (and $z\ll 1$) is as follows:
\begin{equation}
I_{0}(\tau)\approx\frac{2}{\pi^{2}}\frac{\ln\tau}{\tau},\quad I_{1}(\tau)\approx\frac{8}{3\pi^{2}\tau},\quad\tau\gg1,
\end{equation}
leading to
\begin{equation}
I(\tau\gg1)\approx d\, \frac{2^{d+3}}{3\pi^{2d}}\,\frac{\ln^{d-1}\tau}{\tau^{d}}.
\end{equation}
As a consequence, at $z \to 0$, the $\tau$-integral in Eq.~(\ref{Y-int}) logarithmically diverges for $d = 1$ and saturates at a constant of order unity for $d > 1$, yielding Eq.~(\ref{YHF}) of the main text.

The behavior of $Y_{\text{HF}}$ for $d=1,2,3$ is plotted in Fig.~\ref{fig:app:YHF}.

\section{Semiclassical approximation for R{\'e}nyi entropy}
\label{sec:appendix:RenyiGaussian}

In this Appendix, we calculate the $N$-th R{\'e}nyi entropy in the saddle-point approximation (which holds for $g_0 \to \infty$), corresponding to a solution of equations of motion, Eq.~\eqref{eq:Sint:SaddlePointEquations}, subject to the boundary conditions 
\begin{equation}
\hat{U}(\boldsymbol{x}\in A,y=0)=\hat{\mathbb{T}}_{N}
\end{equation}
and
\begin{equation}
\hat{U}(\boldsymbol{x}\notin A,y=0)=\hat{\mathbb{I}}.
\end{equation}
The solution can be sought in the form $\hat{U} = \hat{U}_N \oplus \hat{\mathbb{I}}_{R-N}$, with $\hat{U}_N$ being $N \times N$ unitary matrix. This effectively replaces the number of replicas $R$ by $N$ in all the sums over replica indices. In what follows, we drop the subscript $N$.

We perform a unitary rotation, which diagonalizes the boundary conditions $\hat{U}(\boldsymbol{r})=\hat{T}^{\dagger}\hat{U}^{\prime}(\boldsymbol{r})\hat{T}$ with matrix $\hat{T}$ given by Eq.~\eqref{Trotation}:
\begin{multline}
\partial_{\mu}(\hat{U}^{\prime\dagger}\partial_{\mu}\hat{U}^{\prime})_{rr^{\prime}}=\frac{m^{2}}{2N^{2}}\\
\times \Bigg(\sum_{a_{1,2,3},\,b_{1,2,3},\,r^{\prime\prime}}\bar{\delta}_{a_{1}+a_{3},r+b_{2}}\bar{\delta}_{r^{\prime\prime}+a_{2},b_{1}+b_{3}}\\
\times U_{a_{1}b_{1}}^{\prime\dagger}U_{a_{2}b_{2}}^{\prime}U_{a_{3}b_{3}}^{\prime\dagger}U_{r^{\prime\prime}r^{\prime}}^{\prime}-\text{H.c.}\Bigg).
\label{eq:F3}
\end{multline}
Here, we defined the Kronecker delta-symbol modulo $N$,
\begin{equation}
\bar{\delta}_{a,b}\equiv\sum_{k=-\infty}^{\infty}\delta_{a+kN,b},
\end{equation}
with $\delta_{a,b}$ denoting the ordinary Kronecker symbol.

From Eq.~\eqref{eq:F3}, we see that if its right-hand side consists only of diagonal matrices (implying $a_i = b_i$), then the off-diagonal matrix elements are not generated (i.e., $r = r^\prime$). This observation, together with the diagonal form of boundary conditions for $\hat{U}^\prime(\boldsymbol{r})$, allows us to seek a solution in a diagonal form
\begin{equation}
\hat{U}^{\prime}(\boldsymbol{r})=\diag\left(\left\{ \exp(i\phi_{r}(\boldsymbol{r})\right\} _{r=1}^{N}\right).
\end{equation}
This yields a system of coupled equations for phases $\phi_r(\boldsymbol{r})$:
\begin{equation}
\partial_{\mu}^{2}\phi_{r}=-\frac{m^{2}}{N^{2}}\sum_{a_{1,2,3}}\bar{\delta}_{a_{1}+a_{3},r+a_{2}}\sin\left(\phi_{a_{1}}-\phi_{a_{2}}+\phi_{a_{3}}-\phi_{r}\right),
\label{eq:app:ReplicaSineGordon}
\end{equation}
with the boundary condition $\phi_r(\boldsymbol{x} \in A, y=0) = \lambda_r$
within region $A$ and zero otherwise.

Finally, we note that a single-parameter Ansatz 
\begin{equation}
\phi_{r}(\boldsymbol{r})=\lambda_r \phi(\boldsymbol{r}) / 2 \pi
\label{eq:phi-ta-ansatz}
\end{equation}
is consistent with this system of equations. Indeed, substituting this Ansatz into Eq.~(\ref{eq:app:ReplicaSineGordon}) and expanding $\bar{\delta}$, we obtain for the right-hand side of Eq.~(\ref{eq:app:ReplicaSineGordon}) (that we denote by $\text{RHS}_{\text{F6}}$ for brevity):
\begin{align}
\text{RHS}_{\text{F6}} &=\frac{m^{2}}{N^{2}}\sum_{k=1}^{\infty}\sin(k\phi)
\notag
\\
&\times \sum_{a_{1,2,3}}\left(\delta_{a_{1}+a_{3}+kN,r+a_{2}}-\delta_{a_{1}+a_{3}-kN,r+a_{2}}\right).
\end{align}
Since $a_i \in [1,N]$, we see that only the $k=1$ term contributes; the remaining sums can be calculated and yield
\begin{equation}
\text{RHS}_{\text{F6}}=\frac{\lambda_{r}}{2\pi}m^{2}\sin\phi \,.
\end{equation}
Thus, the Ansatz \eqref{eq:phi-ta-ansatz} is consistent with Eq.~\eqref{eq:app:ReplicaSineGordon} provided $\phi(\boldsymbol{r})$ satisfies the elliptic sine-Gordon equation:
\begin{equation}
\begin{split}
\partial_{\mu}^{2}\phi(\boldsymbol{r})&=m^{2}\sin\phi(\boldsymbol{r}),\\
\phi(\boldsymbol{x},y=0)&=2\pi{\cal I}_{A}(\boldsymbol{x}),
\end{split}
\end{equation}
where  ${\cal I}_{A}(\boldsymbol{x})$ is the indicator function defined in Eq.~\eqref{eq:indicator-I-A}.
Substituting the form \eqref{eq:phi-ta-ansatz} of the configuration into 
 the Lagrangian \eqref{eq:InteractionLagrangian}, we obtain, after some algebra, the sine-Gordon Lagrangian density:
\begin{equation}
\label{eq:app:SineGordonLagrangian}
{\cal L}_{R}=\frac{g}{12}\left(N-\frac{1}{N}\right)\left[\frac{1}{2}(\partial_{\mu}\phi)^{2}+m^{2}\left(1-\cos\phi\right)\right].
\end{equation}

Finally, in order to estimate the ``cost'' of the domain wall per unit surface, it is sufficient to study its transverse structure, which is determined by one-dimensional sine-Gordon equation, whose solution is well-known:
\begin{equation}
\begin{split}
\phi^{\prime\prime}(y)&=m^{2}\sin\phi(y),
\\
\phi(y)&=4\arctan\exp\left[m(y-y_{0})\right].
\end{split}
\end{equation}
Here $y$ is the coordinate perpendicular to the domain wall, and $y_0$ marks the position of the domain wall.
Substituting this solution into Eq.~\eqref{eq:app:SineGordonLagrangian}, we obtain the following action per surface area:
\begin{equation}
S_{R}/||A||=\frac{2}{3}\left(N-\frac{1}{N}\right)\,g\,m.
\end{equation}
This action is related to the R{\'e}nyi entropy via Eqs.~(\ref{eq:RenyiEntanglementEntropy}), (\ref{eq:ObservablesAsPartitionFunction}, and (\ref{eq:SemiclassicalApproximation}):
\begin{equation}
{\cal S}_{A}^{(N)}=\frac{S_{R}}{N-1}=\frac{2}{3}\left(1+\frac{1}{N}\right)\,g\,m\,||A||,
\end{equation}
which leads to Eqs.~(\ref{eq:VolumeLaw}), (\ref{eq:SN:prefactor}) of the main text.

This derivation, yielding the volume law for the symmetry-broken case, should be contrasted with the derivation in the intermediate regime, $\ell_A \ll \ell_{\rm int}$, where the interaction term only weakly modulates the $\mathrm{SU}(R)$ manifold. The semiclassical equations of motion then allow for all possible rotations, and the boundary conditions for the calculation of the entanglement entropy are satisfied without ``tunneling'' between the minima. The calculation then follows the steps of Eqs.~(\ref{eq87})-(\ref{eq:C2A}), yielding the area$\times$log scaling of $\mathcal{S}_A^{(N)}$ for $\ell_A\ll \ell_\text{int}$. The modulation of the manifold can be accounted for in the semiclassical action by perturbation theory, yielding a correction with a relative smallness $\sim (m\ell_A)^2 = (\ell_A / \ell_{\rm int})^2$ as compared to the leading term.

\section{Renormalization group for the interacting NLSM}\label{sec:appendix:RG}

To derive the one-loop RG equations for the model defined by Eq.~\eqref{eq:InteractionLagrangian}, we substitute 
\begin{equation}
\hat{U}=\hat{U}_{0}\hat{U}_{f},\quad\hat{U}_{f}\approx1+i\hat{\Phi}_{f}-\frac{1}{2}\hat{\Phi}_{f}^{2},
\end{equation}
with $\hat{U}_0$ and $\hat{U}_f$ consisting of slow and fast degrees of freedom, respectively. After this substitution, the action acquires the  following form:
\begin{equation}
{\cal L}[\hat{U}]-{\cal L}[\hat{U}_{0}]\approx{\cal L}_{f}^{(2)}+{\cal L}_{1}+{\cal L}_{2},
\end{equation}
with
{\small
\begin{align}
{\cal L}_{f}^{(2)}&=\frac{g}{2}\tr(\partial_{\mu}\Phi_{f})^{2}\label{eq:app:Lf2},\\
{\cal L}_{1}&=-\frac{g}{2}\tr\left(\left[\Phi_{f},\partial_{\mu}\Phi_{f}\right]\partial_{\mu}U_{0}U_{0}^{\dagger}\right),\\
{\cal L}_{2}&=\frac{gm^{2}}{4}\sum_{r_{1}r_{2}}\Bigg[U_{0,r_{1}r_{2}}^{2}(U_{0}^{\dagger}\Phi_{f})_{r_{2}r_{1}}^{2}+(U_{0,r_{2}r_{1}}^{\ast})^{2}(\Phi_{f}U_{0})_{r_{1}r_{2}}^{2}\notag \\
&+|U_{0,r_{1}r_{2}}|^{2}\left(U_{0,r_{1}r_{2}}(U_{0}^{\dagger}\Phi_{f}^{2})_{r_{2}r_{1}}+U_{0,r_{1}r_{2}}^{\ast}(\Phi_{f}^{2}U_{0})_{r_{1}r_{2}}\right)\notag \\
&-4|U_{0,r_{1}r_{2}}|^{2}(\Phi_{f}U_{0})_{r_{1}r_{2}}(U_{0}^{\dagger}\Phi_{f})_{r_{2}r_{1}}\Bigg].
\end{align}
}

Integrating out the fast degrees of freedom, we obtain renormalization of the effective action for slow degrees of freedom:
\begin{equation}
\label{eq:app:EffectiveActionCorrection}
\delta S_{\text{eff}}[\hat{U}_{0}]\approx-\frac{1}{2}\left\langle \left\langle S_{1}^{2}\right\rangle \right\rangle _{f} + \left\langle S_{2}\right\rangle _{f}-\frac{1}{2}\left\langle \left\langle S_{2}^{2}\right\rangle \right\rangle_{f},
\end{equation}
with $\left\langle \ldots\right\rangle _{f}$ denoting an average with respect to the quadratic Lagrangian \eqref{eq:app:Lf2}, which defines the propagator for the fast modes:
\begin{equation}
\label{eq:app:Gf}
\left\langle \Phi_{f,ab}(\boldsymbol{r})\Phi_{f,cd}(\boldsymbol{r}^{\prime})\right\rangle =\left(\delta_{ad}\delta_{bc}-\frac{1}{R}\delta_{ab}\delta_{cd}\right)G_{f}(\boldsymbol{r}-\boldsymbol{r}^{\prime}),
\end{equation}
where the Fourier transform of $G_{f}$ is given by Eq.~(\ref{eq:Gf}).

The first term in Eq. \eqref{eq:app:EffectiveActionCorrection} is responsible for the anomalous dimension of the interaction operator, the second term gives rise to the one-loop renormalization of $g$ in the non-interacting theory for symmetry class AIII, and the third term produces renormalization of $g$ due to interaction. We will study these terms separately, utilizing $\epsilon=(d-1)$-expansion.

\subsection{\texorpdfstring{Interference-induced renormalization of $g$}{Interference-induced renormalization of g}}

This derivation parallels the one presented in Appendix~\ref{sec:appendix:BDIRG} for the symmetry class BDI. Here, we assume no PH symmetry in the model.
From the second order in $\mathcal{L}_1$ we obtain:
\begin{multline}
\delta S_{\text{eff}}^{(1)}\!=\!-\frac{g^{2}}{8}\!\!\int\!\! d^{d+1}\boldsymbol{r}_{1}d^{d+1}\boldsymbol{r}_{2}\Big\llangle\tr\left(\left[\Phi_{f},\partial_{\mu}\Phi_{f}\right]\partial_{\mu}U_{0}U_{0}^{\dagger}\right)_{\boldsymbol{r}_{1}}\\
\times\tr\left(\left[\Phi_{f},\partial_{\nu}\Phi_{f}\right]\partial_{\nu}U_{0}U_{0}^{\dagger}\right)_{\boldsymbol{r}_{2}}\Big\rrangle_{f}.
\end{multline}
With the center-of-mass coordinate ${\boldsymbol{r}=(\boldsymbol{r}_{1}+\boldsymbol{r}_{2})/2}$ and coordinate difference ${\boldsymbol{\rho}=\boldsymbol{r}_{1}-\boldsymbol{r}_{2}}$, utilizing Eq. \eqref{eq:app:Gf} and Wick's theorem, we obtain:
\begin{multline}
\!\!\!\!\!\left\llangle \left[\Phi_{f},\partial_{\mu}\Phi_{f}\right]_{ab,\boldsymbol{r}_{1}}\left[\Phi_{f},\partial_{\nu}\Phi_{f}\right]_{cd,\boldsymbol{r}_{2}}\right\rrangle =-2(R\delta_{ad}\delta_{bc}-\delta_{ab}\delta_{cd})\\
\times\left[\partial_{\nu}G_{f}(\boldsymbol{\rho})\partial_{\mu}G_{f}(\boldsymbol{\rho})-G(\boldsymbol{\rho})\partial_{\mu}\partial_{\nu}G(\boldsymbol{\rho})\right].
\end{multline}
Keeping only the leading term in the expansion in $\boldsymbol{\rho}$, we get:
\begin{multline}
\delta S_{\text{eff}}^{(1)}=-\frac{g^{2}R}{2}\int d^{d+1}\boldsymbol{r}\tr\left(\partial_{\mu}\hat{U}_{0}^{\dagger}\partial_{\nu}\hat{U}_{0}\right)\\
\times \int d^{d+1}\boldsymbol{\rho}\,\partial_{\mu}G_{f}(\boldsymbol{\rho})\partial_{\nu}G_{f}(\boldsymbol{\rho}).
\end{multline}
In the limit $d \to 1$ this yields:
\begin{multline}
\int d^{2}\boldsymbol{\rho}\,\partial_{\mu}G_{f}(\boldsymbol{\rho})\partial_{\nu}G_{f}(\boldsymbol{\rho})=\int\frac{d^{2}\boldsymbol{q}}{(2\pi)^{2}}\frac{q_{\mu}q_{\nu}}{g^{2}q^{4}}\\=\frac{\delta_{\mu\nu}}{4\pi g^{2}}\int_{\ell^{\prime-1}}^{\ell^{-1}}\frac{dq}{q}=\frac{\delta_{\mu\nu}}{4\pi g^{2}}\delta\ln\ell,
\end{multline}
which reproduces the structure of the bare NLSM action and gives rise to the renormalization of the dimensionless coupling constant $G = g \ell^{d-1}$ for the symmetry class AIII, Eq.~(\ref{RG-AIII}):
\begin{equation}
\frac{\delta G}{\delta\ln\ell}=\epsilon\, G-\frac{1}{4\pi},
\label{eqG12}
\end{equation}
where we have taken the limit $R \to 1$.

\subsection{Renormalization of interaction}

Using Eq. \eqref{eq:app:Gf}, we obtain:
\begin{align}
&\left\langle (\Phi_{f}U_{0})_{r_{1}r_{2}}^{2}\right\rangle _{f}=\left(1-\frac{1}{R}\right)U_{0,r_{1}r_{2}}^{2}G_{f}(0),\\
&\left\langle (\hat{U}_{0}^{\dagger}\hat{\Phi}_{f})_{r_{2}r_{1}}^{2}\right\rangle _{f}=\left(1-\frac{1}{R}\right)U_{0,r_{1}r_{2}}^{\ast2}G_{f}(0),\\
&\left\langle (\Phi_{f}^{2}U_{0})_{r_{1}r_{2}}\right\rangle _{f}=\left(R-\frac{1}{R}\right)U_{0,r_{1}r_{2}}G_{f}(0),\\
&\left\langle \left(U_{0}^{\dagger}\Phi_{f}^{2}\right)_{r_{2}r_{1}}\right\rangle =\left(R-\frac{1}{R}\right)U_{0,r_{1}r_{2}}^{\ast}G_{f}(0),\\
&\left\langle (\Phi_{f}U_{0})_{r_{1}r_{2}}(U_{0}^{\dagger}\Phi_{f})_{r_{2}r_{1}}\right\rangle =\left(1-\frac{1}{R}\left|U_{0,r_{1}r_{2}}\right|^{2}\right)G_{f}(0),
\end{align}
and, thus,
\begin{equation}
\delta{\cal L}_{\text{eff}}^{(2)} =-gm^{2}G_{f}(0)\left(R-\frac{1+R}{2}\sum_{r_{1}r_{2}}|U_{0,r_{1}r_{2}}|^{4}\right).
\label{eqG18}
\end{equation}
In the replica limit $R \to 1$, Eq.~\eqref{eqG18} reproduces the structure of the interaction term \eqref{Lint-U} in the original action.
Furthermore, we have:
\begin{equation}
G_{f}(0)\approx\frac{1}{g}\int\frac{d^{d}\boldsymbol{q}}{(2\pi)^{d}}\frac{1}{q^{2}}\approx\frac{1}{2\pi g}\int_{\ell^{\prime-1}}^{\ell^{-1}}\frac{dq}{q}=\frac{\delta\ln\ell}{2\pi g},
\end{equation}
and introducing dimensionless coupling $u \equiv m \ell$, we obtain:
\begin{equation}
\frac{\delta\ln u}{\delta\ln\ell}=1 - \frac{7}{8 \pi G},
\end{equation}
which gives the second RG equation in Eq.~\eqref{eq:RG} of the main text.

\subsection{\texorpdfstring{Renormalization of $g$ due to interactions}{Renormalization of g due to interactions}}
\label{sec:app:g-bkt-like}

We finally focus on the third term in Eq.~(\ref{eq:app:EffectiveActionCorrection}), which, after switching to the center-of-mass coordinates, reads:
\begin{equation}
\delta S_{\text{eff}}^{(3)}=-\frac{1}{2}\int d^{d+1}\boldsymbol{r}d^{d+1}\boldsymbol{\rho}\left\llangle {\cal L}_{2}(\boldsymbol{r}+\boldsymbol{\rho}/2){\cal L}_{2}(\boldsymbol{r}-\boldsymbol{\rho}/2)\right\rrangle_{f}.
\end{equation}
The corresponding term in the Lagrangian is quadratic in fast modes, so that it can be written as:
\begin{equation}
{\cal L}_{2}(\boldsymbol{r})=\frac{g m^2}{4} L_{cd}^{ab}(\boldsymbol{r})\Phi_{ab}(\boldsymbol{r})\Phi_{cd}(\boldsymbol{r}),
\end{equation}
with $L_{ab,cd}(\boldsymbol{r})$ being dimensionless polynomial in $U_0(\boldsymbol{r})$ and $U_0^\dagger(\boldsymbol{r})$ of order $(2,2)$. This allows us to perform Wick's contraction and obtain:
\begin{multline}
\delta S_{\text{eff}}^{(3)}=-\frac{g^2 m^4}{32} P_{cdc^{\prime}d^{\prime}}^{aba^{\prime}b^{\prime}}\int d^{d+1}\boldsymbol{\rho}\,G_{f}^{2}(\boldsymbol{\rho})\\
\times\int d^{d+1}\boldsymbol{r}L_{cd}^{ab}(\boldsymbol{r}+\boldsymbol{\rho}/2)L_{c^{\prime}d^{\prime}}^{a^{\prime}b^{\prime}}(\boldsymbol{r}-\boldsymbol{\rho}/2)
\end{multline}
with tensor $P$ originating from the replica structure of the fast modes propagator, see Eq.~\eqref{eq:app:Gf}.

We proceed by expanding $L$ up to second order in $\boldsymbol{\rho}$; after integrating over $\boldsymbol{r}$ by parts, we obtain the local Lagrangian density:
\begin{equation}
\delta{\cal L}_{\text{eff}}^{(3)}\simeq\frac{g^2 m^4}{64}P_{cdc^{\prime}d^{\prime}}^{aba^{\prime}b^{\prime}}\partial_{\mu}L_{cd}^{ab}\partial_{\nu}L_{c^{\prime}d^{\prime}}^{a^{\prime}b^{\prime}}\int d^{d+1}\boldsymbol{\rho}\,\rho_{\mu}\rho_{\nu}G_{f}^{2}(\boldsymbol{\rho}).
\end{equation}
Since $L$ is polynomial in $U_0$ and $U_0^\dagger$, we can rewrite derivatives explicitly and collect terms ${\propto\partial_\mu U_0^\dagger\partial_\nu U_0}$:
\begin{multline}
\delta\mathcal{L}_{\text{eff}}^{(3)}=\frac{g^{2}m^{4}}{32}\Xi_{r_{1}r_{2},r_{1}^{\prime}r_{2}^{\prime}}[\hat{U}_{0}]\partial_{\mu}U_{0,r_{1}r_{2}}^{\dagger}\partial_{\nu}\hat{U}_{0,r_{1}^{\prime}r_{2}^{\prime}}\\
\times\int d^{d+1}\boldsymbol{\rho}\,\rho_{\mu}\rho_{\nu}G_{f}^{2}(\boldsymbol{\rho}),
\end{multline}
with
\begin{equation}
\Xi_{r_{1}r_{2},r_{1}^{\prime}r_{2}^{\prime}}[\hat{U}_{0}]=P_{cdc^{\prime}d^{\prime}}^{aba^{\prime}b^{\prime}}\frac{\delta L_{cd}^{ab}}{\delta U_{0,r_{1}r_{2}}^{\dagger}}\frac{\delta L_{c^{\prime}d^{\prime}}^{a^{\prime}b^{\prime}}}{\delta U_{0,r_{1}^{\prime}r_{2}^{\prime}}}.
\end{equation}
Tensor $\Xi$ is now a polynomial of power $(3,3)$ in $U_0^\dagger$ and $U_0$.

In order to obtain the renormalization of the NLSM action, we have to split an isotropic harmonic on $\mathrm{SU}(R)$ manifold, which is equivalent to averaging of $\Xi$ over $\mathrm{SU}(R)$ group with Haar measure:
\begin{equation}
\left\langle \Xi_{r_{1}r_{2},r_{1}^{\prime}r_{2}^{\prime}}[\hat{U}_{0}]\right\rangle _{\mathrm{SU}(R)}= \Xi \cdot\delta_{r_{1}r_{2}^{\prime}}\delta_{r_{2}r_{1}^{\prime}},
\end{equation}
with some number $\Xi$. This finally leads to the renormalization of $g$:
\begin{equation}
\delta g=\frac{g^{2}m^{4}}{16}\frac{\Xi}{d+1}\int d^{d+1}\boldsymbol{\rho}\,\rho^{2}G_{f}^{2}(\boldsymbol{\rho})
\end{equation}

The objects that arise in this calculation are cumbersome, and the described procedure was implemented using the computer algebra system. As the result of this calculation, the following value for the constant $\Xi$ was obtained:
\begin{equation}
\Xi = \frac{8 (R-1)(7R+8)}{R(R+1)(R+2)}.
\end{equation}
We conclude that the renormalization of $g$ due to interaction vanishes at one-loop order in the replica limit $R \to 1$.
With Eq.~\eqref{eqG12}, we have thus derived the first RG equation in Eq.~\eqref{eq:RG} of the main text. 

\section{BKT renormalization group}
\label{App:BKT}

In this Appendix, we derive the one-loop renormalization group equations for the BKT-type Lagrangian  Eq.~\eqref{eq:BKTActionDual}. Performing the splitting into ``slow'' and ``fast'' modes $\theta=\theta^{(s)}+\theta^{(f)}$, and expanding ${\cal L}_{R}$ in the ``fast'' modes, we arrive at:
\begin{align}
{\cal L}_{R}[\theta]&={\cal L}_{R}[\theta^{(s)}]+{\cal L}_{R}^{(2)}[\theta^{(f)}]+{\cal L}_{\text{fs}}[\theta^{(s)},\theta^{(f)}],\\
{\cal L}_{\text{fs}}^{(1)}&\approx\frac{\lambda}{2}\sum_{rr^{\prime}}\sin\left[\theta_{r}^{(s)}-\theta_{r^{\prime}}^{(s)}\right]\left(\theta_{r}^{(f)}-\theta_{r^{\prime}}^{(f)}\right),\\
{\cal L}_{\text{fs}}^{(2)}&\approx\frac{\lambda}{4}\sum_{rr^{\prime}}\cos\left[\theta_{r}^{(s)}-\theta_{r^{\prime}}^{(s)}\right]\left(\theta_{r}^{(f)}-\theta_{r^{\prime}}^{(f)}\right)^{2},
\end{align}
with ${\cal L}_{R}^{(2)}[\theta^{(f)}]$ being the quadratic part of Eq.~\eqref{eq:BKTActionDual}, which sets the following propagator for the ``fast'' modes:
\begin{equation}
\left\langle \theta_{r}^{(f)}\theta_{r^{\prime}}^{(f)}\right\rangle _{\boldsymbol{q}}=G_{f}(\boldsymbol{q})\left(\delta_{rr^{\prime}}-\frac{1}{R}\right),\quad G_{f}(\boldsymbol{q})=\frac{4\pi^{2}g}{q^{2}}
\end{equation}
Here, the number of replicas appears from the constraint $$\sum_r \theta_r^{(f)} = 0.$$

The renormalization of $\lambda$ arises from the term ${\cal L}_{\text{fs}}^{(2)}$ in the first order of perturbation theory, and reads:
\begin{equation}
\delta\lambda=-\frac{\lambda}{2}\left\langle \left[\theta_{r}^{(f)}-\theta_{r^{\prime}}^{(f)}\right]^{2}\right\rangle \underset{r\neq r^{\prime}}{=}-\lambda G_{f}(0).
\end{equation}
The Green's function of fast modes at coinciding arguments gives the logarithm:
\begin{equation}
G_{f}(0)=\int_{\sim\ell^{\prime-1}}^{\sim\ell^{-1}}\frac{d^{2}\boldsymbol{q}}{4\pi^{2}}G_{f}(\boldsymbol{q})=2\pi g\, \delta\ln\ell.
\end{equation}
Introducing the dimensionless coupling $\kappa(\ell) = \lambda(\ell) \ell^2$, we arrive at the first RG equation in Eq.~\eqref{eq:BKT:RG}.

The renormalization of $g$ arises in the second order of perturbation theory:
\begin{align}
&\delta S_{R}=-\frac{\lambda^{2}}{32}\int d^{2}\boldsymbol{r}_{1}d^{2}\boldsymbol{r}_{2} 
\notag\\
&\times\!\!\sum_{r_{1}r_{1}^{\prime}r_{2}r_{2}^{\prime}}\!\!\!\cos\left[\theta_{r_{1}}^{(s)}(\boldsymbol{r}_{1})-\theta_{r_{1}^{\prime}}^{(s)}(\boldsymbol{r}_{1})\right]\cos\left[\theta_{r_{2}}^{(s)}(\boldsymbol{r}_{2})-\theta_{r_{2}^{\prime}}^{(s)}(\boldsymbol{r}_{2})\right]\notag\\
&\ \ \times\left\langle \left\langle \left[\theta_{r_{1}}^{(f)}(\boldsymbol{r}_{1})-\theta_{r_{1}^{\prime}}^{(f)}(\boldsymbol{r}_{1})\right]^{2}\left[\theta_{r_{2}}^{(f)}(\boldsymbol{r}_{2})-\theta_{r_{2}^{\prime}}^{(f)}(\boldsymbol{r}_{2})\right]^{2}\right\rangle \right\rangle. 
\end{align}
Repeating the steps similar to calculation in Sec.~\ref{sec:app:g-bkt-like}, we switch to the center-of-mass coordinate $\boldsymbol{r} = (\boldsymbol{r}_1 + \boldsymbol{r}_2) / 2$ and the relative-motion coordinate $\boldsymbol{\rho} = \boldsymbol{r}_1 - \boldsymbol{r}_2$, expand up to the second order in $\boldsymbol{\rho}$, and arrive at the following renormalization of the Lagrangian:
\begin{align}
    \delta{\cal L}_{R}&=\frac{\lambda^{2}}{64}\sum_{r_{1}r_{1}^{\prime}r_{2}r_{2}^{\prime}}\sin\left[\theta_{r_{1}}^{(s)}-\theta_{r_{1}^{\prime}}^{(s)}\right]\sin\left[\theta_{r_{2}}^{(s)}-\theta_{r_{2}^{\prime}}^{(s)}\right]\notag\\    &\times\partial_{\mu}\left[\theta_{r_{1}}^{(s)}-\theta_{r_{1}^{\prime}}^{(s)}\right]\partial_{\nu}\left[\theta_{r_{2}}^{(s)}-\theta_{r_{2}^{\prime}}^{(s)}\right]\notag\\
    &\times\int d^{2}\boldsymbol{\rho}\,\rho_{\mu}\rho_{\nu}\left\langle \left\langle \left[\theta_{r_{1}}^{(f)}-\theta_{r_{1}^{\prime}}^{(f)}\right]^{2}\left[\theta_{r_{2}}^{(f)}-\theta_{r_{2}^{\prime}}^{(f)}\right]^{2}\right\rangle \right\rangle,
\end{align}
where all ``slow'' fields are now taken at the same coordinate. To extract the renormalization of $(\partial_\mu \theta)^2$, we separate the constant term from the product of two sines, equivalent to averaging over corresponding phases:
\begin{equation}
\overline{\sin\left[\theta_{r_{1}}^{(s)}-\theta_{r_{1}^{\prime}}^{(s)}\right]\sin\left[\theta_{r_{2}}^{(s)}-\theta_{r_{2}^{\prime}}^{(s)}\right]}=\frac{1}{2}\left(\delta_{r_{1}r_{2}}\delta_{r_{1}^{\prime}r_{2}^{\prime}}-\delta_{r_{1}r_{2}^{\prime}}\delta_{r_{1}^{\prime}r_{2}}\right)
\end{equation}

The Wick's contraction of fast fields yields:
\begin{align}
\left\langle \left\langle \dots \right\rangle \right\rangle &=2\,G^{2}(\boldsymbol{\rho})\Bigg(\left[\delta_{r_{1}r_{2}}+\delta_{r_{1}r_{2}^{\prime}}+\delta_{r_{1}^{\prime}r_{2}}+\delta_{r_{1}^{\prime}r_{2}^{\prime}}\right]\notag\\
&+2\left[\delta_{r_{1}r_{2}}\delta_{r_{1}^{\prime}r_{2}^{\prime}}+\delta_{r_{1}r_{2}^{\prime}}\delta_{r_{1}^{\prime}r_{2}}\right]\notag\\
&-2\left[\tilde{\delta}_{r_{1}r_{2}r_{1}^{\prime}}+\delta_{r_{1}r_{1}^{\prime}r_{2}^{\prime}}+\delta_{r_{1}r_{2}r_{2}^{\prime}}+\delta_{r_{1}^{\prime}r_{2}r_{2}^{\prime}}\right]\Bigg),
\end{align}
with the generalized Kronecker symbol $\tilde{\delta}_{abc}$ with three indices equal to unity only when all three indices coincide. Finally, performing the summation over replica indices, we arrive at the following renormalization:
\begin{equation}
\delta\left(g^{-1}\right)=\frac{\pi^{2}\lambda^{2}}{16}R\int d^{2}\boldsymbol{\rho}\,\rho^{2}G_{f}^{2}(\boldsymbol{\rho}).
\end{equation}
The integral can be estimated as:
\begin{equation}
\int d^{2}\boldsymbol{\rho}\rho_{\mu}\rho_{\nu}G^{2}(\boldsymbol{\rho})\sim g^{2}\ell^{4}\delta\ln\ell,
\end{equation}
which, for $R=1$, leads to the second equation in Eq.~\eqref{eq:BKT:RG}.

The RG equations (\ref{eq:BKT:RG}) give rise to the following relation between $\kappa$ and $g$: \begin{equation}
    \frac{\kappa^2}{2} + \frac{3 \pi g - 2}{3 g^3} = c,   
    \label{kappa-g}
\end{equation}
where the constant $c$ is determined from the initial condition at $\ell\sim\ell_\text{int}$. The BKT separatrix in the $g$-$\kappa$ plane satisfies Eq.~\eqref{kappa-g} with $c=\pi^3/3$. 
Below the separatrix, integration of the RG equations (\ref{eq:BKT:RG}) yields a line of fixed points for $g$ with $g(\ell\to\infty)>g_c=1/\pi$.
The value of $g_\text{BKT}(\infty)=g(\ell\to\infty)$ is given by the solution of Eq.~\eqref{kappa-g} for $\kappa=0$. 
The flow above the separatrix corresponds to ``localization'', $g\to 0$  (area law for charge fluctuations).

\section{Details on numerical methods}
\label{sec:appendix:numerical-details}

\subsection{Time-dependent variational principle}
\label{appendix:TDVP}

The unitary evolution in numerical TDVP~\cite{tdvp_for_ql,tdvp_mps} data is calculated using the controlled bond-dimension expansion algorithm (CBE-TDVP) proposed in Refs.~\cite{cbe_tdvp,cbe_dmrg}, with modifications according to Ref.~\cite{comment_on_cbe}. Specifically, we use the randomized singular value decomposition (SVD) procedure described in Ref.~\cite{comment_on_cbe} instead of the sequence of SVDs suggested in Refs~\cite{cbe_tdvp, cbe_dmrg} to find the expanded basis at each evolution step. Particle-number conservation is utilized to deal only with charge-conserving blocks of the sparse tensors~\cite{tdvp_u_1_conservation}.
Projective on-site measurements are trivially implemented by applying the bond-dimension one matrix product operators corresponding to click and no-click outcomes according to the Born rule.
At time zero, the system is initialized in a product state in the site space, such that the bonds of the initial MPS are trivial. Usually, in such a situation time evolution with the TDVP is achieved by using the two-site algorithm (2TDVP) described in Ref.~\cite{tdvp_mps} until the desired maximum bond dimensions are reached, and continuing the evolution from then on with the one-site algorithm~\cite{tdvp_mps}, restricting the evolution to the space of MPS with these fixed bond dimensions. However, if a projective measurement is applied to a lattice site, this can result in the vanishing of some of the singular values on the adjacent bonds  (as is obvious in the single-particle case). In this situation, it is desirable to truncate singular values that are  numerically zero and regrow the bond dimension afterward. For this reason, the CBE-TDVP is well suited for our purposes, as it allows for a flexible bond dimension while being computationally more efficient than 2TDVP.  In addition, the CBE-TDVP provides a rough measure for the error due to bond truncation if a maximum bond dimension is imposed, by summing up the truncated weight of the state~\cite{cbe_dmrg}.

For the integration of the TDVP equations we use a time step of \(\delta t=10^{-1}\). While this is not sufficient to simulate exactly each individual quantum trajectory, we only need that the ensemble of quantum trajectories is faithfully reproduced, as we are interested in ensemble-averaged values of observables. We have checked that such average values are correctly reproduced with this time step by performing various tests, including benchmarking to available exact results (see Sec.~\ref{sec:numerics:tdvp}).

For each bond expansion step, we allow for \(k=0.3 \chi\) extra singular vectors, expanding a bond of dimension \(\chi\). This is somewhat larger than \(k = 0.1 \chi\) discussed in Ref.~\cite{comment_on_cbe} in the context of density matrix renormalization group. We found the additional SVD necessary and kept the suggested oversampling parameter of \(p \leq 10\).
If a maximum bond dimension \(\chi_{\rm max}\) is imposed, we still allow for an expansion of any bond beyond \(\chi_{\rm max}\) within these expansion parameters at every time step. If the dimension of the expanded bond exceeds \(\chi_{\rm max}\) it is truncated back to \(\chi_{\rm max}\). This is computationally more expensive but yields the advantage that we can estimate the error due to the restriction of the bond dimension by adding up the weight from discarded singular values. 

\begin{figure}
    \centering
    \includegraphics[width=\columnwidth]{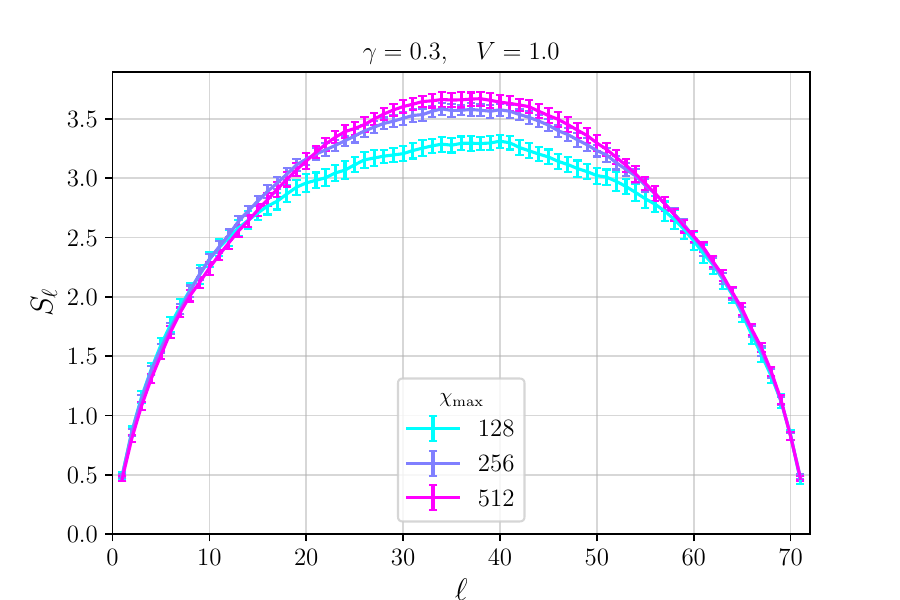}
    \caption{Entanglement entropy $S_\ell$ in a system of size $L = 72$, as a function of cut position $\ell$, for bond dimensions $\chi_{\text{max}} = 128, 256$, and 512, the interaction strength $V = 1$ and the measurement rate $\gamma = 0.3$.}
    \label{fig:entE:bonddim}
\end{figure}

The convergence with respect to the bond dimension is shown in Fig.~\ref{fig:entE:bonddim} where the entropy $S_\ell$ is presented for our largest system size $L=72$, strongest interaction $V=1$, and the weakest measurement rate $\gamma=0.3$, i.e., for the ``worst case'' from the point of view of MPS simulations.

The MPS representation of the state gives cheap access to all R{\'e}nyi entropies, which are calculated from the singular spectra of the corresponding cuts. The second cumulant and covariance can be calculated from expectation values of bond dimension two- and three matrix product operators respectively. The computational effort for calculating these observables is negligible compared to the unitary time evolution which costs \(O(L k\chi^2 t_{\rm max} / \text{min}\{\delta t, t_\gamma\})\)~\cite{cbe_tdvp,comment_on_cbe} where \(t_\gamma = 1 / (L\gamma)\) is the average time between two measurements.

Our CBE-TDVP implementation was written from scratch in the \texttt{Julia} programming language~\cite{bezanson2017julia,JSSv098i16,Distributions.jl-2019,resumablefunctions-jl,PlotsJL} using \texttt{TensorOperations.jl}~\cite{TensorOperations.jl} for pairwise contraction of dense tensor blocks and \texttt{EinExprs.jl}~\cite{sanchezramirez2024einexprscontractionpathstensor} for contraction path optimization.

\subsection{Time-dependent Hartree-Fock method}

The TDHF method~\cite{tdhf_eq_ref,tdhf_wgk,tdhf2} describes the time evolution of the lesser Green's function
\begin{align}
    G^<_{i, j}(t, t') \equiv i \tr\left[\rho_0 c_j^\dagger(t') c_i(t) \right]
\end{align}
in the site space, with initial density matrix \(\rho_0\) and fermionic creation and annihilation operators \(c^\dagger\) and \(c\), by the equations of motion
\begin{align}
    i \partial_t G_{i, j}^<(t, t') = \sum_k \left[H_{0 i, k} - \Sigma^{\rm HF}_{i, k}(t) \right] G_{k, j}^<(t, t'), \label{eq:tdhf}\\
    \Sigma^{\rm HF}_{i, k}(t) \equiv -i \delta_{i, k} \sum_l V_{i, l} G_{l, l}^<(t, t) + i V_{i, k}G^<_{i, k}(t, t).
\end{align}
Here \(V_{i, j}\) is the coupling constant between densities on sites \(i\) and \(j\).
We integrate these non-linear equations numerically with discretization time-steps of a maximum size of \(\delta t = 0.1\).

The entanglement entropy \(\mathcal{S}\), the charge cumulant \(\mathcal{C}^{(2)}\), and the charge covariance $-G$ are calculated from the lesser Green's function at same time arguments~\cite{fgs}, such that it is sufficient to time evolve \(G_{i, j}^<(t, t)\). Since \(H_0\) and \(\Sigma^{\rm HF}\) are tri-diagonal for nearest-neighbor hopping and interactions, unitary evolution can thus be performed at a cost of \(O(L^2 t_{\rm max} / \text{min}\{t_\gamma, \delta t\})\) operations. The measurement updates of the Green's function are performed in the same way as in the non-interacting case~\cite{Poboiko2023a} and contribute \(O(L^2 t_{\rm max} / t_{\gamma})\) operations.

The TDHF implementation was written in \texttt{Python}, using \texttt{NumPy}~\cite{harris2020array} for fast linear algebra, and a \texttt{SciPy}~\cite{2020SciPy-NMeth} interface to \texttt{ODEPACK}~\cite{osti_145724} to integrate Eq.~\eqref{eq:tdhf}.

\bibliography{refs}

\end{document}